\documentclass[runningheads]{llncs}

\usepackage{xcolor}
\definecolor{linkcolor}{RGB}{120,230,120}

\usepackage[LGR,T1]{fontenc}
\usepackage[utf8]{inputenc} 

\newcommand{\textgreek}[1]{\begingroup\fontencoding{LGR}\selectfont#1\endgroup}

\usepackage{fancyvrb}
\usepackage{mathpartir}

\usepackage{stmaryrd} %
\newcommand{\banana}[1]{\llparenthesis\, #1 \,\rrparenthesis}

\DefineVerbatimEnvironment%
  {core}{Verbatim}
  {xleftmargin=\mathindent}

\usepackage{subcaption}
\usepackage{caption}
\usepackage{tikz}
\usepackage{tikz-cd}
\usepackage{csvsimple,booktabs, siunitx, array}
\usepackage{sansmath}

\usepackage{mathtools}

\makeatletter
\@ifundefined{lhs2tex.lhs2tex.sty.read}%
  {\@namedef{lhs2tex.lhs2tex.sty.read}{}%
   \newcommand\SkipToFmtEnd{}%
   \newcommand\EndFmtInput{}%
   \long\def\SkipToFmtEnd#1\EndFmtInput{}%
  }\SkipToFmtEnd

\newcommand\ReadOnlyOnce[1]{\@ifundefined{#1}{\@namedef{#1}{}}\SkipToFmtEnd}
\usepackage{amstext}
\usepackage{amssymb}
\usepackage{stmaryrd}
\DeclareFontFamily{OT1}{cmtex}{}
\DeclareFontShape{OT1}{cmtex}{m}{n}
  {<5><6><7><8>cmtex8
   <9>cmtex9
   <10><10.95><12><14.4><17.28><20.74><24.88>cmtex10}{}
\DeclareFontShape{OT1}{cmtex}{m}{it}
  {<-> ssub * cmtt/m/it}{}

\DeclareFontShape{OT1}{cmtt}{bx}{n}
  {<5><6><7><8>cmtt8
   <9>cmbtt9
   <10><10.95><12><14.4><17.28><20.74><24.88>cmbtt10}{}
\DeclareFontShape{OT1}{cmtex}{bx}{n}
  {<-> ssub * cmtt/bx/n}{}

\newcommand{\Conid}[1]{\mathit{#1}}
\newcommand{\Varid}[1]{\mathit{#1}}
\newcommand{\anonymous}{\kern0.06em \vbox{\hrule\@width.5em}}
\newcommand{\plus}{\mathbin{+\!\!\!+}}
\newcommand{\bind}{\mathbin{>\!\!\!>\mkern-6.7mu=}}
\newcommand{\sequ}{\mathbin{>\!\!\!>}}
\renewcommand{\leq}{\leqslant}
\renewcommand{\geq}{\geqslant}
\usepackage{polytable}

\@ifundefined{mathindent}%
  {\newdimen\mathindent\mathindent\leftmargini}%
  {}%

\def\resethooks{%
  \global\let\SaveRestoreHook\empty
  \global\let\ColumnHook\empty}
\newcommand*{\savecolumns}[1][default]%
  {\g@addto@macro\SaveRestoreHook{\savecolumns[#1]}}
\newcommand*{\restorecolumns}[1][default]%
  {\g@addto@macro\SaveRestoreHook{\restorecolumns[#1]}}
\newcommand*{\aligncolumn}[2]%
  {\g@addto@macro\ColumnHook{\column{#1}{#2}}}

\resethooks

\newcommand{\onelinecommentchars}{\quad-{}- }
\newcommand{\commentbeginchars}{\enskip\{-}
\newcommand{\commentendchars}{-\}\enskip}

\newcommand{\visiblecomments}{%
  \let\onelinecomment=\onelinecommentchars
  \let\commentbegin=\commentbeginchars
  \let\commentend=\commentendchars}

\newcommand{\invisiblecomments}{%
  \let\onelinecomment=\empty
  \let\commentbegin=\empty
  \let\commentend=\empty}

\visiblecomments

\newlength{\blanklineskip}
\newcommand{\hsindent}[1]{\quad}%
\let\hspre\empty
\let\hspost\empty

\EndFmtInput
\makeatother
\ReadOnlyOnce{polycode.fmt}%
\makeatletter

\newcommand{\hsnewpar}[1]%
  {{\parskip=0pt\parindent=0pt\par\vskip #1\noindent}}

\newcommand{\hscodestyle}{}

\newcommand{\sethscode}[1]%
  {\expandafter\let\expandafter\hscode\csname #1\endcsname
   \expandafter\let\expandafter\endhscode\csname end#1\endcsname}

  {\par\noindent
   \advance\leftskip\mathindent
   \hscodestyle
   \let\\=\@normalcr
   \let\hspre\(\let\hspost\)%
   \pboxed}%
  {\endpboxed\)%
   \par\noindent
   \ignorespacesafterend}

  {\hsnewpar\abovedisplayskip
   \advance\leftskip\mathindent
   \hscodestyle
   \let\hspre\(\let\hspost\)%
   \pboxed}%
  {\endpboxed%
   \hsnewpar\belowdisplayskip
   \ignorespacesafterend}

  {\hsnewpar\abovedisplayskip
   \advance\leftskip\mathindent
   \hscodestyle
   \let\\=\@normalcr
   \(\pboxed}%
  {\endpboxed\)%
   \hsnewpar\belowdisplayskip
   \ignorespacesafterend}

\newcommand{\plainhs}{\sethscode{plainhscode}}

\plainhs

  {\hsnewpar\abovedisplayskip
   \advance\leftskip\mathindent
   \hscodestyle
   \let\\=\@normalcr
   \(\parray}%
  {\endparray\)%
   \hsnewpar\belowdisplayskip
   \ignorespacesafterend}

  {\parray}{\endparray}

  {\(\parray}{\endparray\)}

\def\codeframewidth{\arrayrulewidth}
\RequirePackage{calc}

  {\parskip=\abovedisplayskip\par\noindent
   \hscodestyle
   \arrayrulewidth=\codeframewidth
   \tabular{@{}|p{\linewidth-2\arraycolsep-2\arrayrulewidth-2pt}|@{}}%
   \hline\framedhslinecorrect\\{-1.5ex}%
   \let\endoflinesave=\\
   \let\\=\@normalcr
   \(\pboxed}%
  {\endpboxed\)%
   \framedhslinecorrect\endoflinesave{.5ex}\hline
   \endtabular
   \parskip=\belowdisplayskip\par\noindent
   \ignorespacesafterend}

\newcommand{\framedhslinecorrect}[2]%
  {#1[#2]}

  {\(\def\column##1##2{}%
   \let\>\undefined\let\<\undefined\let\\\undefined
   \newcommand\>[1][]{}\newcommand\<[1][]{}\newcommand\\[1][]{}%
   \def\fromto##1##2##3{##3}%
   }{\) }%

  {\let\orighscode=\hscode
   \let\origendhscode=\endhscode
   \def\endhscode{\def\hscode{\endgroup\def\@currenvir{hscode}\\}\begingroup}
   \orighscode\def\hscode{\endgroup\def\@currenvir{hscode}}}%
  {\origendhscode
   \global\let\hscode=\orighscode
   \global\let\endhscode=\origendhscode}%

\makeatother
\EndFmtInput
\ReadOnlyOnce{forall.fmt}%
\makeatletter

\let\HaskellResetHook\empty
\newcommand*{\AtHaskellReset}[1]{%
  \g@addto@macro\HaskellResetHook{#1}}
\newcommand*{\HaskellReset}{\HaskellResetHook}

\newcommand\hsforall{\global\let\hsdot=\hsperiodonce}
\newcommand*\hsperiodonce[2]{#2\global\let\hsdot=\hscompose}
\newcommand*\hscompose[2]{#1}

\AtHaskellReset{\global\let\hsdot=\hscompose}

\HaskellReset

\makeatother
\EndFmtInput

\newcommand{\pair}[2]{\langle #1 , #2 \rangle}

\newcommand{\catname}[1]{{\normalfont\textbf{#1}}}
\newcommand{\disccat}[1]{|#1|}

\spnewtheorem*{corollary*}{Corollary}{\normalfont\bfseries}{\itshape}
\spnewtheorem*{theorem*}{Theorem}{\normalfont\bfseries}{\itshape}
\spnewtheorem*{lemma*}{Lemma}{\normalfont\bfseries}{\itshape}
\spnewtheorem{recscheme}{Recursion Scheme}{\normalfont\bfseries}{}

\usepackage{hyperref}

\usepackage[modulo, mathlines]{lineno}

\usepackage{bbding}

\begin{document}

\title{Fantastic Morphisms and Where to Find Them
\thanks{This work was funded by EPSRC Grant EP/S028129/1.}}
\subtitle{A Guide to Recursion Schemes}
\titlerunning{Fantastic Morphisms and Where to Find Them}

\author{
 Zhixuan Yang {\href{mailto:s.yang20@imperial.ac.uk}{\Envelope}}\ \orcidID{0000-0001-5573-3357}
 \and
 Nicolas Wu \orcidID{0000-0002-4161-985X}
}

\institute{Imperial College London, United Kingdom \\ \email{\{s.yang20,n.wu\}@imperial.ac.uk}}

\authorrunning{Zhixuan Yang and Nicolas Wu}

\maketitle

\begin{abstract}

\emph{Structured recursion schemes} have been widely used in constructing,
optimizing, and reasoning about programs over inductive and coinductive
datatypes. 
Their plain forms, \emph{catamorphisms} and \emph{anamorphisms}, are restricted
in expressivity.
Thus many generalizations have been proposed, which further led to several
unifying frameworks of structured recursion schemes.
However, the existing work on unifying frameworks typically focuses on 
the categorical foundation, and thus is perhaps inaccessible to practitioners
who are willing to apply recursion schemes in practice but are not versed in
category theory.
To fill this gap, this expository paper introduces structured recursion schemes
from a practical point of view:
a variety of recursion schemes are motivated and explained in contexts of 
concrete programming examples.
The categorical duals of these recursion schemes are also explained.

\end{abstract}

\section{Introduction}

Among the wilderness of recursive functions, there exists a taxonomy of tame
functions, each with its own character and behaviour that is more predictable
than the other wilder functions.
The first function to be tamed was the \emph{catamorphism}, so-named by
Meertens in 1988~\cite{Mee88First} who wanted to capture and more closely study
the unique function that arises as a homomorphism from an initial algebra.
Such functions had previously been studied in the context of category theory,
but this identification marked the beginning of the appreciation of
such functions as valuable companions in the menagerie of functional
programmers. They were appreciated for their many benefits to programming: by
expressing recursive programs as a recursion schemes such as a catamorphism,
the structure of a program is made obvious; the recursion is ensured to
terminate; and the program can be reasoned about using the calculational
properties.

These benefits motivated a whole research agenda concerned with identifying
and classifying \emph{structured recursion schemes} that capture the pattern of
many other recursive functions that did not quite fit as catamorphisms.  Just
as with catamorphisms, these structured recursion schemes attracted attention
since they make termination or productivity manifest, and enjoy many useful
calculational properties which would otherwise have to be established afresh
for each new application.

In the early days, and in keeping the Bird-Meertens formalism, also known as Squiggol
due to its lavish use of squiggly notation, the identification of a new species
came both with an exotic name, as well as exotic notation to describe such
recursion schemes. Soon there came a whole zoo of other interesting species,
and this paper attempts to document its main inhabitants.

\subsection{Diversification}

The first variation on the catamorphisms was
paramorphisms~\cite{Mee92Par}, about which Meertens talked at
the 41st IFIP Working Group 2.1 (WG2.1) meeting in Burton, UK (1990).
Paramorphisms describe recursive functions in which the body of structured
recursion has access to not only the (recursively computed) sub-results of the
input, but also the original subterms of the input.

Then came the zoo of morphisms. \emph{Mutumorphisms}~\cite{Fokkinga90:Tupling},
which are pairs of mutually recursive functions; \emph{zygomorphisms}~\cite{Mal90Alg},
which consist of a main recursive function and an auxiliary one on which it
depends;
\emph{monadic catamorphisms}~\cite{Fok94Monadic}, which are recursive functions that
also cause computational effects;
so-called \emph{generalized folds}~\cite{BiP98Gen}, which use polymorphic recursion
to handle nested datatypes;
\emph{histomorphisms}~\cite{UuV99Pri}, in which the body has access to the
recursive images of all subterms, not just the immediate ones;
and then there were \emph{generic accumulations}~\cite{Par02Gen}, which keep
intermediate results in additional parameters for later stages in the
computation.

While catamorphisms focused on terminating programs based on initial algebra,
the theory also generalized in the dual direction: \emph{anamorphisms}. These
describe productive programs based on final coalgebras, that is, programs that
progressively output structure, perhaps indefinitely.
As variations on anamorphisms, there are \emph{apomorphisms}~\cite{VeU98Fun}, which may
generate subterms monolithically rather than
step by step; \emph{futumorphisms}~\cite{UuV99Pri}, which may generate
multiple levels of a subterm in a single step, rather than just one;
and many other anonymous schemes that dualize better known
inductive patterns of recursion. %

Recursion schemes that combined the features of inductive and coinductive
datatypes were also considered. A \emph{hylomorphism}~\cite{MFP91Fun} arises
when an anamorphism is followed by a catamorphism, and a
\emph{metamorphism}~\cite{Gibbons05Meta}
is when they are the other way around. A more sophisticated recursion scheme gives
\emph{dynamorphisms}~\cite{KV06Rec} which encodes dynamic programming algorithms, where a
lookup table is coinductively constructed in an inductive computation over the
input.

\subsection{Unification}

The many divergent generalizations of catamorphisms can be bewildering
to the uninitiated, and there have been attempts to unify them.  One
approach is the identification of recursion schemes from comonads (\textsc{rsfc}s for
short) by~Uustalu et al.\ \cite{UVP01Rec}. Comonads capture the general
idea of `evaluation in context'~\cite{Uustalu&Vene2008:Comonadic}, and
this scheme makes contextual information available to the body of the recursion.
It was used to subsume both zygomorphisms and histomorphisms.

Another attempt by~Hinze \cite{Hin12Adj} used adjunctions as the common
thread.  Adjoint folds arise by inserting a left adjoint functor into
the recursive characterization, thereby adapting the form of the
recursion; they subsume accumulating folds, mutumorphisms, zygomorphisms, and
generalized folds.
Later, it was observed that adjoint folds could be used to
subsume \textsc{rsfc}s~\cite{HWG13USRS}.

Thus far, the unifications had dealt largely with generalizations of
catamorphisms and anamorphisms separately. The job of putting combinations of
these together and covering complex beasts such as dynamorphisms was
first achieved by Hinze and Wu \cite{HiWu13Histo}, which was then generalized
by~\mbox{Hinze et al.'s} \emph{conjugate hylomorphisms} \cite{HWG15Conj}, which
WG2.1 dubbed \emph{mamamorphisms}.
This worked by viewing all recursion schemes as specialized forms of
hylomorphisms, and showing that they are well-defined hylomorphisms using
adjunctions and conjugate natural transformations.

\subsection{Overview}
The existing literature \cite{HWG13USRS,HWG15Conj,Uustalu&Vene2008:Comonadic} on
unifying accounts to structured recursion schemes has focused on the
categorical foundation of recursion schemes rather than their motivations or
applications, and thus is perhaps not quite useful for practitioners who would
like to learn about recursion schemes and apply them in practice.
To fill the gap, this paper introduces the zoo of recursion schemes by putting 
them in programming contexts.
This paper provides a survey of many recursion schemes that have
been explored, and is organized as follows.

\begin{itemize}
\item \autoref{sec:background} explains the idea of modelling (co)inductive
datatypes as fixed points of functors, which makes generic recursion schemes
possible. 

\item \autoref{sec:cata} explains the three fundamental recursion schemes:
\emph{catamorphisms}, which compute values by consuming inductive data; 
\emph{anamorphisms}, which build coinductive data from values;
and their common generalization, \emph{hylomorphisms}, which build data
from values and consume them.

\item \autoref{sec:curry} introduces structured recursion with an accumulating
parameter.

\item \autoref{sec:mutu} is about mutual recursion on inductive datatypes,
known as \emph{mutumorphisms}, and their duals \emph{comutumorphisms}, which
build mutually defined coinductive datatypes from a single value.

\item \autoref{sec:para} talks about primitive recursion, known as
\emph{paramorphisms}, featuring the ability to access both the original subterms 
and the corresponding output in the recursive function.
Their corecursive counterpart, \emph{apomorphisms}, and a generalization,
\emph{zygomorphisms}, are also shown.

\item \autoref{sec:histo} discusses the so-called course-of-values recursion,
\emph{histomorphisms}, featuring the ability to access the results of all direct
and indirect subterms in the body of recursive function, which is typically
necessary in dynamic programming.
Several related schemes, \emph{futumorphisms}, \emph{dynamorphisms}, and
\emph{chronomorphisms} are briefly discussed.

\item \autoref{sec:monadic} introduces recursion schemes that 
cause computational effects.

\item \autoref{sec:GADTs} explains recursion schemes on nested datatypes and 
\textsc{gadt}s.

\item \autoref{sec:calc} briefly demonstrates how one can do equational reasoning
about programs using calculational properties of recursion schemes.

\item Finally, \autoref{sec:more} discusses two general recipes for finding more recursion 
schemes and concludes.
\end{itemize}

The recursion schemes that we will see in this paper are summarized in \autoref{tab}.
Sections \ref{sec:cata}--\ref{sec:GADTs} 
are loosely ordered by their complexity, rather than by the order in which they
appeared in the literature, and these sections are mutually independent so can
be read in an arbitrary order.
A common pattern in these sections is that we start with a concrete programming
example, from which we distil a recursion scheme, followed by more examples.
Where appropriate, we also consider their dual corecursion scheme and
hylomorphic generalization.

{
\hypersetup{linkbordercolor={linkcolor}}
\begin{table}[!ht]
\renewcommand{\arraystretch}{1.2}
\setlength{\tabcolsep}{0.5em}
\caption{Recursion schemes explored in this paper}
\label{tab}
\center
\begin{tabular}{@{}p{2.3cm} p{4.9cm} p{4.4cm} @{}}
\toprule
Scheme & Type Signature & Usage  \\ \midrule
\hyperref[scm:cata]{Catamorphism}  & \ensuremath{(\Varid{f}\;\Varid{a}\to \Varid{a})\to \mu\;\Varid{f}\to \Varid{a}} & Consume inductive data\\ 
\hyperref[scm:ana]{Anamorphism}    & \ensuremath{(\Varid{c}\to \Varid{f}\;\Varid{c})\to \Varid{c}\to \nu\;\Varid{f}} & Generate coinductive data\\ 
\hyperref[scm:hylo]{Hylomorphism}  & \ensuremath{(\Varid{f}\;\Varid{a}\to \Varid{a})\to (\Varid{c}\to \Varid{f}\;\Varid{c})\to \Varid{c}\to \Varid{a}} & Generate then consume data\\ \midrule
\hyperref[scm:accu]{Accumulation} & 
\ensuremath{(\forall \Varid{x}\hsforall \hsdot{\circ }{.}\Varid{f}\;\Varid{x}\to \Varid{p}\to \Varid{f}\;(\Varid{x},\Varid{p}))\to } \newline\phantom{m} \ensuremath{(\Varid{f}\;\Varid{a}\to \Varid{p}\to \Varid{a})\to \mu\;\Varid{f}\to \Varid{p}\to \Varid{a}}
 & Recursion with an accumulating parameter \\ \midrule
\hyperref[scm:mutu]{Mutumorphism} & \ensuremath{(\Varid{f}\;(\Varid{a},\Varid{b})\to \Varid{a})\to (\Varid{f}\;(\Varid{a},\Varid{b})\to \Varid{b})}\newline\phantom{m}\ensuremath{\to (\mu\;\Varid{f}\to \Varid{a},\mu\;\Varid{f}\to \Varid{b})} & Mutual recursion on inductive data \\
\hyperref[scm:comutu]{Comutumorphism} & \ensuremath{(\Varid{c}\to \Varid{f}\;\Varid{c}\;\Varid{c})\to (\Varid{c}\to \Varid{g}\;\Varid{c}\;\Varid{c})} \newline\phantom{m} \ensuremath{\to \Varid{c}\to (\nu_1\;\Varid{f}\;\Varid{g},\nu_2\;\Varid{f}\;\Varid{g})}  & Generate mutually defined coinductive data \\ \midrule
\hyperref[scm:para]{Paramorphism} & \ensuremath{(\Varid{f}\;(\mu\;\Varid{f},\Varid{a})\to \Varid{a})\to \mu\;\Varid{f}\to \Varid{a}} & Primitive recursion, i.e.\ access to original input \\
\hyperref[scm:apo]{Apomorphism} & \ensuremath{(\Varid{c}\to \Varid{f}\;(\Conid{Either}\;(\nu\;\Varid{f})\;\Varid{c}))}\newline\phantom{m}\ensuremath{\to \Varid{c}\to \nu\;\Varid{f}} & Early termination of generation \\ 
\hyperref[scm:zygo]{Zygomorphism} & \ensuremath{(\Varid{f}\;(\Varid{a},\Varid{b})\to \Varid{a})\to (\Varid{f}\;\Varid{b}\to \Varid{b})}\newline\phantom{m}\ensuremath{\to \mu\;\Varid{f}\to \Varid{a}} & Recursion with auxiliary information \\ \midrule
\hyperref[scm:histo]{Histomorphism} & \ensuremath{(\Varid{f}\;(\Conid{Cofree}\;\Varid{f}\;\Varid{a})\to \Varid{a})\to \mu\;\Varid{f}\to \Varid{a}} & Access to all sub-results  \\
\hyperref[scm:dyna]{Dynamorphism} & \ensuremath{(\Varid{f}\;(\Conid{Cofree}\;\Varid{f}\;\Varid{a})\to \Varid{a})}\newline\phantom{m} \ensuremath{\to (\Varid{c}\to \Varid{f}\;\Varid{c})\to \Varid{c}\to \Varid{a}} & Dynamic programing  \\
\hyperref[scm:futu]{Futumorphism} & \ensuremath{(\Varid{c}\to \Varid{f}\;(\Conid{Free}\;\Varid{f}\;\Varid{c}))\to \Varid{c}\to \nu\;\Varid{f}} & Generate multiple layers  \\ \midrule
\hyperref[scm:mcata]{Monadic \newline catamorphism} & \ensuremath{(\forall \Varid{x}\hsforall \hsdot{\circ }{.}\Varid{f}\;(\Varid{m}\;\Varid{x})\to \Varid{m}\;(\Varid{f}\;\Varid{x}))}\newline \phantom{m}\ensuremath{\to (\Varid{f}\;\Varid{a}\to \Varid{m}\;\Varid{a})\to \mu\;\Varid{f}\to \Varid{m}\;\Varid{a}} & Recursion causing computational effects \\\midrule
\hyperref[scm:icata]{Indexed \newline catamorphism} & \ensuremath{(\Varid{f}\;\Varid{a}\;\dot{\rightarrow}\;\Varid{a})\to \dot{\mu}\;\Varid{f}\;\dot{\rightarrow}\;\Varid{a}} & Consume nested datatypes and \textsc{gadt}s \\
\bottomrule
\end{tabular}
\end{table}
}

\section{Datatypes and Fixed Points}\label{sec:background}
This paper assumes basic familiarity with Haskell as we use it to present all examples
and recursion schemes, but we do not assume any knowledge of category theory.
In this section, we briefly review the prerequisite of recursion schemes---recursive
datatypes, viewed as fixed points of functors.

\subsubsection{Datatypes}
\emph{Algebraic data types} (\textsc{adt}s) in Haskell allow the programmer to create new
datatypes from existing ones.
For example, the type \ensuremath{\Conid{List}\;\Varid{a}} of lists of elements of type \ensuremath{\Varid{a}} can be declared
as follows:
\begin{equation}
\label{eq:list}
\ensuremath{\mathbf{data}\;\Conid{List}\;\Varid{a}\mathrel{=}\Conid{Nil}\mid \Conid{Cons}\;\Varid{a}\;(\Conid{List}\;\Varid{a})}
\end{equation}
which means that an element of \ensuremath{\Conid{List}\;\Varid{a}} is exactly \ensuremath{\Conid{Nil}} or \ensuremath{\Conid{Cons}\;\Varid{x}\;\Varid{xs}} for all
\ensuremath{\Varid{x}\mathbin{::}\Varid{a}} and \ensuremath{\Varid{xs}\mathbin{::}\Conid{List}\;\Varid{a}}.
Similarly, the type \ensuremath{\Conid{Tree}\;\Varid{a}} of binary trees whose nodes are labelled with \ensuremath{\Varid{a}}-elements
can be declared as follows:
\begin{equation}\label{eq:tree}
\ensuremath{\mathbf{data}\;\Conid{Tree}\;\Varid{a}\mathrel{=}\Conid{Empty}\mid \Conid{Node}\;(\Conid{Tree}\;\Varid{a})\;\Varid{a}\;(\Conid{Tree}\;\Varid{a})}
\end{equation}
In definitions like \ensuremath{\Conid{List}\;\Varid{a}} and \ensuremath{\Conid{Tree}\;\Varid{a}}, the datatypes being defined also appear
on the right-hand side of the declaration, so they are \emph{recursive types}.
Moreover, \ensuremath{\Conid{List}\;\Varid{a}} and \ensuremath{\Conid{Tree}\;\Varid{a}} are among a special family of recursive types,
called \emph{inductive datatypes}, meaning that they are \emph{least} fixed points of
{functors}.

\subsubsection{Functors and Algebras}
Let us recall how endofunctors, or simply functors, in Haskell are type constructors \ensuremath{\Varid{f}\mathbin{::}\mathbin{*}\to \mathbin{*}} instantiating the following type class:
\begin{hscode}\SaveRestoreHook
\column{B}{@{}>{\hspre}l<{\hspost}@{}}%
\column{E}{@{}>{\hspre}l<{\hspost}@{}}%
\>[B]{}\mathbf{class}\;\Conid{Functor}\;\Varid{f}\;\mathbf{where}\;\Varid{fmap}\mathbin{::}(\Varid{a}\to \Varid{b})\to \Varid{f}\;\Varid{a}\to \Varid{f}\;\Varid{b}{}\<[E]%
\ColumnHook
\end{hscode}\resethooks
Additionally, \ensuremath{\Varid{fmap}} is expected to satisfy two functor laws:
\[
\ensuremath{\Varid{fmap}\;\Varid{id}\mathrel{=}\Varid{id}}  \hspace{1cm}  \ensuremath{\Varid{fmap}\;(\Varid{h}\hsdot{\circ }{.}\Varid{g})\mathrel{=}\Varid{fmap}\;\Varid{h}\hsdot{\circ }{.}\Varid{fmap}\;\Varid{g}}
\]
for all functions \ensuremath{\Varid{g}\mathbin{::}\Varid{a}\to \Varid{b}} and \ensuremath{\Varid{h}\mathbin{::}\Varid{b}\to \Varid{c}}.

Given a functor \ensuremath{\Varid{f}}, we call a function of type \ensuremath{\Varid{f}\;\Varid{a}\to \Varid{a}}, for some type \ensuremath{\Varid{a}},
an \emph{\ensuremath{\Varid{f}}-algebra}, and a function of type \ensuremath{\Varid{a}\to \Varid{f}\;\Varid{a}} an
\emph{\ensuremath{\Varid{f}}-coalgebra}.  In either case, type \ensuremath{\Varid{a}} is called the \emph{carrier} of
the (co)algebra.

\subsubsection{Fixed Points}
Given a functor \ensuremath{\Varid{f}}, a fixed point for \ensuremath{\Varid{f}} is a type \ensuremath{\Varid{p}} such that \ensuremath{\Varid{p}} is isomorphic to
\ensuremath{\Varid{f}\;\Varid{p}}.
In the set theoretic semantics, a functor may have more than one fixed point:
the \emph{least fixed point}, denoted by \ensuremath{\mu\;\Varid{f}}, is the set of \ensuremath{\Varid{f}}-branching
trees of \emph{finite} depths, while the \emph{greatest fixed point}, denoted
by \ensuremath{\nu\;\Varid{f}}, is intuitively the set of \ensuremath{\Varid{f}}-branching trees of \emph{possibly
infinite} depths.

However, due to the fact that Haskell is a lazy language with general
recursion, the least and greatest fixed points of a Haskell functor \ensuremath{\Varid{f}}
\emph{coincide} as the following datatype of possibly infinite \ensuremath{\Varid{f}}-branching trees:
\begin{hscode}\SaveRestoreHook
\column{B}{@{}>{\hspre}l<{\hspost}@{}}%
\column{E}{@{}>{\hspre}l<{\hspost}@{}}%
\>[B]{}\mathbf{newtype}\;\Conid{Fix}\;\Varid{f}\mathrel{=}\Conid{In}\;\{\mskip1.5mu \Varid{out}\mathbin{::}\Varid{f}\;(\Conid{Fix}\;\Varid{f})\mskip1.5mu\}{}\<[E]%
\ColumnHook
\end{hscode}\resethooks
This notation introduces the constructor \ensuremath{\Conid{In}\mathbin{::}\Varid{f}\;(\Conid{Fix}\;\Varid{f})\to \Conid{Fix}\;\Varid{f}} to create
fixed point, and its inverse \ensuremath{\Varid{out}\mathbin{::}\Conid{Fix}\;\Varid{f}\to \Varid{f}\;(\Conid{Fix}\;\Varid{f})}.

Although Haskell allows general recursion, the point of using structural recursion 
is precisely avoiding general recursion whenever possible, since general recursion
is typically tricky to reason about.
Hence in this paper we use Haskell as if it is a total programming language, by
making sure all recursive functions that we use are structurally recursive as
much as possible.
Thus we distinguish the least and greatest fixed points
as two datatypes:
\begin{equation*}
\ensuremath{\mathbf{newtype}\;\mu\;\Varid{f}\mathrel{=}\Conid{In}\;(\Varid{f}\;(\mu\;\Varid{f}))} 
\hspace{1cm}
\ensuremath{\mathbf{newtype}\;\nu\;\Varid{f}\mathrel{=}\Varid{Out}^\circ\;(\Varid{f}\;(\nu\;\Varid{f}))}
\end{equation*}
While these two datatypes \ensuremath{\mu\;\Varid{f}} and \ensuremath{\nu\;\Varid{f}} are the same datatype declaration,
we mentally understand \ensuremath{\mu\;\Varid{f}} as the type of \emph{finite} \ensuremath{\Varid{f}}-branching trees, and
\ensuremath{\nu\;\Varid{f}} as the type of \emph{possibly infinite} ones, as in the set-theoretic semantics. 
Making such a nominal distinction is not entirely pointless:
the type system at least ensures that we never accidentally misuse an element
of \ensuremath{\nu\;\Varid{f}} as an element of \ensuremath{\mu\;\Varid{f}}, unless we make an explicit conversion.
But it is our own responsibility to make sure that we never construct an infinite
element in \ensuremath{\mu\;\Varid{f}} using general recursion.

\begin{example}\label{ex:listF:treeF}
The datatypes (\ref{eq:list}) and (\ref{eq:tree}) that we saw earlier are
isomorphic to fixed points of functors \ensuremath{\Conid{ListF}} and \ensuremath{\Conid{TreeF}} defined as follows
(with the evident \ensuremath{\Varid{fmap}} that can be derived by \textsc{ghc}
automatically\footnote{It requires the \texttt{DeriveFunctor} extension of
\textsc{ghc} to derive functors automatically.}):
\begin{hscode}\SaveRestoreHook
\column{B}{@{}>{\hspre}l<{\hspost}@{}}%
\column{17}{@{}>{\hspre}l<{\hspost}@{}}%
\column{26}{@{}>{\hspre}l<{\hspost}@{}}%
\column{41}{@{}>{\hspre}l<{\hspost}@{}}%
\column{E}{@{}>{\hspre}l<{\hspost}@{}}%
\>[B]{}\mathbf{data}\;\Conid{ListF}\;\Varid{a}\;\Varid{x}{}\<[17]%
\>[17]{}\mathrel{=}\Conid{Nil}{}\<[26]%
\>[26]{}\mid \Conid{Cons}\;\Varid{a}\;\Varid{x}\;{}\<[41]%
\>[41]{}\mathbf{deriving}\;\Conid{Functor}{}\<[E]%
\\
\>[B]{}\mathbf{data}\;\Conid{TreeF}\;\Varid{a}\;\Varid{x}{}\<[17]%
\>[17]{}\mathrel{=}\Conid{Empty}{}\<[26]%
\>[26]{}\mid \Conid{Node}\;\Varid{x}\;\Varid{a}\;\Varid{x}\;{}\<[41]%
\>[41]{}\mathbf{deriving}\;\Conid{Functor}{}\<[E]%
\ColumnHook
\end{hscode}\resethooks
The type \ensuremath{\mu\;(\Conid{ListF}\;\Varid{a})} represents finite lists of \ensuremath{\Varid{a}} elements and \ensuremath{\mu\;(\Conid{TreeF}\;\Varid{a})} represents finite binary trees carrying \ensuremath{\Varid{a}} elements.
Correspondingly, \ensuremath{\nu\;(\Conid{ListF}\;\Varid{a})} and \ensuremath{\nu\;(\Conid{TreeF}\;\Varid{a})} are possibly infinite lists and trees respectively.

As an example, the correspondence between \ensuremath{\mu\;(\Conid{ListF}\;\Varid{a})} and finite elements of
\ensuremath{[\mskip1.5mu \Varid{a}\mskip1.5mu]} is evidenced by the following isomorphism.
{
\setlength{\mathindent}{0cm}
\begin{hscode}\SaveRestoreHook
\column{B}{@{}>{\hspre}l<{\hspost}@{}}%
\column{16}{@{}>{\hspre}l<{\hspost}@{}}%
\column{44}{@{}>{\hspre}l<{\hspost}@{}}%
\column{55}{@{}>{\hspre}l<{\hspost}@{}}%
\column{82}{@{}>{\hspre}l<{\hspost}@{}}%
\column{E}{@{}>{\hspre}l<{\hspost}@{}}%
\>[B]{}\Varid{conv}_{\mu}\mathbin{::}[\mskip1.5mu \Varid{a}\mskip1.5mu]{}\<[16]%
\>[16]{}\to \mu\;(\Conid{ListF}\;\Varid{a})\;{}\<[44]%
\>[44]{}\hspace{1cm}\;{}\<[55]%
\>[55]{}\Varid{conv}^{\circ}_{\mu}\mathbin{::}\mu\;(\Conid{ListF}\;\Varid{a}){}\<[82]%
\>[82]{}\to [\mskip1.5mu \Varid{a}\mskip1.5mu]{}\<[E]%
\\
\>[B]{}\Varid{conv}_{\mu}\;[\mskip1.5mu \mskip1.5mu]{}\<[16]%
\>[16]{}\mathrel{=}\Conid{In}\;\Conid{Nil}\;{}\<[55]%
\>[55]{}\Varid{conv}^{\circ}_{\mu}\;(\Conid{In}\;\Conid{Nil}){}\<[82]%
\>[82]{}\mathrel{=}[\mskip1.5mu \mskip1.5mu]{}\<[E]%
\\
\>[B]{}\Varid{conv}_{\mu}\;(\Varid{a}\mathbin{:}\Varid{as}){}\<[16]%
\>[16]{}\mathrel{=}\Conid{In}\;(\Conid{Cons}\;\Varid{a}\;(\Varid{conv}_{\mu}\;\Varid{as}))\;{}\<[55]%
\>[55]{}\Varid{conv}^{\circ}_{\mu}\;(\Conid{In}\;(\Conid{Cons}\;\Varid{a}\;\Varid{as})){}\<[82]%
\>[82]{}\mathrel{=}\Varid{a}\mathbin{:}\Varid{conv}^{\circ}_{\mu}\;\Varid{as}{}\<[E]%
\ColumnHook
\end{hscode}\resethooks
}
Supposing that there is a function computing the length of a list,
\[\ensuremath{\Varid{length}\mathbin{::}\mu\;(\Conid{ListF}\;\Varid{a})\to \Conid{Integer}}\]
The type checker of Haskell will then ensure that we never pass a value of \ensuremath{\nu\;(\Conid{ListF}\;\Varid{a})} to this function.
\end{example}

\subsubsection{Initial and Final (Co)Algebra}
The constructor \ensuremath{\Conid{In}\mathbin{::}\Varid{f}\;(\mu\;\Varid{f})\to \mu\;\Varid{f}} is an \ensuremath{\Varid{f}}-algebra with carrier \ensuremath{\mu\;\Varid{f}},
and it has an inverse \ensuremath{\Varid{in}^\circ\mathbin{::}\mu\;\Varid{f}\to \Varid{f}\;(\mu\;\Varid{f})} defined as
\begin{hscode}\SaveRestoreHook
\column{B}{@{}>{\hspre}l<{\hspost}@{}}%
\column{E}{@{}>{\hspre}l<{\hspost}@{}}%
\>[B]{}\Varid{in}^\circ\;(\Conid{In}\;\Varid{x})\mathrel{=}\Varid{x}{}\<[E]%
\ColumnHook
\end{hscode}\resethooks
which is an \ensuremath{\Varid{f}}-coalgebra thta witnesses the isomorphism between \ensuremath{\mu\;\Varid{f}} and \ensuremath{\Varid{f}\;(\mu\;\Varid{f})}, a fact known as Lambek's Lemma~\cite{Lam68Fix}.
Conversely, the constructor \ensuremath{\Varid{Out}^\circ\mathbin{::}\Varid{f}\;(\nu\;\Varid{f})\to \nu\;\Varid{f}} is an \ensuremath{\Varid{f}}-algebra with carrier \ensuremath{\nu\;\Varid{f}}, and its inverse
\ensuremath{\Varid{out}\mathbin{::}\nu\;\Varid{f}\to \Varid{f}\;(\nu\;\Varid{f})} defined as
\begin{hscode}\SaveRestoreHook
\column{B}{@{}>{\hspre}l<{\hspost}@{}}%
\column{E}{@{}>{\hspre}l<{\hspost}@{}}%
\>[B]{}\Varid{out}\;(\Varid{Out}^\circ\;\Varid{x})\mathrel{=}\Varid{x}{}\<[E]%
\ColumnHook
\end{hscode}\resethooks
is an \ensuremath{\Varid{f}}-coalgebra with carrier \ensuremath{\nu\;\Varid{f}}.

What is special with \ensuremath{\Conid{In}} and \ensuremath{\Varid{out}} is that \ensuremath{\Conid{In}} is the so-called \emph{initial algebra}
of \ensuremath{\Varid{f}}, in the sense that it has the nice property that for any \ensuremath{\Varid{f}}-algebra
\ensuremath{\Varid{alg}\mathbin{::}\Varid{f}\;\Varid{a}\to \Varid{a}}, there is exactly one function \ensuremath{\Varid{h}\mathbin{::}\mu\;\Varid{f}\to \Varid{a}} such that
\begin{equation}\label{eq:initial}
\ensuremath{\Varid{h}\hsdot{\circ }{.}\Conid{In}\mathrel{=}\Varid{alg}\hsdot{\circ }{.}\Varid{fmap}\;\Varid{h}}
\end{equation}
Dually, \ensuremath{\Varid{out}} is called the \emph{final coalgebra} of \ensuremath{\Varid{f}} since for any
\ensuremath{\Varid{f}}-coalgebra \ensuremath{\Varid{coalg}\mathbin{::}\Varid{c}\to \Varid{f}\;\Varid{c}}, there is exactly one function \ensuremath{\Varid{h}\mathbin{::}\Varid{c}\to \nu\;\Varid{f}} such that
\begin{equation}\label{eq:final}
\ensuremath{\Varid{out}\hsdot{\circ }{.}\Varid{h}\mathrel{=}\Varid{fmap}\;\Varid{h}\hsdot{\circ }{.}\Varid{coalg}}
\end{equation}
The \ensuremath{\Varid{h}}'s in (\ref{eq:initial}) and (\ref{eq:final}) are precisely the two
fundamental recursion schemes, \emph{catamorphisms} and \emph{anamorphisms},
which we will talk about in the next section.

\section{Fundamental Recursion Schemes}
\label{sec:cata}

Most if not all programs are about processing data, and as Hoare \cite{Hoare72Note} noted, `there are certain close analogies between the methods used for structuring data and the methods for structuring a program which processes that data.'
In essence, \emph{data structure determines program structure} \cite{FFR18How,Gibbons2021}.
The determination is abstracted as recursion schemes for programs processing recursive datatypes.

In this section, we look at the three fundamental recursion schemes: catamorphisms, in which the program is structured by its input; anamorphisms, in which the program is structured by its output; 
and hylomorphisms, in which the program is structured by an internal recursive call structure.

\subsection{Catamorphisms}
We start our journey with programs whose structure follows their input.
As the first example, consider the program computing the length of a list:
\begin{hscode}\SaveRestoreHook
\column{B}{@{}>{\hspre}l<{\hspost}@{}}%
\column{12}{@{}>{\hspre}l<{\hspost}@{}}%
\column{20}{@{}>{\hspre}l<{\hspost}@{}}%
\column{E}{@{}>{\hspre}l<{\hspost}@{}}%
\>[B]{}\Varid{length}\mathbin{::}{}\<[12]%
\>[12]{}[\mskip1.5mu \Varid{a}\mskip1.5mu]{}\<[20]%
\>[20]{}\to \Conid{Integer}{}\<[E]%
\\
\>[B]{}\Varid{length}\;{}\<[12]%
\>[12]{}[\mskip1.5mu \mskip1.5mu]{}\<[20]%
\>[20]{}\mathrel{=}\mathrm{0}{}\<[E]%
\\
\>[B]{}\Varid{length}\;{}\<[12]%
\>[12]{}(\Varid{x}\mathbin{:}\Varid{xs}){}\<[20]%
\>[20]{}\mathrel{=}\mathrm{1}\mathbin{+}\Varid{length}\;\Varid{xs}{}\<[E]%
\ColumnHook
\end{hscode}\resethooks
In Haskell, a list is either the empty list \ensuremath{[\mskip1.5mu \mskip1.5mu]} or \ensuremath{\Varid{x}\mathbin{:}\Varid{xs}}, an element \ensuremath{\Varid{x}} prepended to list \ensuremath{\Varid{xs}}.
This structure of lists is closely reflected by the program \ensuremath{\Varid{length}}, which is defined by two cases too, one for the empty list \ensuremath{[\mskip1.5mu \mskip1.5mu]} and one for the recursive case \ensuremath{\Varid{x}\mathbin{:}\Varid{xs}}.
Additionally, in the recursive case \ensuremath{\Varid{length}\;(\Varid{x}\mathbin{:}\Varid{xs})} is solely determined by \ensuremath{\Varid{length}\;\Varid{xs}} without further usage of \ensuremath{\Varid{xs}}.

\subsubsection{List Folds}
The pattern in \ensuremath{\Varid{length}} is called \emph{structural recursion} and is expressed by the function \ensuremath{\Varid{foldr}} in Haskell:
\begin{hscode}\SaveRestoreHook
\column{B}{@{}>{\hspre}l<{\hspost}@{}}%
\column{19}{@{}>{\hspre}l<{\hspost}@{}}%
\column{E}{@{}>{\hspre}l<{\hspost}@{}}%
\>[B]{}\Varid{foldr}\mathbin{::}(\Varid{a}\to \Varid{b}\to \Varid{b})\to \Varid{b}\to [\mskip1.5mu \Varid{a}\mskip1.5mu]\to \Varid{b}{}\<[E]%
\\
\>[B]{}\Varid{foldr}\;\Varid{f}\;\Varid{e}\;[\mskip1.5mu \mskip1.5mu]{}\<[19]%
\>[19]{}\mathrel{=}\Varid{e}{}\<[E]%
\\
\>[B]{}\Varid{foldr}\;\Varid{f}\;\Varid{e}\;(\Varid{x}\mathbin{:}\Varid{xs}){}\<[19]%
\>[19]{}\mathrel{=}\Varid{f}\;\Varid{x}\;(\Varid{foldr}\;\Varid{f}\;\Varid{e}\;\Varid{xs}){}\<[E]%
\ColumnHook
\end{hscode}\resethooks
which is very useful in list processing.
As a fold, \ensuremath{\Varid{length}\mathrel{=}\Varid{foldr}\;(\lambda \anonymous \;\Varid{l}\to \mathrm{1}\mathbin{+}\Varid{l})\;\mathrm{0}}.
The frequently used function \ensuremath{\Varid{map}} is also a fold:
\begin{hscode}\SaveRestoreHook
\column{B}{@{}>{\hspre}l<{\hspost}@{}}%
\column{E}{@{}>{\hspre}l<{\hspost}@{}}%
\>[B]{}\Varid{map}\mathbin{::}(\Varid{a}\to \Varid{b})\to [\mskip1.5mu \Varid{a}\mskip1.5mu]\to [\mskip1.5mu \Varid{b}\mskip1.5mu]{}\<[E]%
\\
\>[B]{}\Varid{map}\;\Varid{f}\mathrel{=}\Varid{foldr}\;(\lambda \Varid{x}\;\Varid{xs}\to \Varid{f}\;\Varid{x}\mathbin{:}\Varid{xs})\;[\mskip1.5mu \mskip1.5mu]{}\<[E]%
\ColumnHook
\end{hscode}\resethooks
Another example is the function flattening a list of lists into a list:
\begin{hscode}\SaveRestoreHook
\column{B}{@{}>{\hspre}l<{\hspost}@{}}%
\column{E}{@{}>{\hspre}l<{\hspost}@{}}%
\>[B]{}\Varid{concat}\mathbin{::}[\mskip1.5mu [\mskip1.5mu \Varid{a}\mskip1.5mu]\mskip1.5mu]\to [\mskip1.5mu \Varid{a}\mskip1.5mu]{}\<[E]%
\\
\>[B]{}\Varid{concat}\mathrel{=}\Varid{foldr}\;(\plus )\;[\mskip1.5mu \mskip1.5mu]{}\<[E]%
\ColumnHook
\end{hscode}\resethooks
By expressing structural recursive functions as folds, their structure becomes
clearer, similarly in spirit to the well accepted practice of structuring
programs with if-conditionals and for-/while-loops in imperative languages.

\begin{recscheme}[\ensuremath{\Varid{cata}}]\label{scm:cata}
Folds on lists can be readily generalized to the generic setting, where the
shape of the datatype is determined by a functor
\cite{Mal90Dat,Fok92Law,Hag87Cat}.
Such functions are called \emph{catamorphisms},
and come from the following recursion scheme:
\begin{hscode}\SaveRestoreHook
\column{B}{@{}>{\hspre}l<{\hspost}@{}}%
\column{E}{@{}>{\hspre}l<{\hspost}@{}}%
\>[B]{}\Varid{cata}\mathbin{::}\Conid{Functor}\;\Varid{f}\Rightarrow (\Varid{f}\;\Varid{a}\to \Varid{a})\to \mu\;\Varid{f}\to \Varid{a}{}\<[E]%
\\
\>[B]{}\Varid{cata}\;\Varid{alg}\mathrel{=}\Varid{alg}\hsdot{\circ }{.}\Varid{fmap}\;(\Varid{cata}\;\Varid{alg})\hsdot{\circ }{.}\Varid{in}^\circ{}\<[E]%
\ColumnHook
\end{hscode}\resethooks
\end{recscheme}
Intuitively, \ensuremath{\Varid{cata}\;\Varid{alg}} gradually breaks down the inductively defined input
data, computing the result by replacing constructors with the given algebra
\ensuremath{\Varid{alg}}.

The name \ensuremath{\Varid{cata}} dubbed by Meertens \cite{Mee88First} is from Greek \textgreek{κατά}
meaning `downwards along' or `according to'.
A notation for \ensuremath{\Varid{cata}\;\Varid{alg}} is the so-called banana bracket $\banana{\ensuremath{\Varid{alg}}}$ but
we will not use this style of notation in this paper, as there will not be
enough squiggly brackets for all the different recursion schemes.

\begin{example}
By converting the builtin list type \ensuremath{[\mskip1.5mu \Varid{a}\mskip1.5mu]} to the initial algebra of \ensuremath{\Conid{ListF}} as in
\autoref{ex:listF:treeF}, we can recover \ensuremath{\Varid{foldr}} from \ensuremath{\Varid{cata}} as follows: 
\begin{hscode}\SaveRestoreHook
\column{B}{@{}>{\hspre}l<{\hspost}@{}}%
\column{3}{@{}>{\hspre}l<{\hspost}@{}}%
\column{19}{@{}>{\hspre}l<{\hspost}@{}}%
\column{E}{@{}>{\hspre}l<{\hspost}@{}}%
\>[B]{}\Varid{foldr}\mathbin{::}(\Varid{a}\to \Varid{b}\to \Varid{b})\to \Varid{b}\to [\mskip1.5mu \Varid{a}\mskip1.5mu]\to \Varid{b}{}\<[E]%
\\
\>[B]{}\Varid{foldr}\;\Varid{f}\;\Varid{e}\mathrel{=}\Varid{cata}\;\Varid{alg}\hsdot{\circ }{.}\Varid{conv}_{\mu}\;\mathbf{where}{}\<[E]%
\\
\>[B]{}\hsindent{3}{}\<[3]%
\>[3]{}\Varid{alg}\;\Conid{Nil}{}\<[19]%
\>[19]{}\mathrel{=}\Varid{e}{}\<[E]%
\\
\>[B]{}\hsindent{3}{}\<[3]%
\>[3]{}\Varid{alg}\;(\Conid{Cons}\;\Varid{a}\;\Varid{x}){}\<[19]%
\>[19]{}\mathrel{=}\Varid{f}\;\Varid{a}\;\Varid{x}{}\<[E]%
\ColumnHook
\end{hscode}\resethooks
Now we can also fold datatypes other than lists, such as binary trees:
\begin{hscode}\SaveRestoreHook
\column{B}{@{}>{\hspre}l<{\hspost}@{}}%
\column{3}{@{}>{\hspre}l<{\hspost}@{}}%
\column{21}{@{}>{\hspre}l<{\hspost}@{}}%
\column{E}{@{}>{\hspre}l<{\hspost}@{}}%
\>[B]{}\Varid{size}\mathbin{::}\mu\;(\Conid{TreeF}\;\Varid{e})\to \Conid{Integer}{}\<[E]%
\\
\>[B]{}\Varid{size}\mathrel{=}\Varid{cata}\;\Varid{alg}\;\mathbf{where}{}\<[E]%
\\
\>[B]{}\hsindent{3}{}\<[3]%
\>[3]{}\Varid{alg}\mathbin{::}\Conid{TreeF}\;\Varid{a}\;\Conid{Integer}\to \Conid{Integer}{}\<[E]%
\\
\>[B]{}\hsindent{3}{}\<[3]%
\>[3]{}\Varid{alg}\;\Conid{Empty}{}\<[21]%
\>[21]{}\mathrel{=}\mathrm{0}{}\<[E]%
\\
\>[B]{}\hsindent{3}{}\<[3]%
\>[3]{}\Varid{alg}\;(\Conid{Node}\;\Varid{l}\;\Varid{e}\;\Varid{r}){}\<[21]%
\>[21]{}\mathrel{=}\Varid{l}\mathbin{+}\mathrm{1}\mathbin{+}\Varid{r}{}\<[E]%
\ColumnHook
\end{hscode}\resethooks
\end{example}

\begin{example}[Interpreting \textsc{dsl}s]
\label{ex:dsl}
The `killer application' of catamorphisms is using them to implement \emph{domain-specific
languages} (\textsc{dsl}s) \cite{Hutton98,GibbonsW14}.
The abstract syntax of a \textsc{dsl} can usually be modelled as an inductive datatype,
and then the (denotational) semantics of the \textsc{dsl} can be given as a catamorphism.
The semantics given in this way is \emph{compositional}, meaning that
the semantics of a program is determined by the semantics of its immediate
sub-parts---exactly the pattern of catamorphisms.

As a small example here, consider a mini language of mutable memory consisting
of three language constructs: 
\ensuremath{\Conid{Put}\;(\Varid{i},\Varid{x})\;\Varid{k}} writes value \ensuremath{\Varid{x}} to memory cell of address \ensuremath{\Varid{i}} and then executes program \ensuremath{\Varid{k}};
\ensuremath{\Conid{Get}\;\Varid{i}\;\Varid{k}} reads memory cell \ensuremath{\Varid{i}}, letting the result be \ensuremath{\Varid{s}}, and then executes program \ensuremath{\Varid{k}\;\Varid{s}};
and \ensuremath{\Conid{Ret}\;\Varid{a}} terminates the execution with return value \ensuremath{\Varid{a}}.
The abstract syntax of the language can be modelled as the initial algebra \ensuremath{\mu\;(\Conid{ProgF}\;\Varid{s}\;\Varid{a})}
of the following functor:
\begin{hscode}\SaveRestoreHook
\column{B}{@{}>{\hspre}l<{\hspost}@{}}%
\column{19}{@{}>{\hspre}l<{\hspost}@{}}%
\column{E}{@{}>{\hspre}l<{\hspost}@{}}%
\>[B]{}\mathbf{data}\;\Conid{ProgF}\;\Varid{s}\;\Varid{a}\;\Varid{x}{}\<[19]%
\>[19]{}\mathrel{=}\Conid{Ret}\;\Varid{a}\mid \Conid{Put}\;(\Conid{Int},\Varid{s})\;\Varid{x}\mid \Conid{Get}\;\Conid{Int}\;(\Varid{s}\to \Varid{x}){}\<[E]%
\ColumnHook
\end{hscode}\resethooks
where \ensuremath{\Varid{s}} is the type of values stored by memory cells and \ensuremath{\Varid{a}} is the type of values finally returned.
An example of a program in this language is
\begin{hscode}\SaveRestoreHook
\column{B}{@{}>{\hspre}l<{\hspost}@{}}%
\column{E}{@{}>{\hspre}l<{\hspost}@{}}%
\>[B]{}p_1\mathbin{::}\mu\;(\Conid{ProgF}\;\Conid{Int}\;\Conid{Int}){}\<[E]%
\\
\>[B]{}p_1\mathrel{=}\Conid{In}\;(\Conid{Get}\;\mathrm{0}\;(\lambda \Varid{s}\to (\Conid{In}\;(\Conid{Put}\;(\mathrm{0},\Varid{s}\mathbin{+}\mathrm{1})\;(\Conid{In}\;(\Conid{Ret}\;\Varid{s})))))){}\<[E]%
\ColumnHook
\end{hscode}\resethooks
which reads the 0-th cell, increments it, and returns the old value.
The syntax is admittedly clumsy because of the repeating \ensuremath{\Conid{In}} constructors, but they can be eliminated if `smart constructors' such as \ensuremath{\Varid{ret}\mathrel{=}\Conid{In}\hsdot{\circ }{.}\Conid{Ret}} are defined.

The semantics of a program in this mini language can be given as a value of type \ensuremath{\Conid{Map}\;\Conid{Int}\;\Varid{s}\to \Varid{a}}, and the interpretation is a catamorphisms:
\begin{hscode}\SaveRestoreHook
\column{B}{@{}>{\hspre}l<{\hspost}@{}}%
\column{3}{@{}>{\hspre}l<{\hspost}@{}}%
\column{25}{@{}>{\hspre}l<{\hspost}@{}}%
\column{E}{@{}>{\hspre}l<{\hspost}@{}}%
\>[B]{}\Varid{interp}\mathbin{::}\mu\;(\Conid{ProgF}\;\Varid{s}\;\Varid{a})\to (\Conid{Map}\;\Conid{Int}\;\Varid{s}\to \Varid{a}){}\<[E]%
\\
\>[B]{}\Varid{interp}\mathrel{=}\Varid{cata}\;\Varid{handle}\;\mathbf{where}{}\<[E]%
\\
\>[B]{}\hsindent{3}{}\<[3]%
\>[3]{}\Varid{handle}\;(\Conid{Ret}\;\Varid{a}){}\<[25]%
\>[25]{}\mathrel{=}\lambda \anonymous \to \Varid{a}{}\<[E]%
\\
\>[B]{}\hsindent{3}{}\<[3]%
\>[3]{}\Varid{handle}\;(\Conid{Put}\;(\Varid{i},\Varid{x})\;\Varid{k}){}\<[25]%
\>[25]{}\mathrel{=}\lambda \Varid{m}\to \Varid{k}\;(\Varid{update}\;\Varid{m}\;\Varid{i}\;\Varid{x}){}\<[E]%
\\
\>[B]{}\hsindent{3}{}\<[3]%
\>[3]{}\Varid{handle}\;(\Conid{Get}\;\Varid{i}\;\Varid{k}){}\<[25]%
\>[25]{}\mathrel{=}\lambda \Varid{m}\to \Varid{k}\;(\Varid{m}\mathbin{!}\Varid{i})\;\Varid{m}{}\<[E]%
\ColumnHook
\end{hscode}\resethooks
where \ensuremath{\Varid{update}\;\Varid{m}\;\Varid{i}\;\Varid{x}} is the map \ensuremath{\Varid{m}} with the value at \ensuremath{\Varid{i}} changed to \ensuremath{\Varid{x}}, and \ensuremath{\Varid{m}\mathbin{!}\Varid{i}} looks up \ensuremath{\Varid{i}} in \ensuremath{\Varid{m}}.
Then we can use it to run programs:
\begin{hscode}\SaveRestoreHook
\column{B}{@{}>{\hspre}l<{\hspost}@{}}%
\column{E}{@{}>{\hspre}l<{\hspost}@{}}%
\>[B]{}*\!>\ \Varid{interp}\;p_1\;(\Varid{fromList}\;[\mskip1.5mu (\mathrm{0},\mathrm{100})\mskip1.5mu])\mbox{\onelinecomment  outputs 100}{}\<[E]%
\ColumnHook
\end{hscode}\resethooks

\end{example}

\subsection{Anamorphisms}\label{sec:ana}

In catamorphisms, the structure of a program mimics the structure of the input.
Needless to say, this pattern is insufficient to cover all programs in the wild.
Imagine a program returning a record:
\begin{hscode}\SaveRestoreHook
\column{B}{@{}>{\hspre}l<{\hspost}@{}}%
\column{E}{@{}>{\hspre}l<{\hspost}@{}}%
\>[B]{}\mathbf{data}\;\Conid{Person}\mathrel{=}\Conid{Person}\;\{\mskip1.5mu \Varid{name}\mathbin{::}\Conid{String},\Varid{addr}\mathbin{::}\Conid{String},\Varid{phone}\mathbin{::}[\mskip1.5mu \Conid{Int}\mskip1.5mu]\mskip1.5mu\}{}\<[E]%
\\[\blanklineskip]%
\>[B]{}\Varid{mkEntry}\mathbin{::}\Conid{StaffInfo}\to \Conid{Person}{}\<[E]%
\\
\>[B]{}\Varid{mkEntry}\;\Varid{i}\mathrel{=}\Conid{Person}\;\Varid{n}\;\Varid{a}\;\Varid{p}\;\mathbf{where}\;\Varid{n}\mathrel{=}\mathbin{...};\Varid{a}\mathrel{=}\mathbin{...};\Varid{p}\mathrel{=}\mathbin{...}{}\<[E]%
\ColumnHook
\end{hscode}\resethooks
The structure of the program more resembles the structure of its output---each field of the output is computed by a corresponding part of the program.
Similarly, when the output is a recursive datatype, a natural pattern is that the program generates the output recursively, called \emph{(structural) corecursion} \cite{Gibbons2021}.
Consider the following program generating evenly spaced numbers over an interval.
\begin{hscode}\SaveRestoreHook
\column{B}{@{}>{\hspre}l<{\hspost}@{}}%
\column{3}{@{}>{\hspre}l<{\hspost}@{}}%
\column{5}{@{}>{\hspre}l<{\hspost}@{}}%
\column{18}{@{}>{\hspre}l<{\hspost}@{}}%
\column{E}{@{}>{\hspre}l<{\hspost}@{}}%
\>[B]{}\Varid{linspace}\mathbin{::}\Conid{RealFrac}\;\Varid{a}\Rightarrow \Varid{a}\to \Varid{a}\to \Conid{Integer}\to [\mskip1.5mu \Varid{a}\mskip1.5mu]{}\<[E]%
\\
\>[B]{}\Varid{linspace}\;\Varid{s}\;\Varid{e}\;\Varid{n}\mathrel{=}\Varid{gen}\;\Varid{s}\;\mathbf{where}{}\<[E]%
\\
\>[B]{}\hsindent{3}{}\<[3]%
\>[3]{}\Varid{step}\mathrel{=}(\Varid{e}\mathbin{-}\Varid{s})\mathbin{/}\Varid{fromIntegral}\;(\Varid{n}\mathbin{+}\mathrm{1}){}\<[E]%
\\
\>[B]{}\hsindent{3}{}\<[3]%
\>[3]{}\Varid{gen}\;\Varid{i}{}\<[E]%
\\
\>[3]{}\hsindent{2}{}\<[5]%
\>[5]{}\mid \Varid{i}\mathbin{<}\Varid{e}{}\<[18]%
\>[18]{}\mathrel{=}\Varid{i}\mathbin{:}\Varid{gen}\;(\Varid{i}\mathbin{+}\Varid{step}){}\<[E]%
\\
\>[3]{}\hsindent{2}{}\<[5]%
\>[5]{}\mid \Varid{otherwise}{}\<[18]%
\>[18]{}\mathrel{=}[\mskip1.5mu \mskip1.5mu]{}\<[E]%
\ColumnHook
\end{hscode}\resethooks
The program \ensuremath{\Varid{gen}} does not mirror the structure of its numeric input at all, but it follows the structure of its output, which is a list:
for the two cases of a list, \ensuremath{[\mskip1.5mu \mskip1.5mu]} and \ensuremath{(\mathbin{:})}, \ensuremath{\Varid{gen}} has a corresponding branch generating it.

\subsubsection{List Unfolds}
The pattern of generating a list in the example above is abstracted as the Haskell function \ensuremath{\Varid{unfoldr}}:
\begin{hscode}\SaveRestoreHook
\column{B}{@{}>{\hspre}l<{\hspost}@{}}%
\column{3}{@{}>{\hspre}l<{\hspost}@{}}%
\column{19}{@{}>{\hspre}l<{\hspost}@{}}%
\column{E}{@{}>{\hspre}l<{\hspost}@{}}%
\>[B]{}\Varid{unfoldr}\mathbin{::}(\Varid{b}\to \Conid{Maybe}\;(\Varid{a},\Varid{b}))\to \Varid{b}\to [\mskip1.5mu \Varid{a}\mskip1.5mu]{}\<[E]%
\\
\>[B]{}\Varid{unfoldr}\;\Varid{g}\;\Varid{s}\mathrel{=}\mathbf{case}\;\Varid{g}\;\Varid{s}\;\mathbf{of}{}\<[E]%
\\
\>[B]{}\hsindent{3}{}\<[3]%
\>[3]{}(\Conid{Just}\;(\Varid{a},\Varid{s'})){}\<[19]%
\>[19]{}\to \Varid{a}\mathbin{:}\Varid{unfoldr}\;\Varid{g}\;\Varid{s'}{}\<[E]%
\\
\>[B]{}\hsindent{3}{}\<[3]%
\>[3]{}\Conid{Nothing}{}\<[19]%
\>[19]{}\to [\mskip1.5mu \mskip1.5mu]{}\<[E]%
\ColumnHook
\end{hscode}\resethooks
in which \ensuremath{\Varid{g}} either produces \ensuremath{\Conid{Nothing}} indicating the end of the output or produces from a seed \ensuremath{\Varid{s}} the next element \ensuremath{\Varid{a}} of the output together with a new seed \ensuremath{\Varid{s'}} for generating the rest of the output.
Thus we can rewrite \ensuremath{\Varid{linspace}} as
\begin{hscode}\SaveRestoreHook
\column{B}{@{}>{\hspre}l<{\hspost}@{}}%
\column{3}{@{}>{\hspre}l<{\hspost}@{}}%
\column{E}{@{}>{\hspre}l<{\hspost}@{}}%
\>[B]{}\Varid{linspace}\;\Varid{s}\;\Varid{e}\;\Varid{n}\mathrel{=}\Varid{unfoldr}\;\Varid{gen}\;\Varid{s}\;\mathbf{where}{}\<[E]%
\\
\>[B]{}\hsindent{3}{}\<[3]%
\>[3]{}\Varid{step}\mathrel{=}(\Varid{e}\mathbin{-}\Varid{s})\mathbin{/}\Varid{fromIntegral}\;(\Varid{n}\mathbin{+}\mathrm{1}){}\<[E]%
\\
\>[B]{}\hsindent{3}{}\<[3]%
\>[3]{}\Varid{gen}\;\Varid{i}\mathrel{=}\mathbf{if}\;\Varid{i}\mathbin{<}\Varid{e}\;\mathbf{then}\;\Conid{Just}\;(\Varid{i},\Varid{i}\mathbin{+}\Varid{step})\;\mathbf{else}\;\Conid{Nothing}{}\<[E]%
\ColumnHook
\end{hscode}\resethooks
Note that the list produced by \ensuremath{\Varid{unfoldr}} is not necessarily finite.
For example,
\begin{hscode}\SaveRestoreHook
\column{B}{@{}>{\hspre}l<{\hspost}@{}}%
\column{E}{@{}>{\hspre}l<{\hspost}@{}}%
\>[B]{}\Varid{from}\mathbin{::}\Conid{Integer}\to [\mskip1.5mu \Conid{Integer}\mskip1.5mu]{}\<[E]%
\\
\>[B]{}\Varid{from}\mathrel{=}\Varid{unfoldr}\;(\lambda \Varid{n}\to \Conid{Just}\;(\Varid{n},\Varid{n}\mathbin{+}\mathrm{1})){}\<[E]%
\ColumnHook
\end{hscode}\resethooks
generates the infinite list of all integers from \ensuremath{\Varid{n}}.

\begin{recscheme}[\ensuremath{\Varid{ana}}]\label{scm:ana}
In the same way that \ensuremath{\Varid{cata}} generalizes \ensuremath{\Varid{foldr}}, \ensuremath{\Varid{unfoldr}} can be generalized
from lists to arbitrary coinductive datatypes. Functions arising from this
(co)recursion scheme are called \emph{anamorphisms}:
\begin{hscode}\SaveRestoreHook
\column{B}{@{}>{\hspre}l<{\hspost}@{}}%
\column{E}{@{}>{\hspre}l<{\hspost}@{}}%
\>[B]{}\Varid{ana}\mathbin{::}\Conid{Functor}\;\Varid{f}\Rightarrow (\Varid{c}\to \Varid{f}\;\Varid{c})\to \Varid{c}\to \nu\;\Varid{f}{}\<[E]%
\\
\>[B]{}\Varid{ana}\;\Varid{coalg}\mathrel{=}\Varid{Out}^\circ\hsdot{\circ }{.}\Varid{fmap}\;(\Varid{ana}\;\Varid{coalg})\hsdot{\circ }{.}\Varid{coalg}{}\<[E]%
\ColumnHook
\end{hscode}\resethooks
The name %
\emph{ana} is from the Greek word
\textgreek{ανά} means `upwards', dual to \emph{cata} meaning `downwards'.
\end{recscheme}

\begin{example}
Modulo the isomorphism between \ensuremath{[\mskip1.5mu \Varid{a}\mskip1.5mu]} and \ensuremath{\nu\;(\Conid{ListF}\;\Varid{a})},
\ensuremath{\Varid{unfoldr}} produces an anamorphism:
\begin{hscode}\SaveRestoreHook
\column{B}{@{}>{\hspre}l<{\hspost}@{}}%
\column{3}{@{}>{\hspre}l<{\hspost}@{}}%
\column{26}{@{}>{\hspre}l<{\hspost}@{}}%
\column{41}{@{}>{\hspre}l<{\hspost}@{}}%
\column{E}{@{}>{\hspre}l<{\hspost}@{}}%
\>[B]{}\Varid{unfoldr}\mathbin{::}(\Varid{b}\to \Conid{Maybe}\;(\Varid{a},\Varid{b}))\to \Varid{b}\to [\mskip1.5mu \Varid{a}\mskip1.5mu]{}\<[E]%
\\
\>[B]{}\Varid{unfoldr}\;\Varid{g}\mathrel{=}\Varid{conv}^{\circ}_{\nu}\hsdot{\circ }{.}\Varid{ana}\;\Varid{coalg}\;\mathbf{where}{}\<[E]%
\\
\>[B]{}\hsindent{3}{}\<[3]%
\>[3]{}\Varid{coalg}\;\Varid{b}\mathrel{=}\mathbf{case}\;\Varid{g}\;\Varid{b}\;\mathbf{of}\;{}\<[26]%
\>[26]{}\Conid{Nothing}{}\<[41]%
\>[41]{}\to \Conid{Nil}{}\<[E]%
\\
\>[26]{}(\Conid{Just}\;(\Varid{a},\Varid{b})){}\<[41]%
\>[41]{}\to \Conid{Cons}\;\Varid{a}\;\Varid{b}{}\<[E]%
\ColumnHook
\end{hscode}\resethooks
\end{example}

\begin{example}
A more interesting example of anamorphisms is merging a pair of ordered lists:
\begin{hscode}\SaveRestoreHook
\column{B}{@{}>{\hspre}l<{\hspost}@{}}%
\column{3}{@{}>{\hspre}l<{\hspost}@{}}%
\column{5}{@{}>{\hspre}l<{\hspost}@{}}%
\column{18}{@{}>{\hspre}l<{\hspost}@{}}%
\column{29}{@{}>{\hspre}l<{\hspost}@{}}%
\column{E}{@{}>{\hspre}l<{\hspost}@{}}%
\>[B]{}\Varid{merge}\mathbin{::}\Conid{Ord}\;\Varid{a}\Rightarrow (\nu\;(\Conid{ListF}\;\Varid{a}),\nu\;(\Conid{ListF}\;\Varid{a}))\to \nu\;(\Conid{ListF}\;\Varid{a}){}\<[E]%
\\
\>[B]{}\Varid{merge}\mathrel{=}\Varid{ana}\;\Varid{c}\;\mathbf{where}{}\<[E]%
\\
\>[B]{}\hsindent{3}{}\<[3]%
\>[3]{}\Varid{c}\;(\Varid{x},\Varid{y}){}\<[E]%
\\
\>[3]{}\hsindent{2}{}\<[5]%
\>[5]{}\mid \Varid{null}_{\nu}\;\Varid{x}\mathrel{\wedge}\Varid{null}_{\nu}\;\Varid{y}{}\<[29]%
\>[29]{}\mathrel{=}\Conid{Nil}{}\<[E]%
\\
\>[3]{}\hsindent{2}{}\<[5]%
\>[5]{}\mid \Varid{null}_{\nu}\;\Varid{y}\mathrel{\vee}\Varid{head}_{\nu}\;\Varid{x}\mathbin{<}\Varid{head}_{\nu}\;\Varid{y}{}\<[E]%
\\
\>[5]{}\hsindent{13}{}\<[18]%
\>[18]{}\mathrel{=}\Conid{Cons}\;(\Varid{head}_{\nu}\;\Varid{x})\;(\Varid{tail}_{\nu}\;\Varid{x},\Varid{y}){}\<[E]%
\\
\>[3]{}\hsindent{2}{}\<[5]%
\>[5]{}\mid \Varid{otherwise}{}\<[18]%
\>[18]{}\mathrel{=}\Conid{Cons}\;(\Varid{head}_{\nu}\;\Varid{y})\;(\Varid{x},\Varid{tail}_{\nu}\;\Varid{y}){}\<[E]%
\ColumnHook
\end{hscode}\resethooks
where \ensuremath{\Varid{null}_{\nu}}, \ensuremath{\Varid{head}_{\nu}} and \ensuremath{\Varid{tail}_{\nu}} are the corresponding list functions for \ensuremath{\nu\;(\Conid{ListF}\;\Varid{a})}.
\end{example}

\subsection{Hylomorphisms}

Catamorphisms consume data and anamorphisms produce data, but some algorithms are more complex than playing a single role---they produce and consume data at the same time.
Taking the quicksort algorithm for example, a (not-in-place, worst-case complexity $\mathcal{O}(n^2)$) implementation is:
\begin{hscode}\SaveRestoreHook
\column{B}{@{}>{\hspre}l<{\hspost}@{}}%
\column{3}{@{}>{\hspre}l<{\hspost}@{}}%
\column{6}{@{}>{\hspre}l<{\hspost}@{}}%
\column{15}{@{}>{\hspre}l<{\hspost}@{}}%
\column{E}{@{}>{\hspre}l<{\hspost}@{}}%
\>[B]{}\Varid{qsort}\mathbin{::}\Conid{Ord}\;\Varid{a}\Rightarrow [\mskip1.5mu \Varid{a}\mskip1.5mu]\to [\mskip1.5mu \Varid{a}\mskip1.5mu]{}\<[E]%
\\
\>[B]{}\Varid{qsort}\;[\mskip1.5mu \mskip1.5mu]{}\<[15]%
\>[15]{}\mathrel{=}[\mskip1.5mu \mskip1.5mu]{}\<[E]%
\\
\>[B]{}\Varid{qsort}\;(\Varid{a}\mathbin{:}\Varid{as}){}\<[15]%
\>[15]{}\mathrel{=}\Varid{qsort}\;\Varid{l}\plus [\mskip1.5mu \Varid{a}\mskip1.5mu]\plus \Varid{qsort}\;\Varid{r}\;\mathbf{where}{}\<[E]%
\\
\>[B]{}\hsindent{3}{}\<[3]%
\>[3]{}\Varid{l}{}\<[6]%
\>[6]{}\mathrel{=}[\mskip1.5mu \Varid{b}\mid \Varid{b}\leftarrow \Varid{as},\Varid{b}\mathbin{<}\Varid{a}\mskip1.5mu]{}\<[E]%
\\
\>[B]{}\hsindent{3}{}\<[3]%
\>[3]{}\Varid{r}{}\<[6]%
\>[6]{}\mathrel{=}[\mskip1.5mu \Varid{b}\mid \Varid{b}\leftarrow \Varid{as},\Varid{b}\geq \Varid{a}\mskip1.5mu]{}\<[E]%
\ColumnHook
\end{hscode}\resethooks
Although the input \ensuremath{[\mskip1.5mu \Varid{a}\mskip1.5mu]} is an inductive datatype, \ensuremath{\Varid{qsort}} is not a catamorphism as the recursion is not performed on the sub-list \ensuremath{\Varid{as}}.
Neither is it an anamorphism, since the output is not produced in the head-and-recursion manner.

Felleisen et al.\ \cite{FFR18How} referred to this form of recursive programs as
\emph{generative recursion} since the input \ensuremath{\Varid{a}\mathbin{:}\Varid{as}} is used to generate a set of
sub-problems, namely \ensuremath{\Varid{l}} and \ensuremath{\Varid{r}}, which are recursively solved, and their
solutions are combined to solve the overall problem \ensuremath{\Varid{a}\mathbin{:}\Varid{as}}.
The structure of computing \ensuremath{\Varid{qsort}} is manifested in the following rewrite
of \ensuremath{\Varid{qsort}}:
\begin{hscode}\SaveRestoreHook
\column{B}{@{}>{\hspre}l<{\hspost}@{}}%
\column{19}{@{}>{\hspre}l<{\hspost}@{}}%
\column{23}{@{}>{\hspre}l<{\hspost}@{}}%
\column{27}{@{}>{\hspre}l<{\hspost}@{}}%
\column{E}{@{}>{\hspre}l<{\hspost}@{}}%
\>[B]{}\Varid{qsort'}\mathbin{::}\Conid{Ord}\;\Varid{a}\Rightarrow [\mskip1.5mu \Varid{a}\mskip1.5mu]\to [\mskip1.5mu \Varid{a}\mskip1.5mu]{}\<[E]%
\\
\>[B]{}\Varid{qsort'}\mathrel{=}\Varid{combine}\hsdot{\circ }{.}\Varid{fmap}\;\Varid{qsort'}\hsdot{\circ }{.}\Varid{partition}{}\<[E]%
\\[\blanklineskip]%
\>[B]{}\Varid{partition}\mathbin{::}\Conid{Ord}\;\Varid{a}\Rightarrow [\mskip1.5mu \Varid{a}\mskip1.5mu]\to \Conid{TreeF}\;\Varid{a}\;[\mskip1.5mu \Varid{a}\mskip1.5mu]{}\<[E]%
\\
\>[B]{}\Varid{partition}\;[\mskip1.5mu \mskip1.5mu]{}\<[19]%
\>[19]{}\mathrel{=}\Conid{Empty}{}\<[E]%
\\
\>[B]{}\Varid{partition}\;(\Varid{a}\mathbin{:}\Varid{as}){}\<[19]%
\>[19]{}\mathrel{=}\Conid{Node}\;{}\<[27]%
\>[27]{}[\mskip1.5mu \Varid{b}\mid \Varid{b}\leftarrow \Varid{as},\Varid{b}\mathbin{<}\Varid{a}\mskip1.5mu]\;\Varid{a}\;[\mskip1.5mu \Varid{b}\mid \Varid{b}\leftarrow \Varid{as},\Varid{b}\geq \Varid{a}\mskip1.5mu]{}\<[E]%
\\[\blanklineskip]%
\>[B]{}\Varid{combine}\mathbin{::}\Conid{TreeF}\;\Varid{a}\;[\mskip1.5mu \Varid{a}\mskip1.5mu]\to [\mskip1.5mu \Varid{a}\mskip1.5mu]{}\<[E]%
\\
\>[B]{}\Varid{combine}\;\Conid{Empty}{}\<[23]%
\>[23]{}\mathrel{=}[\mskip1.5mu \mskip1.5mu]{}\<[E]%
\\
\>[B]{}\Varid{combine}\;(\Conid{Node}\;\Varid{l}\;\Varid{x}\;\Varid{r}){}\<[23]%
\>[23]{}\mathrel{=}\Varid{l}\plus [\mskip1.5mu \Varid{x}\mskip1.5mu]\plus \Varid{r}{}\<[E]%
\ColumnHook
\end{hscode}\resethooks
The functor $\ensuremath{\Conid{TreeF}\;\Varid{a}\;\Varid{x}\mathrel{=}\Conid{Empty}\mid \Conid{Node}\;\Varid{x}\;\Varid{a}\;\Varid{x}}$ governs the recursive call structure, which is a binary tree.
The \ensuremath{(\Conid{TreeF}\;\Varid{a})}-coalgebra \ensuremath{\Varid{partition}} divides a problem (if not
trivial) into two sub-problems, and the \ensuremath{(\Conid{TreeF}\;\Varid{a})}-algebra \ensuremath{\Varid{combine}}
concatenates the results of sub-problems to form a solution to the
whole problem.

It is worth noting that, quicksort, as well as many other sorting
algorithms such as merge sort can be understood as the combination of
catamorphisms and anamorphisms in various ways, leading to numerous
dualities between various sorting
algorithms~\cite{HinzeJHWM12,HinzeMW13}, but we will not explore that
further here.

\begin{recscheme}[\ensuremath{\Varid{hylo}}]\label{scm:hylo}
Abstracting the pattern of divide-and-conquer algorithms like \ensuremath{\Varid{qsort}} results
in the recursion scheme for \emph{hylomorphisms}:
\begin{hscode}\SaveRestoreHook
\column{B}{@{}>{\hspre}l<{\hspost}@{}}%
\column{20}{@{}>{\hspre}l<{\hspost}@{}}%
\column{E}{@{}>{\hspre}l<{\hspost}@{}}%
\>[B]{}\Varid{hylo}\mathbin{::}\Conid{Functor}\;\Varid{f}{}\<[20]%
\>[20]{}\Rightarrow (\Varid{f}\;\Varid{a}\to \Varid{a})\to (\Varid{c}\to \Varid{f}\;\Varid{c})\to \Varid{c}\to \Varid{a}{}\<[E]%
\\
\>[B]{}\Varid{hylo}\;\Varid{a}\;\Varid{c}\mathrel{=}\Varid{a}\hsdot{\circ }{.}\Varid{fmap}\;(\Varid{hylo}\;\Varid{a}\;\Varid{c})\hsdot{\circ }{.}\Varid{c}{}\<[E]%
\ColumnHook
\end{hscode}\resethooks
The name is due to Meijer et al.\ \cite{MFP91Fun} and is a term from Aristotelian philosophy
that objects are compounded of matter and form, where the prefix hylo- (Greek
\textgreek{ὕλη-}) means `matter'.
\end{recscheme}

Hylomorphisms are highly expressive.
In fact, all recursion schemes in this paper can be defined as special cases of
hylomorphisms,
and Hu et al.\ \cite{Hu99} showed a mechanical way to transform almost all recursive
functions in practice into hylomorphisms.
In particular, hylomorphisms subsume both catamorphisms and anamorphisms:
for all \ensuremath{\Varid{alg}\mathbin{::}\Varid{f}\;\Varid{a}\to \Varid{a}} and \ensuremath{\Varid{coalg}\mathbin{::}\Varid{c}\to \Varid{f}\;\Varid{c}}, we have
\[
\ensuremath{\Varid{cata}\;\Varid{alg}\mathrel{=}\Varid{hylo}\;\Varid{alg}\;\Varid{in}^\circ}  \qquad\text{and}\qquad \ensuremath{\Varid{ana}\;\Varid{coalg}\mathrel{=}\Varid{hylo}\;\Varid{Out}^\circ\;\Varid{coalg}}.
\]

However, the expressiveness of \ensuremath{\Varid{hylo}} comes at a cost: even when both \ensuremath{\Varid{alg}\mathbin{::}\Varid{f}\;\Varid{a}\to \Varid{a}} and \ensuremath{\Varid{coalg}\mathbin{::}\Varid{c}\to \Varid{f}\;\Varid{c}} are total functions, \ensuremath{\Varid{hylo}\;\Varid{alg}\;\Varid{coalg}} may not be total (in contrast, \ensuremath{\Varid{cata}\;\Varid{alg}} and \ensuremath{\Varid{ana}\;\Varid{coalg}} are always total whenever \ensuremath{\Varid{alg}} and \ensuremath{\Varid{coalg}} are).
Intuitively, it is because the coalgebra \ensuremath{\Varid{coalg}} may infinitely generate sub-problems while the algebra \ensuremath{\Varid{alg}} may require all subproblems solved to solve the whole problem.
\begin{example}
As an instance of the problematic situation, consider a coalgebra
\begin{hscode}\SaveRestoreHook
\column{B}{@{}>{\hspre}l<{\hspost}@{}}%
\column{E}{@{}>{\hspre}l<{\hspost}@{}}%
\>[B]{}\Varid{geo}\mathbin{::}\Conid{Integer}\to \Conid{ListF}\;\Conid{Double}\;\Conid{Integer}{}\<[E]%
\\
\>[B]{}\Varid{geo}\;\Varid{n}\mathrel{=}\Conid{Cons}\;(\mathrm{1}\mathbin{/}\Varid{fromIntegral}\;\Varid{n})\;(\mathrm{2}\mathbin{*}\Varid{n}){}\<[E]%
\ColumnHook
\end{hscode}\resethooks
which generates the geometric sequence $[\frac{1}{n},\ \frac{1}{2n},\ \frac{1}{4n},\ \frac{1}{8n},\ ..]$, and an algebra
\begin{hscode}\SaveRestoreHook
\column{B}{@{}>{\hspre}l<{\hspost}@{}}%
\column{17}{@{}>{\hspre}l<{\hspost}@{}}%
\column{E}{@{}>{\hspre}l<{\hspost}@{}}%
\>[B]{}\Varid{sum}\mathbin{::}\Conid{ListF}\;\Conid{Double}\;\Conid{Double}\to \Conid{Double}{}\<[E]%
\\
\>[B]{}\Varid{sum}\;\Conid{Nil}{}\<[17]%
\>[17]{}\mathrel{=}\mathrm{0}{}\<[E]%
\\
\>[B]{}\Varid{sum}\;(\Conid{Cons}\;\Varid{n}\;\Varid{p}){}\<[17]%
\>[17]{}\mathrel{=}\Varid{n}\mathbin{+}\Varid{p}{}\<[E]%
\ColumnHook
\end{hscode}\resethooks
which sums a sequence.
Both \ensuremath{\Varid{geo}} and \ensuremath{\Varid{sum}} are total Haskell functions, but the function \ensuremath{\Varid{zeno}\mathrel{=}\Varid{hylo}\;\Varid{sum}\;\Varid{geo}} diverges for all input \ensuremath{\Varid{i}\mathbin{::}\Conid{Integer}}.
(It does not mean that Achilles can never overtake the tortoise---\ensuremath{\Varid{zeno}} diverges because it really tries to add up an infinite sequence rather than taking the limit.)
\end{example}

\subsubsection{Recover Totality}
One way to tame the well-definedness of \ensuremath{\Varid{hylo}} is to consider
coalgebras \ensuremath{\Varid{coalg}\mathbin{::}\Varid{c}\to \Varid{f}\;\Varid{c}} with the special properties that the equation
\begin{equation}\label{eq:hylo:eq}
\ensuremath{\Varid{h}\mathrel{=}\Varid{alg}\hsdot{\circ }{.}\Varid{fmap}\;\Varid{h}\hsdot{\circ }{.}\Varid{coalg}}
\end{equation}
has a unique solution \ensuremath{\Varid{h}\mathbin{::}\Varid{c}\to \Varid{a}} for \emph{all} algebras \ensuremath{\Varid{alg}\mathbin{::}\Varid{f}\;\Varid{a}\to \Varid{a}}.
Such coalgebras are called \emph{recursive coalgebras}.
Dually, one can also consider \emph{corecursive algebras} \ensuremath{\Varid{alg}} that make
(\ref{eq:hylo:eq}) have a unique solution for all \ensuremath{\Varid{coalg}}.
For example, the coalgebra \ensuremath{\Varid{in}^\circ\mathbin{::}\mu\;\Varid{f}\to \Varid{f}\;(\mu\;\Varid{f})} is recursive, since 
the equation
\[
\ensuremath{\Varid{h}\mathrel{=}\Varid{alg}\hsdot{\circ }{.}\Varid{fmap}\;\Varid{h}\hsdot{\circ }{.}\Varid{in}^\circ} \quad\iff\quad \ensuremath{\Varid{h}\hsdot{\circ }{.}\Conid{In}\mathrel{=}\Varid{alg}\hsdot{\circ }{.}\Varid{fmap}\;\Varid{h}}
\]
has a unique solution by property (\ref{eq:initial}) of the initial algebra.
Dually, \ensuremath{\Varid{Out}^\circ\mathbin{::}\Varid{f}\;(\nu\;\Varid{f})\to \nu\;\Varid{f}} is a corecursive algebra by (\ref{eq:final}).

Besides these two basic examples, quite some effort has been made in searching 
for more recursive coalgebras (and corecursive algebras):
Capretta et al.~\cite{capretta06reccoalg} first show that it is possible to construct new recursive
coalgebras from existing ones using comonads, and later 
Hinze et al.\ \cite{HWG15Conj} show a more general technique using adjunctions and conjugate 
pairs. 
With these techniques, all recursion schemes on (co)inductive datatypes
presented in this paper can be uniformly understood as hylomorphisms with a
recursive coalgebra or corecursive algebra.
However, we shall not emphasize this perspective in this paper since it sometimes
involves non-trivial category theory to massage a recursion scheme into a
hylomorphism with a recursive coalgebra (or a corecursive algebra).

\begin{example}
The coalgebra \ensuremath{\Varid{partition}\mathbin{::}[\mskip1.5mu \Varid{a}\mskip1.5mu]\to \Conid{TreeF}\;\Varid{a}\;[\mskip1.5mu \Varid{a}\mskip1.5mu]} above is recursive (when only
finite lists are allowed as input).
This can be proved by an easy inductive argument:
for any total \ensuremath{\Varid{alg}\mathbin{::}\Conid{TreeF}\;\Varid{a}\;\Varid{b}\to \Varid{b}}, suppose that \ensuremath{\Varid{h}\mathbin{::}[\mskip1.5mu \Varid{a}\mskip1.5mu]\to \Varid{b}} satisfies
\begin{equation}\label{eq:hylo:part}
\ensuremath{\Varid{h}\mathrel{=}\Varid{alg}\hsdot{\circ }{.}\Varid{fmap}\;\Varid{h}\hsdot{\circ }{.}\Varid{partition}}.
\end{equation}
Given any finite list \ensuremath{\Varid{xs}}, we show \ensuremath{\Varid{h}\;\Varid{xs}} is determined by \ensuremath{\Varid{alg}} by an induction
on \ensuremath{\Varid{xs}}.
For the base case \ensuremath{\Varid{xs}\mathrel{=}[\mskip1.5mu \mskip1.5mu]}, we have
\[
\ensuremath{\Varid{h}\;[\mskip1.5mu \mskip1.5mu]\mathrel{=}\Varid{alg}\;(\Varid{fmap}\;\Varid{h}\;(\Varid{partition}\;[\mskip1.5mu \mskip1.5mu]))\mathrel{=}\Varid{alg}\;(\Varid{fmap}\;\Varid{h}\;\Conid{Empty})\mathrel{=}\Varid{alg}\;\Conid{Empty}}.
\]
For the inductive case \ensuremath{\Varid{xs}\mathrel{=}\Varid{y}\mathbin{:}\Varid{ys}}, we have
\begin{align*}
\ensuremath{\Varid{h}\;(\Varid{y}\mathbin{:}\Varid{ys})} & = \ensuremath{\Varid{alg}\;(\Varid{fmap}\;\Varid{h}\;(\Varid{partition}\;(\Varid{y}\mathbin{:}\Varid{ys})))} \\
           & = \ensuremath{\Varid{alg}\;(\Varid{fmap}\;\Varid{h}\;(\Conid{Node}\;\Varid{ls}\;\Varid{y}\;\Varid{rs}))} \\
           & = \ensuremath{\Varid{alg}\;(\Conid{Node}\;(\Varid{h}\;\Varid{ls})\;\Varid{y}\;(\Varid{h}\;\Varid{rs}))}
\end{align*}
where \ensuremath{\Varid{ls}\mathrel{=}[\mskip1.5mu \Varid{l}\mid \Varid{l}\leftarrow \Varid{ys},\Varid{l}\mathbin{<}\Varid{y}\mskip1.5mu]} and \ensuremath{\Varid{rs}\mathrel{=}[\mskip1.5mu \Varid{r}\mid \Varid{r}\leftarrow \Varid{ys},\Varid{r}\geq \Varid{a}\mskip1.5mu]} are
strictly smaller than \ensuremath{\Varid{xs}\mathrel{=}\Varid{y}\mathbin{:}\Varid{ys}}, and thus \ensuremath{\Varid{h}\;\Varid{ls}} and \ensuremath{\Varid{h}\;\Varid{rs}} are uniquely
determined by \ensuremath{\Varid{alg}}. Consequently, \ensuremath{\Varid{h}\;(\Varid{y}\mathbin{:}\Varid{ys})} is uniquely determined by \ensuremath{\Varid{alg}}.
Thus we conclude that \ensuremath{\Varid{h}} satisfying the hylo equation (\ref{eq:hylo:part}) is
unique.
\end{example}

\subsubsection{Aside: Metamorphisms}
If we separate the producing and consuming phases of a hylomorphism \ensuremath{\Varid{hylo}\;\Varid{alg}\;\Varid{coalg}} for some recursive \ensuremath{\Varid{coalg}}, we have the following equations (both follow
from the uniqueness of the solution to hylomorphism equations with {recursive}
coalgebra \ensuremath{\Varid{coalg}}):
\begin{align*}
\ensuremath{\Varid{hylo}\;\Varid{alg}\;\Varid{coalg}} &= \ensuremath{\Varid{cata}\;\Varid{alg}\hsdot{\circ }{.}\Varid{hylo}\;\Conid{In}\;\Varid{coalg}} \\
                 &= \ensuremath{\Varid{cata}\;\Varid{alg}\hsdot{\circ }{.}\nu 2 \mu\hsdot{\circ }{.}\Varid{ana}\;\Varid{coalg}}
\end{align*}
where \ensuremath{\nu 2 \mu\mathrel{=}\Varid{hylo}\;\Conid{In}\;\Varid{out}\mathbin{::}\nu\;\Varid{f}\to \mu\;\Varid{f}} is the \emph{partial} function
that converts the subset of finite elements of a coinductive datatype into its
inductive counterpart.
Thus, loosely speaking, a \ensuremath{\Varid{hylo}} is a \ensuremath{\Varid{cata}} after an \ensuremath{\Varid{ana}}.
The opposite direction of composition can also be considered:
\begin{hscode}\SaveRestoreHook
\column{B}{@{}>{\hspre}l<{\hspost}@{}}%
\column{7}{@{}>{\hspre}l<{\hspost}@{}}%
\column{E}{@{}>{\hspre}l<{\hspost}@{}}%
\>[B]{}\Varid{meta}{}\<[7]%
\>[7]{}\mathbin{::}(\Conid{Functor}\;\Varid{f},\Conid{Functor}\;\Varid{g})\Rightarrow (\Varid{c}\to \Varid{g}\;\Varid{c})\to (\Varid{f}\;\Varid{c}\to \Varid{c})\to \mu\;\Varid{f}\to \nu\;\Varid{g}{}\<[E]%
\\
\>[B]{}\Varid{meta}\;\Varid{coalg}\;\Varid{alg}\mathrel{=}\Varid{ana}\;\Varid{coalg}\hsdot{\circ }{.}\Varid{cata}\;\Varid{alg}{}\<[E]%
\ColumnHook
\end{hscode}\resethooks
which produces functions called
\emph{metamorphisms} by Gibbons~\cite{Gibbons05Meta}
because they \emph{metamorphose} data represented
by functor \ensuremath{\Varid{f}} to \ensuremath{\Varid{g}}.
Unlike hylomorphisms, the producing and consuming phases in metamorphisms cannot be straightforwardly fused into a single recursive function.
Gibbons \cite{Gibbons05Meta,Gib19Cod} gives conditions for doing this when \ensuremath{\Varid{f}} is \ensuremath{\Conid{ListF}}, but we will not expand on this in this paper.

\section{Accumulations}
\label{sec:curry}

Accumulating parameters are a well known technique for optimizing recursive functions.
An example is optimizing the following \ensuremath{\Varid{reverse}} function:
\begin{hscode}\SaveRestoreHook
\column{B}{@{}>{\hspre}l<{\hspost}@{}}%
\column{13}{@{}>{\hspre}l<{\hspost}@{}}%
\column{23}{@{}>{\hspre}l<{\hspost}@{}}%
\column{E}{@{}>{\hspre}l<{\hspost}@{}}%
\>[B]{}\Varid{reverse}\mathbin{::}{}\<[13]%
\>[13]{}[\mskip1.5mu \Varid{a}\mskip1.5mu]\to [\mskip1.5mu \Varid{a}\mskip1.5mu]{}\<[E]%
\\
\>[B]{}\Varid{reverse}\;{}\<[13]%
\>[13]{}[\mskip1.5mu \mskip1.5mu]{}\<[23]%
\>[23]{}\mathrel{=}[\mskip1.5mu \mskip1.5mu]{}\<[E]%
\\
\>[B]{}\Varid{reverse}\;{}\<[13]%
\>[13]{}(\Varid{x}\mathbin{:}\Varid{xs}){}\<[23]%
\>[23]{}\mathrel{=}\Varid{reverse}\;\Varid{xs}\plus [\mskip1.5mu \Varid{x}\mskip1.5mu]{}\<[E]%
\ColumnHook
\end{hscode}\resethooks
This can be transformed from running in quadratic time
(due to the fact that \ensuremath{\Varid{xs}\plus \Varid{ys}} runs in $\mathcal{O}(\ensuremath{\Varid{length}\;\Varid{xs}})$ time)
to linear time by first generalizing the function with an additional
parameter---an \emph{accumulating parameter} \ensuremath{\Varid{ys}}:
\begin{hscode}\SaveRestoreHook
\column{B}{@{}>{\hspre}l<{\hspost}@{}}%
\column{21}{@{}>{\hspre}l<{\hspost}@{}}%
\column{E}{@{}>{\hspre}l<{\hspost}@{}}%
\>[B]{}\Varid{revCat}\mathbin{::}[\mskip1.5mu \Varid{a}\mskip1.5mu]\to [\mskip1.5mu \Varid{a}\mskip1.5mu]\to [\mskip1.5mu \Varid{a}\mskip1.5mu]{}\<[E]%
\\
\>[B]{}\Varid{revCat}\;\Varid{ys}\;[\mskip1.5mu \mskip1.5mu]{}\<[21]%
\>[21]{}\mathrel{=}\Varid{ys}{}\<[E]%
\\
\>[B]{}\Varid{revCat}\;\Varid{ys}\;(\Varid{x}\mathbin{:}\Varid{xs}){}\<[21]%
\>[21]{}\mathrel{=}\Varid{revCat}\;(\Varid{x}\mathbin{:}\Varid{ys})\;\Varid{xs}{}\<[E]%
\ColumnHook
\end{hscode}\resethooks
This specializes to \ensuremath{\Varid{reverse}} by letting \ensuremath{\Varid{reverse}\mathrel{=}\Varid{recCat}\;[\mskip1.5mu \mskip1.5mu]}.
This pattern of scanning a list from left to right and accumulating a parameter
at the same time is abstracted as the Haskell function \ensuremath{\Varid{foldl}}:
\begin{hscode}\SaveRestoreHook
\column{B}{@{}>{\hspre}l<{\hspost}@{}}%
\column{19}{@{}>{\hspre}l<{\hspost}@{}}%
\column{E}{@{}>{\hspre}l<{\hspost}@{}}%
\>[B]{}\Varid{foldl}\mathbin{::}(\Varid{b}\to \Varid{a}\to \Varid{b})\to \Varid{b}\to [\mskip1.5mu \Varid{a}\mskip1.5mu]\to \Varid{b}{}\<[E]%
\\
\>[B]{}\Varid{foldl}\;\Varid{f}\;\Varid{e}\;[\mskip1.5mu \mskip1.5mu]{}\<[19]%
\>[19]{}\mathrel{=}\Varid{e}{}\<[E]%
\\
\>[B]{}\Varid{foldl}\;\Varid{f}\;\Varid{e}\;(\Varid{x}\mathbin{:}\Varid{xs}){}\<[19]%
\>[19]{}\mathrel{=}\Varid{foldl}\;\Varid{f}\;(\Varid{f}\;\Varid{e}\;\Varid{x})\;\Varid{xs}{}\<[E]%
\ColumnHook
\end{hscode}\resethooks
which specializes to \ensuremath{\Varid{revCat}} for \ensuremath{\Varid{f}\mathrel{=}\lambda \Varid{ys}\;\Varid{x}\to \Varid{x}\mathbin{:}\Varid{ys}}.
Similar to \ensuremath{\Varid{foldr}}, \ensuremath{\Varid{foldl}} follows the structure of the input---a base
case for \ensuremath{[\mskip1.5mu \mskip1.5mu]} and an inductive case for \ensuremath{\Varid{x}\mathbin{:}\Varid{xs}}.
What differs is that \ensuremath{\Varid{foldl}} has an argument \ensuremath{\Varid{e}} varied during the recursion.

The pattern of accumulation is not limited to lists. 
For example, consider writing a program that transforms a binary tree labelled
with integers to the tree whose nodes are relabelled with the \emph{sum} of the
labels along the path from the root in the original tree.
A natural idea is to keep an accumulating parameter for the sum of labels from the root:
\begin{hscode}\SaveRestoreHook
\column{B}{@{}>{\hspre}l<{\hspost}@{}}%
\column{3}{@{}>{\hspre}l<{\hspost}@{}}%
\column{29}{@{}>{\hspre}l<{\hspost}@{}}%
\column{32}{@{}>{\hspre}l<{\hspost}@{}}%
\column{E}{@{}>{\hspre}l<{\hspost}@{}}%
\>[B]{}\Varid{relabel}\mathbin{::}\mu\;(\Conid{TreeF}\;\Conid{Integer})\to \Conid{Integer}\to \mu\;(\Conid{TreeF}\;\Conid{Integer}){}\<[E]%
\\
\>[B]{}\Varid{relabel}\;(\Conid{In}\;\Conid{Empty})\;{}\<[29]%
\>[29]{}\Varid{s}{}\<[32]%
\>[32]{}\mathrel{=}\Conid{In}\;\Conid{Empty}{}\<[E]%
\\
\>[B]{}\Varid{relabel}\;(\Conid{In}\;(\Conid{Node}\;\Varid{l}\;\Varid{e}\;\Varid{r}))\;{}\<[29]%
\>[29]{}\Varid{s}{}\<[32]%
\>[32]{}\mathrel{=}\Conid{In}\;(\Conid{Node}\;(\Varid{relabel}\;\Varid{l}\;\Varid{s'})\;\Varid{s'}\;(\Varid{relabel}\;\Varid{r}\;\Varid{s'})){}\<[E]%
\\
\>[B]{}\hsindent{3}{}\<[3]%
\>[3]{}\mathbf{where}\;\Varid{s'}\mathrel{=}\Varid{s}\mathbin{+}\Varid{e}{}\<[E]%
\ColumnHook
\end{hscode}\resethooks
In the \ensuremath{\Conid{Node}} case, the current accumulating parameter \ensuremath{\Varid{s}} is updated to \ensuremath{\Varid{s'}} for
both of the subtrees, but we can certainly accumulate the parameter for the
subtrees using other accumulating strategies.
In general, an accumulating strategy can be captured as a function of type 
\[\ensuremath{\forall \Varid{x}\hsforall \hsdot{\circ }{.}\Conid{TreeF}\;\Conid{Integer}\;\Varid{x}\to \Conid{Integer}\to \Conid{TreeF}\;\Conid{Integer}\;(\Varid{x},\Conid{Integer})}\]
For example, the strategy for \ensuremath{\Varid{relabel}} is
\begin{hscode}\SaveRestoreHook
\column{B}{@{}>{\hspre}l<{\hspost}@{}}%
\column{27}{@{}>{\hspre}l<{\hspost}@{}}%
\column{E}{@{}>{\hspre}l<{\hspost}@{}}%
\>[B]{}\Varid{st}_{\Varid{relabel}}\;\Conid{Empty}\;\Varid{s}{}\<[27]%
\>[27]{}\mathrel{=}\Conid{Empty}{}\<[E]%
\\
\>[B]{}\Varid{st}_{\Varid{relabel}}\;(\Conid{Node}\;\Varid{l}\;\Varid{e}\;\Varid{r})\;\Varid{s}{}\<[27]%
\>[27]{}\mathrel{=}\Conid{Node}\;(\Varid{l},\Varid{s'})\;\Varid{e}\;(\Varid{r},\Varid{s'})\;\mathbf{where}\;\Varid{s'}\mathrel{=}\Varid{s}\mathbin{+}\Varid{e}{}\<[E]%
\ColumnHook
\end{hscode}\resethooks
Notice that the type above is polymorphic over \ensuremath{\Varid{x}}, which means that the
accumulation cannot depend on the subtrees.
This is not strictly necessary, but it reflects the pattern of most
accumulations in practice.

\begin{recscheme}[\ensuremath{\Varid{accu}}]\label{scm:accu}
Abstracting the idea for a generic initial
algebra, we obtain the recursion
scheme for \emph{accumulations}
\cite{HHT99Cal,Par02Gen,Gib00Gen}\footnote{The
recursion
scheme requires the \textsc{ghc} extension \texttt{RankNTypes} since the first argument
involves a polymorphic function.}:
\begin{hscode}\SaveRestoreHook
\column{B}{@{}>{\hspre}l<{\hspost}@{}}%
\column{20}{@{}>{\hspre}l<{\hspost}@{}}%
\column{E}{@{}>{\hspre}l<{\hspost}@{}}%
\>[B]{}\Varid{accu}\mathbin{::}\Conid{Functor}\;\Varid{f}{}\<[20]%
\>[20]{}\Rightarrow (\forall \Varid{x}\hsforall \hsdot{\circ }{.}\Varid{f}\;\Varid{x}\to \Varid{p}\to \Varid{f}\;(\Varid{x},\Varid{p})){}\<[E]%
\\
\>[20]{}\to (\Varid{f}\;\Varid{a}\to \Varid{p}\to \Varid{a})\to \mu\;\Varid{f}\to \Varid{p}\to \Varid{a}{}\<[E]%
\\
\>[B]{}\Varid{accu}\;\Varid{st}\;\Varid{alg}\;(\Conid{In}\;\Varid{t})\;\Varid{p}\mathrel{=}\Varid{alg}\;(\Varid{fmap}\;(\Varid{uncurry}\;(\Varid{accu}\;\Varid{st}\;\Varid{alg}))\;(\Varid{st}\;\Varid{t}\;\Varid{p}))\;\Varid{p}{}\<[E]%
\ColumnHook
\end{hscode}\resethooks
\end{recscheme}
\noindent Using the recursion scheme, the \ensuremath{\Varid{relabel}} function can be rewritten as
\begin{hscode}\SaveRestoreHook
\column{B}{@{}>{\hspre}l<{\hspost}@{}}%
\column{3}{@{}>{\hspre}l<{\hspost}@{}}%
\column{21}{@{}>{\hspre}l<{\hspost}@{}}%
\column{E}{@{}>{\hspre}l<{\hspost}@{}}%
\>[B]{}\Varid{relabel'}\mathbin{::}\mu\;(\Conid{TreeF}\;\Conid{Integer})\to \Conid{Integer}\to \mu\;(\Conid{TreeF}\;\Conid{Integer}){}\<[E]%
\\
\>[B]{}\Varid{relabel'}\mathrel{=}\Varid{accu}\;\Varid{st}_{\Varid{relabel}}\;\Varid{alg}\;\mathbf{where}{}\<[E]%
\\
\>[B]{}\hsindent{3}{}\<[3]%
\>[3]{}\Varid{alg}\;\Conid{Empty}\;{}\<[21]%
\>[21]{}\Varid{s}\mathrel{=}\Conid{In}\;\Conid{Empty}{}\<[E]%
\\
\>[B]{}\hsindent{3}{}\<[3]%
\>[3]{}\Varid{alg}\;(\Conid{Node}\;\Varid{l}\;\anonymous \;\Varid{r})\;{}\<[21]%
\>[21]{}\Varid{s}\mathrel{=}\Conid{In}\;(\Conid{Node}\;\Varid{l}\;\Varid{s}\;\Varid{r}){}\<[E]%
\ColumnHook
\end{hscode}\resethooks

\begin{example}
In \autoref{ex:dsl}, the semantics function \ensuremath{\Varid{interp}} is written as a catamorphism
into a function type \ensuremath{\Conid{Map}\;\Conid{Int}\;\Varid{s}\to \Varid{a}}.
With a closer look, we can see that the \ensuremath{\Conid{Map}\;\Conid{Int}\;\Varid{s}} parameter is an accumulating 
parameter, so we can more accurately express \ensuremath{\Varid{interp}} using \ensuremath{\Varid{accu}}:
\begin{hscode}\SaveRestoreHook
\column{B}{@{}>{\hspre}l<{\hspost}@{}}%
\column{3}{@{}>{\hspre}l<{\hspost}@{}}%
\column{18}{@{}>{\hspre}l<{\hspost}@{}}%
\column{21}{@{}>{\hspre}l<{\hspost}@{}}%
\column{E}{@{}>{\hspre}l<{\hspost}@{}}%
\>[B]{}\Varid{interp'}\mathbin{::}\mu\;(\Conid{ProgF}\;\Varid{s}\;\Varid{a})\to \Conid{Map}\;\Conid{Int}\;\Varid{s}\to \Varid{a}{}\<[E]%
\\
\>[B]{}\Varid{interp'}\mathrel{=}\Varid{accu}\;\Varid{st}\;\Varid{alg}\;\mathbf{where}{}\<[E]%
\\
\>[B]{}\hsindent{3}{}\<[3]%
\>[3]{}\Varid{st}\mathbin{::}\Conid{ProgF}\;\Varid{s}\;\Varid{a}\;\Varid{x}\to \Conid{Map}\;\Conid{Int}\;\Varid{s}\to \Conid{ProgF}\;\Varid{s}\;\Varid{a}\;(\Varid{x},\Conid{Map}\;\Conid{Int}\;\Varid{s}){}\<[E]%
\\
\>[B]{}\hsindent{3}{}\<[3]%
\>[3]{}\Varid{st}\;(\Conid{Ret}\;\Varid{a})\;{}\<[21]%
\>[21]{}\Varid{m}\mathrel{=}\Conid{Ret}\;\Varid{a}{}\<[E]%
\\
\>[B]{}\hsindent{3}{}\<[3]%
\>[3]{}\Varid{st}\;(\Conid{Put}\;(\Varid{i},\Varid{x})\;\Varid{k})\;{}\<[21]%
\>[21]{}\Varid{m}\mathrel{=}\Conid{Put}\;(\Varid{i},\Varid{x})\;(\Varid{k},\Varid{update}\;\Varid{m}\;\Varid{i}\;\Varid{x}){}\<[E]%
\\
\>[B]{}\hsindent{3}{}\<[3]%
\>[3]{}\Varid{st}\;(\Conid{Get}\;\Varid{i}\;\Varid{k})\;{}\<[21]%
\>[21]{}\Varid{m}\mathrel{=}\Conid{Get}\;\Varid{i}\;(\lambda \Varid{x}\to (\Varid{k}\;\Varid{x},\Varid{m})){}\<[E]%
\\[\blanklineskip]%
\>[B]{}\hsindent{3}{}\<[3]%
\>[3]{}\Varid{alg}\mathbin{::}\Conid{ProgF}\;\Varid{s}\;\Varid{a}\;\Varid{a}\to \Conid{Map}\;\Conid{Int}\;\Varid{s}\to \Varid{a}{}\<[E]%
\\
\>[B]{}\hsindent{3}{}\<[3]%
\>[3]{}\Varid{alg}\;(\Conid{Ret}\;\Varid{a})\;{}\<[18]%
\>[18]{}\Varid{m}\mathrel{=}\Varid{a}{}\<[E]%
\\
\>[B]{}\hsindent{3}{}\<[3]%
\>[3]{}\Varid{alg}\;(\Conid{Put}\;\anonymous \;\Varid{k})\;{}\<[18]%
\>[18]{}\Varid{m}\mathrel{=}\Varid{k}{}\<[E]%
\\
\>[B]{}\hsindent{3}{}\<[3]%
\>[3]{}\Varid{alg}\;(\Conid{Get}\;\Varid{i}\;\Varid{k})\;{}\<[18]%
\>[18]{}\Varid{m}\mathrel{=}\Varid{k}\;(\Varid{m}\mathbin{!}\Varid{i}){}\<[E]%
\ColumnHook
\end{hscode}\resethooks
Compared to the previous version \ensuremath{\Varid{interp}} in \autoref{ex:dsl}, this version \ensuremath{\Varid{interp'}}
singles out \ensuremath{\Varid{st}}, which controls how the memory \ensuremath{\Varid{m}} is altered by each operation,
whereas \ensuremath{\Varid{alg}} shows how each operation continues.
\end{example}

\section{Mutual Recursion}
\label{sec:mutu}

This section is about \emph{mutual recursion} in two forms: mutually recursive
functions and mutually recursive datatypes.
Mutually recursive functions are called mutumorphisms, and we will discuss their
categorical dual, which turns out to be corecursion generating elements of mutually
recursive datatypes.
\subsection{Mutumorphisms}
In Haskell, function definitions can not only be recursive but also be
\emph{mutually} recursive---two or more functions are defined in terms of each
other.
A simple example is \ensuremath{\Varid{isOdd}} and \ensuremath{\Varid{isEven}} determining the parity of a natural number:
\begin{minipage}[t]{.5\textwidth}
\vspace{-\abovedisplayskip}
\begin{hscode}\SaveRestoreHook
\column{B}{@{}>{\hspre}l<{\hspost}@{}}%
\column{23}{@{}>{\hspre}l<{\hspost}@{}}%
\column{E}{@{}>{\hspre}l<{\hspost}@{}}%
\>[B]{}\mathbf{data}\;\Conid{NatF}\;\Varid{a}\mathrel{=}\Conid{Zero}\mid \Conid{Succ}\;\Varid{a}{}\<[E]%
\\[\blanklineskip]%
\>[B]{}\Varid{isEven}\mathbin{::}\Conid{Nat}\to \Conid{Bool}{}\<[E]%
\\
\>[B]{}\Varid{isEven}\;(\Conid{In}\;\Conid{Zero}){}\<[23]%
\>[23]{}\mathrel{=}\Conid{True}{}\<[E]%
\\
\>[B]{}\Varid{isEven}\;(\Conid{In}\;(\Conid{Succ}\;\Varid{n})){}\<[23]%
\>[23]{}\mathrel{=}\Varid{isOdd}\;\Varid{n}{}\<[E]%
\ColumnHook
\end{hscode}\resethooks
\end{minipage}%
\begin{minipage}[t]{.5\textwidth}
\vspace{-\abovedisplayskip}
\begin{hscode}\SaveRestoreHook
\column{B}{@{}>{\hspre}l<{\hspost}@{}}%
\column{22}{@{}>{\hspre}l<{\hspost}@{}}%
\column{E}{@{}>{\hspre}l<{\hspost}@{}}%
\>[B]{}\mathbf{type}\;\Conid{Nat}\mathrel{=}\mu\;\Conid{NatF}{}\<[E]%
\\[\blanklineskip]%
\>[B]{}\Varid{isOdd}\mathbin{::}\Conid{Nat}\to \Conid{Bool}{}\<[E]%
\\
\>[B]{}\Varid{isOdd}\;(\Conid{In}\;\Conid{Zero}){}\<[22]%
\>[22]{}\mathrel{=}\Conid{False}{}\<[E]%
\\
\>[B]{}\Varid{isOdd}\;(\Conid{In}\;(\Conid{Succ}\;\Varid{n})){}\<[22]%
\>[22]{}\mathrel{=}\Varid{isEven}\;\Varid{n}{}\<[E]%
\ColumnHook
\end{hscode}\resethooks
\end{minipage}

\noindent
Here we are using an inductive definition of natural numbers: \ensuremath{\Conid{Zero}} is a natural number and \ensuremath{\Conid{Succ}\;\Varid{n}} is a natural number whenever \ensuremath{\Varid{n}} is.
Both \ensuremath{\Varid{isEven}} and \ensuremath{\Varid{isOdd}} are very much like a catamorphism: they have a
non-recursive definition for the base case \ensuremath{\Conid{Zero}}, and a recursive definition
for the inductive case \ensuremath{\Conid{Succ}\;\Varid{n}} in terms of the substructure \ensuremath{\Varid{n}}, except that
their recursive definitions depend on the recursive result for \ensuremath{\Varid{n}} of the
other function, instead of their own, making them not a catamorphism.

Another example of mutual recursion is the following way of computing the
\mbox{Fibonacci} number $F_i$ (i.e.\ $F_0 = 0$, $F_1 = 1$, and $F_n = F_{n-1} +
F_{n-2}$ for $n \geq 2$):
\begin{minipage}[t]{.5\textwidth}
\vspace{-\abovedisplayskip}
\begin{hscode}\SaveRestoreHook
\column{B}{@{}>{\hspre}l<{\hspost}@{}}%
\column{E}{@{}>{\hspre}l<{\hspost}@{}}%
\>[B]{}\Varid{fib}\mathbin{::}\Conid{Nat}\to \Conid{Integer}{}\<[E]%
\\
\>[B]{}\Varid{fib}\;(\Conid{In}\;\Conid{Zero})\mathrel{=}\mathrm{0}{}\<[E]%
\\
\>[B]{}\Varid{fib}\;(\Conid{In}\;(\Conid{Succ}\;\Varid{n}))\mathrel{=}\Varid{fib}\;\Varid{n}\mathbin{+}\Varid{aux}\;\Varid{n}{}\<[E]%
\ColumnHook
\end{hscode}\resethooks
\end{minipage}%
\begin{minipage}[t]{.5\textwidth}
\vspace{-\abovedisplayskip}
\begin{hscode}\SaveRestoreHook
\column{B}{@{}>{\hspre}l<{\hspost}@{}}%
\column{E}{@{}>{\hspre}l<{\hspost}@{}}%
\>[B]{}\Varid{aux}\mathbin{::}\Conid{Nat}\to \Conid{Integer}{}\<[E]%
\\
\>[B]{}\Varid{aux}\;(\Conid{In}\;\Conid{Zero})\mathrel{=}\mathrm{1}{}\<[E]%
\\
\>[B]{}\Varid{aux}\;(\Conid{In}\;(\Conid{Succ}\;\Varid{n}))\mathrel{=}\Varid{fib}\;\Varid{n}{}\<[E]%
\ColumnHook
\end{hscode}\resethooks
\end{minipage}
The value \ensuremath{\Varid{aux}\;\Varid{n}} is defined to be equal to the $(\ensuremath{\Varid{n}}-1)$-th Fibonacci number
$F_{n-1}$ for $\ensuremath{\Varid{n}} \geq 1$, and $\ensuremath{\Varid{aux}\;\mathrm{0}}$ is chosen to be $F_1 - F_0 = 1$.
Consequently, $\ensuremath{\Varid{fib}\;\mathrm{0}} = F_0$, 
\[\ensuremath{\Varid{fib}\;\mathrm{1}} = \ensuremath{\Varid{fib}\;\mathrm{0}\mathbin{+}\Varid{aux}\;\mathrm{1}} = F_0 + (F_1 - F_0) = F_1,\] and $\ensuremath{\Varid{fib}\;\Varid{n}} = \ensuremath{\Varid{fib}\;(\Varid{n}\mathbin{-}\mathrm{1})} + \ensuremath{\Varid{fib}\;(\Varid{n}\mathbin{-}\mathrm{2})}$ for $n >= 2$, which matches the definition of Fibonacci sequence.

\subsubsection{Well-Definedness} 
The recursive definitions of the examples above are well-defined, in the sense
that there is a unique solution to each group of recursive definitions regarded
as a system of equations.
For the example of \ensuremath{\Varid{fib}} and \ensuremath{\Varid{aux}}, the values at \ensuremath{\Conid{Zero}} are uniquely
determined for both functions:
\begin{align*}
\pair{\ensuremath{\Varid{fib}\;\mathrm{0}}}{\ensuremath{\Varid{aux}\;\mathrm{0}}} = \pair{0}{1}
\end{align*}
Then the values at \ensuremath{\Conid{Succ}\;\Conid{Zero}} are uniquely determined for both functions too,
according to their inductive cases:
$ \pair{\ensuremath{\Varid{fib}\;\mathrm{1}}}{\ensuremath{\Varid{aux}\;\mathrm{1}}} = \pair{1}{0} $, and so on for all inputs:
\begin{align*}
\pair{\ensuremath{\Varid{fib}\;\mathrm{2}}}{\ensuremath{\Varid{aux}\;\mathrm{2}}} = \pair{1}{1}, \ 
\pair{\ensuremath{\Varid{fib}\;\mathrm{3}}}{\ensuremath{\Varid{aux}\;\mathrm{3}}} = \pair{2}{1}, \ 
\pair{\ensuremath{\Varid{fib}\;\mathrm{4}}}{\ensuremath{\Varid{aux}\;\mathrm{4}}} = \pair{3}{2}, \ 
\dots
\end{align*}
The same line of reasoning applies too when we generalize this pattern to mutual
recursion on a generic inductive datatype.

\begin{recscheme}\label{scm:mutu}
Two mutually recursive functions on an inductive datatype
are called \emph{mutumorphisms}~\cite{Fok92Law},
and arise from the recursion scheme that defines them
both at once:%
\footnote{The name \emph{mutumorphism} is a bit special in the zoo of recursion schemes:
the prefix mutu- is from Latin rather than Greek.}
\begin{hscode}\SaveRestoreHook
\column{B}{@{}>{\hspre}l<{\hspost}@{}}%
\column{3}{@{}>{\hspre}l<{\hspost}@{}}%
\column{10}{@{}>{\hspre}l<{\hspost}@{}}%
\column{17}{@{}>{\hspre}l<{\hspost}@{}}%
\column{20}{@{}>{\hspre}l<{\hspost}@{}}%
\column{E}{@{}>{\hspre}l<{\hspost}@{}}%
\>[B]{}\Varid{mutu}\mathbin{::}\Conid{Functor}\;\Varid{f}{}\<[20]%
\>[20]{}\Rightarrow (\Varid{f}\;(\Varid{a},\Varid{b})\to \Varid{a})\to (\Varid{f}\;(\Varid{a},\Varid{b})\to \Varid{b})\to (\mu\;\Varid{f}\to \Varid{a},\mu\;\Varid{f}\to \Varid{b}){}\<[E]%
\\
\>[B]{}\Varid{mutu}\;\Varid{alg}_1\;\Varid{alg}_2\mathrel{=}(\Varid{fst}\hsdot{\circ }{.}\Varid{h},\Varid{snd}\hsdot{\circ }{.}\Varid{h}){}\<[E]%
\\
\>[B]{}\hsindent{3}{}\<[3]%
\>[3]{}\mathbf{where}\;{}\<[10]%
\>[10]{}\Varid{h}{}\<[17]%
\>[17]{}\mathrel{=}\Varid{cata}\;\Varid{alg}{}\<[E]%
\\
\>[10]{}\Varid{alg}\;\Varid{x}{}\<[17]%
\>[17]{}\mathrel{=}(\Varid{alg}_1\;\Varid{x},\Varid{alg}_2\;\Varid{x}){}\<[E]%
\ColumnHook
\end{hscode}\resethooks
in which \ensuremath{\Varid{alg}\mathbin{::}\Varid{f}\;(\Varid{a},\Varid{b})\to (\Varid{a},\Varid{b})} makes use of
\ensuremath{\Varid{alg}_1} and \ensuremath{\Varid{alg}_2} to compute the results of the two functions
being defined, from the sub-results of both functions.
\end{recscheme}

For example, using \ensuremath{\Varid{mutu}}, \ensuremath{\Varid{fib}} and \ensuremath{\Varid{aux}} can be expressed as

\savecolumns
\begin{hscode}\SaveRestoreHook
\column{B}{@{}>{\hspre}l<{\hspost}@{}}%
\column{E}{@{}>{\hspre}l<{\hspost}@{}}%
\>[B]{}\Varid{fib},\Varid{aux}\mathbin{::}\Conid{Nat}\to \Conid{Integer}{}\<[E]%
\\
\>[B]{}(\Varid{fib},\Varid{aux})\mathrel{=}\Varid{mutu}\;\Varid{f}\;\Varid{g}\;\mathbf{where}{}\<[E]%
\ColumnHook
\end{hscode}\resethooks
\begin{minipage}[t]{.5\textwidth}
\restorecolumns
\begin{hscode}\SaveRestoreHook
\column{B}{@{}>{\hspre}l<{\hspost}@{}}%
\column{3}{@{}>{\hspre}l<{\hspost}@{}}%
\column{E}{@{}>{\hspre}l<{\hspost}@{}}%
\>[3]{}\Varid{f}\;\Conid{Zero}\mathrel{=}\mathrm{0}{}\<[E]%
\\
\>[3]{}\Varid{f}\;(\Conid{Succ}\;(\Varid{n},\Varid{m}))\mathrel{=}\Varid{n}\mathbin{+}\Varid{m}{}\<[E]%
\ColumnHook
\end{hscode}\resethooks
\end{minipage}%
\begin{minipage}[t]{.5\textwidth}
\begin{hscode}\SaveRestoreHook
\column{B}{@{}>{\hspre}l<{\hspost}@{}}%
\column{3}{@{}>{\hspre}l<{\hspost}@{}}%
\column{E}{@{}>{\hspre}l<{\hspost}@{}}%
\>[3]{}\Varid{g}\;\Conid{Zero}\mathrel{=}\mathrm{1}{}\<[E]%
\\
\>[3]{}\Varid{g}\;(\Conid{Succ}\;(\Varid{n},\anonymous ))\mathrel{=}\Varid{n}{}\<[E]%
\ColumnHook
\end{hscode}\resethooks
\end{minipage}

\noindent In the unifying theory of recursion schemes of conjugate hylomorphisms,
a mutumorphism \ensuremath{\Varid{mutu}\;\Varid{alg}_1\;\Varid{alg}_2\mathbin{::}(\mu\;\Varid{f}\to \Varid{a},\mu\;\Varid{f}\to \Varid{b})} is the left-adjunct
of a catamorphism of type \ensuremath{\mu\;\Varid{f}\to (\Varid{a},\Varid{b})} via the adjunction $\Delta \dashv
\times$ between the product category $C \times C$ and some base category
$C$~\cite{HW16USR} (In the setting of this paper, $C = \catname{Set}$).  The
same adjunction also underlies a dual corecursion scheme that we explain below.

\subsection{Dual of Mutumorphisms}
Since mutumorphisms are two or more mutually recursive functions folding one
inductive datatype, we can consider the dual situation---unfolding a seed to two or more
mutually-defined coinductive datatypes.
An instructive example is recovering an expression from a G\"odel number that encodes the expression.
Consider the grammar of a simple family of arithmetic expressions:
\begin{hscode}\SaveRestoreHook
\column{B}{@{}>{\hspre}l<{\hspost}@{}}%
\column{12}{@{}>{\hspre}l<{\hspost}@{}}%
\column{E}{@{}>{\hspre}l<{\hspost}@{}}%
\>[B]{}\mathbf{data}\;\Conid{Expr}{}\<[12]%
\>[12]{}\mathrel{=}\Conid{Add}\;\Conid{Expr}\;\Conid{Term}\mid \Conid{Minus}\;\Conid{Expr}\;\Conid{Term}\mid \Conid{FromT}\;\Conid{Term}{}\<[E]%
\\[\blanklineskip]%
\>[B]{}\mathbf{data}\;\Conid{Term}{}\<[12]%
\>[12]{}\mathrel{=}\Conid{Lit}\;\Conid{Integer}\mid \Conid{Neg}\;\Conid{Term}\mid \Conid{Paren}\;\Conid{Expr}{}\<[E]%
\ColumnHook
\end{hscode}\resethooks
which is a pair of mutually-recursive datatypes.
A G\"odel numbering of this grammar \emph{invertibly} maps an \ensuremath{\Conid{Expr}} or a \ensuremath{\Conid{Term}} to a natural number, for example:
\begin{align*}
\ensuremath{\Varid{encE}\;(\Conid{Add}\;\Varid{e}\;\Varid{t})} &= 2^{\ensuremath{\Varid{encE}\;\Varid{e}}} * 3^{\ensuremath{\Varid{encT}\;\Varid{t}}}     \quad  &\ensuremath{\Varid{encT}\;(\Conid{Lit}\;\Varid{n})} &= 2^{\ensuremath{\Varid{encLit}\;\Varid{n}}}\\
\ensuremath{\Varid{encE}\;(\Conid{Minus}\;\Varid{e}\;\Varid{t})} &= 5^{\ensuremath{\Varid{encE}\;\Varid{e}}} * 7^{\ensuremath{\Varid{encT}\;\Varid{t}}}   \quad  &\ensuremath{\Varid{encT}\;(\Conid{Neg}\;\Varid{t})} &= 3^{\ensuremath{\Varid{encT}\;\Varid{t}}}\\
\ensuremath{\Varid{encE}\;(\Conid{FromT}\;\Varid{t})} &= 11^{\ensuremath{\Varid{encT}\;\Varid{t}}}                \quad  &\ensuremath{\Varid{h}\;(\Conid{Paren}\;\Varid{e})} &= 5^{\ensuremath{\Varid{encE}\;\Varid{e}}}
\end{align*}
where \ensuremath{\Varid{encLit}\;\Varid{n}\mathrel{=}\mathbf{if}\;\Varid{n}\geq \mathrm{0}\;\mathbf{then}\;\mathrm{2}\mathbin{*}\Varid{n}\mathbin{+}\mathrm{1}\;\mathbf{else}\;\mathrm{2}\mathbin{*}(\mathbin{-}\Varid{n})} invertibly maps any integer to a positive integer.
Although the encoding functions $encE$ and $encT$ clearly hint at a recursion
scheme (of folding mutually-recursive datatypes to the same type), in this
section we are interested in the opposite decoding direction:
\begin{hscode}\SaveRestoreHook
\column{B}{@{}>{\hspre}l<{\hspost}@{}}%
\column{3}{@{}>{\hspre}l<{\hspost}@{}}%
\column{8}{@{}>{\hspre}l<{\hspost}@{}}%
\column{15}{@{}>{\hspre}l<{\hspost}@{}}%
\column{16}{@{}>{\hspre}l<{\hspost}@{}}%
\column{23}{@{}>{\hspre}l<{\hspost}@{}}%
\column{24}{@{}>{\hspre}l<{\hspost}@{}}%
\column{E}{@{}>{\hspre}l<{\hspost}@{}}%
\>[B]{}\Varid{decE}\mathbin{::}\Conid{Integer}\to \Conid{Expr}{}\<[E]%
\\
\>[B]{}\Varid{decE}\;\Varid{n}\mathrel{=}\mathbf{let}\;(\Varid{e}_2,\Varid{e}_3,\Varid{e}_5,\Varid{e}_7,\Varid{e}_{11})\mathrel{=}\Varid{factorize11}\;\Varid{n}{}\<[E]%
\\
\>[B]{}\hsindent{3}{}\<[3]%
\>[3]{}\mathbf{in}\;\mathbf{if}\;\Varid{e}_2\mathbin{>}\mathrm{0}\mathrel{\vee}\Varid{e}_3\mathbin{>}\mathrm{0}\;\mathbf{then}\;\Conid{Add}\;(\Varid{decE}\;\Varid{e}_2)\;(\Varid{decT}\;\Varid{e}_3){}\<[E]%
\\
\>[3]{}\hsindent{13}{}\<[16]%
\>[16]{}\mathbf{else}\;{}\<[23]%
\>[23]{}\mathbf{if}\;\Varid{e}_5\mathbin{>}\mathrm{0}\mathrel{\vee}\Varid{e}_7\mathbin{>}\mathrm{0}{}\<[E]%
\\
\>[23]{}\mathbf{then}\;\Conid{Minus}\;(\Varid{decE}\;\Varid{e}_5)\;(\Varid{decT}\;\Varid{e}_7){}\<[E]%
\\
\>[23]{}\hsindent{1}{}\<[24]%
\>[24]{}\mathbf{else}\;\Conid{FromT}\;(\Varid{decT}\;\Varid{e}_{11}){}\<[E]%
\\[\blanklineskip]%
\>[B]{}\Varid{decT}\mathbin{::}\Conid{Integer}\to \Conid{Term}{}\<[E]%
\\
\>[B]{}\Varid{decT}\;\Varid{n}\mathrel{=}\mathbf{let}\;(\Varid{e}_2,\Varid{e}_3,\Varid{e}_5,\anonymous ,\anonymous )\mathrel{=}\Varid{factorize11}\;\Varid{n}{}\<[E]%
\\
\>[B]{}\hsindent{3}{}\<[3]%
\>[3]{}\mathbf{in}\;\mathbf{if}\;\Varid{e}_2\mathbin{>}\mathrm{0}\;\mathbf{then}\;\Conid{Lit}\;(\Varid{decLit}\;\Varid{e}_2){}\<[E]%
\\
\>[3]{}\hsindent{5}{}\<[8]%
\>[8]{}\mathbf{else}\;{}\<[15]%
\>[15]{}\mathbf{if}\;\Varid{e}_3\mathbin{>}\mathrm{0}\;\mathbf{then}\;\Conid{Neg}\;(\Varid{decT}\;\Varid{e}_3){}\<[E]%
\\
\>[15]{}\mathbf{else}\;\Conid{Paren}\;(\Varid{decE}\;\Varid{e}_5){}\<[E]%
\ColumnHook
\end{hscode}\resethooks
where \ensuremath{\Varid{factorize11}\;\Varid{n}} computes the exponents for $2, 3, 5, 7$ and $11$ in the prime factorization of $n$, and \ensuremath{\Varid{decLit}} is the inverse of \ensuremath{\Varid{encLit}}.
Functions \ensuremath{\Varid{decT}} and \ensuremath{\Varid{decE}} can correctly recover the encoded expression/term because of the fundamental theorem of arithmetic (i.e.\ the unique-prime-factorization theorem).

In the definitions of \ensuremath{\Varid{decE}} and \ensuremath{\Varid{decT}}, the choice of \ensuremath{\Varid{decE}} or \ensuremath{\Varid{decT}} when making a recursive call must match the type of the substructure at that position.
It would be convenient, if the correct choice (of \ensuremath{\Varid{decE}} or \ensuremath{\Varid{decT}}) can be automatically made based on the types---we can let a recursion scheme do the job for us.

For the generality of our recursion scheme, let us first generalize \ensuremath{\Conid{Expr}} and
\ensuremath{\Conid{Term}} to an arbitrary pair of mutually recursive datatypes, which we model as
fixed points of two bifunctors \ensuremath{\Varid{f}} and \ensuremath{\Varid{g}} given by the \ensuremath{\Conid{Bifunctor}} class:
\begin{hscode}\SaveRestoreHook
\column{B}{@{}>{\hspre}l<{\hspost}@{}}%
\column{3}{@{}>{\hspre}l<{\hspost}@{}}%
\column{E}{@{}>{\hspre}l<{\hspost}@{}}%
\>[B]{}\mathbf{class}\;\Conid{Bifunctor}\;\Varid{f}\;\mathbf{where}{}\<[E]%
\\
\>[B]{}\hsindent{3}{}\<[3]%
\>[3]{}\Varid{bimap}\mathbin{::}(\Varid{a}\to \Varid{a'})\to (\Varid{b}\to \Varid{b'})\to \Varid{f}\;\Varid{a}\;\Varid{b}\to \Varid{f}\;\Varid{a'}\;\Varid{b'}{}\<[E]%
\ColumnHook
\end{hscode}\resethooks
A bifunctor \ensuremath{\Varid{f}\;\Varid{a}\;\Varid{b}} can be understood as a functor whose domain is the product
category. A valid instance must respect the following laws:
\begin{hscode}\SaveRestoreHook
\column{B}{@{}>{\hspre}l<{\hspost}@{}}%
\column{E}{@{}>{\hspre}l<{\hspost}@{}}%
\>[B]{}\Varid{bimap}\;\Varid{id}\;\Varid{id}\mathrel{=}\Varid{id}{}\<[E]%
\\
\>[B]{}\Varid{bimap}\;(\Varid{h}\hsdot{\circ }{.}\Varid{g})\;(\Varid{k}\hsdot{\circ }{.}\Varid{j})\mathrel{=}\Varid{bimap}\;\Varid{h}\;\Varid{k}\hsdot{\circ }{.}\Varid{bimap}\;\Varid{g}\;\Varid{j}{}\<[E]%
\ColumnHook
\end{hscode}\resethooks
These laws correspond to the usual identity and composition laws of functors.

The least fixed point models finite inductive data, and the greatest
fixed point models possibly infinite coinductive data. Here we are interested
in the latter, specialized to bifunctors:

\begin{minipage}[t]{.4\textwidth}
\vspace{-\abovedisplayskip}
{
\setlength{\mathindent}{0pt}
\begin{hscode}\SaveRestoreHook
\column{B}{@{}>{\hspre}l<{\hspost}@{}}%
\column{3}{@{}>{\hspre}l<{\hspost}@{}}%
\column{E}{@{}>{\hspre}l<{\hspost}@{}}%
\>[B]{}\mathbf{newtype}\;\nu_1\;\Varid{f}\;\Varid{g}\;\mathbf{where}{}\<[E]%
\\
\>[B]{}\hsindent{3}{}\<[3]%
\>[3]{}\Varid{Out}^{\circ}_1\mathbin{::}\Varid{f}\;(\nu_1\;\Varid{f}\;\Varid{g})\;(\nu_2\;\Varid{f}\;\Varid{g})\to \nu_1\;\Varid{f}\;\Varid{g}{}\<[E]%
\ColumnHook
\end{hscode}\resethooks
}
\end{minipage}
\ \ 
\begin{minipage}[t]{.5\textwidth}
\vspace{-\abovedisplayskip}
\begin{hscode}\SaveRestoreHook
\column{B}{@{}>{\hspre}l<{\hspost}@{}}%
\column{3}{@{}>{\hspre}l<{\hspost}@{}}%
\column{E}{@{}>{\hspre}l<{\hspost}@{}}%
\>[B]{}\mathbf{newtype}\;\nu_2\;\Varid{f}\;\Varid{g}\;\mathbf{where}{}\<[E]%
\\
\>[B]{}\hsindent{3}{}\<[3]%
\>[3]{}\Varid{Out}^{\circ}_2\mathbin{::}\Varid{g}\;(\nu_1\;\Varid{f}\;\Varid{g})\;(\nu_2\;\Varid{f}\;\Varid{g})\to \nu_2\;\Varid{f}\;\Varid{g}{}\<[E]%
\ColumnHook
\end{hscode}\resethooks
\end{minipage}

\noindent For instance, \ensuremath{\Conid{Expr}} is isomorphic to \ensuremath{\nu_1\;\Conid{ExprF}\;\Conid{TermF}} and \ensuremath{\Conid{Term}} is isomorphic to \ensuremath{\nu_2\;\Conid{ExprF}\;\Conid{TermF}}:
\begin{hscode}\SaveRestoreHook
\column{B}{@{}>{\hspre}l<{\hspost}@{}}%
\column{17}{@{}>{\hspre}l<{\hspost}@{}}%
\column{E}{@{}>{\hspre}l<{\hspost}@{}}%
\>[B]{}\mathbf{data}\;\Conid{ExprF}\;\Varid{e}\;\Varid{t}{}\<[17]%
\>[17]{}\mathrel{=}\Conid{Add'}\;\Varid{e}\;\Varid{t}\mid \Conid{Minus'}\;\Varid{e}\;\Varid{t}\mid \Conid{FromT'}\;\Varid{t}{}\<[E]%
\\
\>[B]{}\mathbf{data}\;\Conid{TermF}\;\Varid{e}\;\Varid{t}{}\<[17]%
\>[17]{}\mathrel{=}\Conid{Lit'}\;\Conid{Int}\mid \Conid{Neg'}\;\Varid{t}\mid \Conid{Paren'}\;\Varid{e}{}\<[E]%
\ColumnHook
\end{hscode}\resethooks

\begin{recscheme}[\ensuremath{\Varid{comutu}}]\label{scm:comutu}
Now we can define a recursion scheme that generates a pair of elements of
mutually recursive datatypes from a single seed:
\begin{hscode}\SaveRestoreHook
\column{B}{@{}>{\hspre}l<{\hspost}@{}}%
\column{3}{@{}>{\hspre}l<{\hspost}@{}}%
\column{9}{@{}>{\hspre}l<{\hspost}@{}}%
\column{E}{@{}>{\hspre}l<{\hspost}@{}}%
\>[B]{}\Varid{comutu}{}\<[9]%
\>[9]{}\mathbin{::}(\Conid{Bifunctor}\;\Varid{f},\Conid{Bifunctor}\;\Varid{g})\Rightarrow (\Varid{c}\to \Varid{f}\;\Varid{c}\;\Varid{c})\to (\Varid{c}\to \Varid{g}\;\Varid{c}\;\Varid{c}){}\<[E]%
\\
\>[9]{}\to \Varid{c}\to (\nu_1\;\Varid{f}\;\Varid{g},\nu_2\;\Varid{f}\;\Varid{g}){}\<[E]%
\\
\>[B]{}\Varid{comutu}\;\Varid{c}_1\;\Varid{c}_2\;\Varid{s}\mathrel{=}(\Varid{x}\;\Varid{s},\Varid{y}\;\Varid{s})\;\mathbf{where}{}\<[E]%
\\
\>[B]{}\hsindent{3}{}\<[3]%
\>[3]{}\Varid{x}\mathrel{=}\Varid{Out}^{\circ}_1\hsdot{\circ }{.}\Varid{bimap}\;\Varid{x}\;\Varid{y}\hsdot{\circ }{.}\Varid{c}_1{}\<[E]%
\\
\>[B]{}\hsindent{3}{}\<[3]%
\>[3]{}\Varid{y}\mathrel{=}\Varid{Out}^{\circ}_2\hsdot{\circ }{.}\Varid{bimap}\;\Varid{x}\;\Varid{y}\hsdot{\circ }{.}\Varid{c}_2{}\<[E]%
\ColumnHook
\end{hscode}\resethooks
which remains unnamed in the literature, and so we will call this the recursion scheme for
\emph{comutumorphisms}, because of its relationship to mutumorphisms.
\end{recscheme}

\begin{example}
The \ensuremath{\Varid{comutu}} scheme renders our decoding example to become
\begin{hscode}\SaveRestoreHook
\column{B}{@{}>{\hspre}l<{\hspost}@{}}%
\column{5}{@{}>{\hspre}l<{\hspost}@{}}%
\column{7}{@{}>{\hspre}l<{\hspost}@{}}%
\column{12}{@{}>{\hspre}l<{\hspost}@{}}%
\column{19}{@{}>{\hspre}l<{\hspost}@{}}%
\column{21}{@{}>{\hspre}l<{\hspost}@{}}%
\column{E}{@{}>{\hspre}l<{\hspost}@{}}%
\>[B]{}\Varid{decExprTerm}\mathbin{::}\Conid{Integer}\to (\nu_1\;\Conid{ExprF}\;\Conid{TermF},\nu_2\;\Conid{ExprF}\;\Conid{TermF}){}\<[E]%
\\
\>[B]{}\Varid{decExprTerm}\mathrel{=}\Varid{comutu}\;\Varid{genExpr}\;\Varid{genTerm}\;\mathbf{where}{}\<[E]%
\\[\blanklineskip]%
\>[B]{}\hsindent{5}{}\<[5]%
\>[5]{}\Varid{genExpr}\mathbin{::}\Conid{Integer}\to \Conid{ExprF}\;\Conid{Integer}\;\Conid{Integer}{}\<[E]%
\\
\>[B]{}\hsindent{5}{}\<[5]%
\>[5]{}\Varid{genExpr}\;\Varid{n}\mathrel{=}{}\<[E]%
\\
\>[5]{}\hsindent{2}{}\<[7]%
\>[7]{}\mathbf{let}\;(\Varid{e}_2,\Varid{e}_3,\Varid{e}_5,\Varid{e}_7,\Varid{e}_{11})\mathrel{=}\Varid{factorize11}\;\Varid{n}{}\<[E]%
\\
\>[5]{}\hsindent{2}{}\<[7]%
\>[7]{}\mathbf{in}\;\mathbf{if}\;\Varid{e}_2\mathbin{>}\mathrm{0}\mathrel{\vee}\Varid{e}_3\mathbin{>}\mathrm{0}\;\mathbf{then}\;\Conid{Add'}\;\Varid{e}_2\;\Varid{e}_3{}\<[E]%
\\
\>[7]{}\hsindent{5}{}\<[12]%
\>[12]{}\mathbf{else}\;{}\<[19]%
\>[19]{}\mathbf{if}\;\Varid{e}_5\mathbin{>}\mathrm{0}\mathrel{\vee}\Varid{e}_7\mathbin{>}\mathrm{0}{}\<[E]%
\\
\>[19]{}\hsindent{2}{}\<[21]%
\>[21]{}\mathbf{then}\;\Conid{Minus'}\;\Varid{e}_5\;\Varid{e}_7\;\mathbf{else}\;\Conid{FromT'}\;\Varid{e}_{11}{}\<[E]%
\\[\blanklineskip]%
\>[B]{}\hsindent{5}{}\<[5]%
\>[5]{}\Varid{genTerm}\mathbin{::}\Conid{Integer}\to \Conid{TermF}\;\Conid{Integer}\;\Conid{Integer}{}\<[E]%
\\
\>[B]{}\hsindent{5}{}\<[5]%
\>[5]{}\Varid{genTerm}\;\Varid{n}\mathrel{=}{}\<[E]%
\\
\>[5]{}\hsindent{2}{}\<[7]%
\>[7]{}\mathbf{let}\;(\Varid{e}_2,\Varid{e}_3,\Varid{e}_5,\anonymous ,\anonymous )\mathrel{=}\Varid{factorize11}\;\Varid{n}{}\<[E]%
\\
\>[5]{}\hsindent{2}{}\<[7]%
\>[7]{}\mathbf{in}\;\mathbf{if}\;\Varid{e}_2\mathbin{>}\mathrm{0}\;\mathbf{then}\;\Conid{Lit'}\;(\Varid{decLit}\;\Varid{e}_2){}\<[E]%
\\
\>[7]{}\hsindent{5}{}\<[12]%
\>[12]{}\mathbf{else}\;\mathbf{if}\;\Varid{e}_3\mathbin{>}\mathrm{0}\;\mathbf{then}\;\Conid{Neg'}\;\Varid{e}_3\;\mathbf{else}\;\Conid{Paren'}\;\Varid{e}_5{}\<[E]%
\ColumnHook
\end{hscode}\resethooks
Comparing to the direct definitions of \ensuremath{\Varid{decE}} and \ensuremath{\Varid{decT}}, \ensuremath{\Varid{genTerm}} and \ensuremath{\Varid{genExpr}} are simpler as they just generate a new seed for each recursive position and recursive calls of the correct type is invoked by the recursion scheme \ensuremath{\Varid{comutu}}.
\end{example}

Theoretically, \ensuremath{\Varid{comutu}} is the adjoint unfold from the adjunction $\Delta \dashv \times$:
\ensuremath{\Varid{comutu}\;\Varid{c}_1\;\Varid{c}_2\mathbin{::}\Varid{c}\to (\nu_1\;\Varid{f}\;\Varid{g},\nu_2\;\Varid{f}\;\Varid{g})} is the right-adjunct of an anamorphism of type \ensuremath{(\Varid{c}\to \nu_1\;\Varid{f}\;\Varid{g},\Varid{c}\to \nu_2\;\Varid{f}\;\Varid{g})} in the product category $C \times C$.
A closely related adjunction ${+} \dashv \Delta$ also gives two recursion schemes for mutual recursion.
One is an adjoint fold that consumes mutually recursive datatypes, of which an example is the encoding function of G\"odel numbering discussed above, and dually an adjoint unfold that generates \ensuremath{\nu\;\Varid{f}} from seed \ensuremath{\Conid{Either}\;\Varid{c}_1\;\Varid{c}_2}, which captures \emph{mutual corecursion}.
Although attractive and practically important, we forgo an exhibition of these two recursion schemes here.

\section{Primitive (Co)Recursion}
\label{sec:para}

In this section, we investigate the pattern in recursive programs in which the original input is directly involved besides the recursively computed results, resulting in a generalization of catamorphisms---\emph{paramorphisms}.
We also discuss a generalization, \emph{zygomorphisms}, and the categorical dual \emph{apomorphisms}.

\subsection{Paramorphisms}

A wide family of recursive functions that are not directly covered by
catamorphisms are those in which the original substructures are
directly used in addition to their images under the function being defined.
An example is one of the most frequently demonstrated recursive function \ensuremath{\Varid{factorial}},
where \ensuremath{\Conid{Nat}} has been given a suitable \ensuremath{\Conid{Num}} instance:
\begin{hscode}\SaveRestoreHook
\column{B}{@{}>{\hspre}l<{\hspost}@{}}%
\column{26}{@{}>{\hspre}l<{\hspost}@{}}%
\column{E}{@{}>{\hspre}l<{\hspost}@{}}%
\>[B]{}\Varid{factorial}\mathbin{::}\Conid{Nat}\to \Conid{Nat}{}\<[E]%
\\
\>[B]{}\Varid{factorial}\;(\Conid{In}\;\Conid{Zero}){}\<[26]%
\>[26]{}\mathrel{=}\mathrm{1}{}\<[E]%
\\
\>[B]{}\Varid{factorial}\;(\Conid{In}\;(\Conid{Succ}\;\Varid{n})){}\<[26]%
\>[26]{}\mathrel{=}\Conid{In}\;(\Conid{Succ}\;\Varid{n})\mathbin{*}\Varid{factorial}\;\Varid{n}{}\<[E]%
\ColumnHook
\end{hscode}\resethooks
In the second case, besides the recursively computed result \ensuremath{\Varid{factorial}\;\Varid{n}}, the substructure \ensuremath{\Varid{n}} itself is also used, but it is not directly provided by \ensuremath{\Varid{cata}}.
A slightly more practical example is counting the number of words (more accurately, maximal sub-sequences of non-space characters) in a list of characters:
\begin{hscode}\SaveRestoreHook
\column{B}{@{}>{\hspre}l<{\hspost}@{}}%
\column{3}{@{}>{\hspre}l<{\hspost}@{}}%
\column{22}{@{}>{\hspre}l<{\hspost}@{}}%
\column{E}{@{}>{\hspre}l<{\hspost}@{}}%
\>[B]{}\Varid{wc}\mathbin{::}\mu\;(\Conid{ListF}\;\Conid{Char})\to \Conid{Integer}{}\<[E]%
\\
\>[B]{}\Varid{wc}\;(\Conid{In}\;\Conid{Nil}){}\<[22]%
\>[22]{}\mathrel{=}\mathrm{0}{}\<[E]%
\\
\>[B]{}\Varid{wc}\;(\Conid{In}\;(\Conid{Cons}\;\Varid{c}\;\Varid{cs})){}\<[22]%
\>[22]{}\mathrel{=}\mathbf{if}\;\Varid{isNewWord}\;\mathbf{then}\;\Varid{wc}\;\Varid{cs}\mathbin{+}\mathrm{1}\;\mathbf{else}\;\Varid{wc}\;\Varid{cs}{}\<[E]%
\\
\>[B]{}\hsindent{3}{}\<[3]%
\>[3]{}\mathbf{where}\;\Varid{isNewWord}\mathrel{=}\neg \;(\Varid{isSpace}\;\Varid{c})\mathrel{\wedge}(\Varid{null}\;\Varid{cs}\mathrel{\vee}\Varid{isSpace}\;(\Varid{head}\;\Varid{cs})){}\<[E]%
\ColumnHook
\end{hscode}\resethooks
Again in the second case, \ensuremath{\Varid{cs}} is used besides \ensuremath{\Varid{wc}\;\Varid{cs}}, making it not a direct instance of catamorphisms either.

To express \ensuremath{\Varid{factorial}} and \ensuremath{\Varid{wc}} with a structural recursion scheme, we can use
mutumorphisms by understanding \ensuremath{\Varid{factorial}} and \ensuremath{\Varid{wc}} as mutually defined with 
with the identity function.
For example,
\begin{hscode}\SaveRestoreHook
\column{B}{@{}>{\hspre}l<{\hspost}@{}}%
\column{3}{@{}>{\hspre}l<{\hspost}@{}}%
\column{24}{@{}>{\hspre}l<{\hspost}@{}}%
\column{E}{@{}>{\hspre}l<{\hspost}@{}}%
\>[B]{}\Varid{factorial'}\mathrel{=}\Varid{fst}\;(\Varid{mutu}\;\Varid{alg}\;\Varid{alg}_{\Varid{id}})\;\mathbf{where}{}\<[E]%
\\
\>[B]{}\hsindent{3}{}\<[3]%
\>[3]{}\Varid{alg}\;\Conid{Zero}{}\<[24]%
\>[24]{}\mathrel{=}\mathrm{1}{}\<[E]%
\\
\>[B]{}\hsindent{3}{}\<[3]%
\>[3]{}\Varid{alg}\;(\Conid{Succ}\;(\Varid{fn},\Varid{n})){}\<[24]%
\>[24]{}\mathrel{=}(\Conid{In}\;(\Conid{Succ}\;\Varid{n}))\mathbin{*}\Varid{fn}{}\<[E]%
\\[\blanklineskip]%
\>[B]{}\hsindent{3}{}\<[3]%
\>[3]{}\Varid{alg}_{\Varid{id}}\;\Conid{Zero}{}\<[24]%
\>[24]{}\mathrel{=}\Conid{In}\;\Conid{Zero}{}\<[E]%
\\
\>[B]{}\hsindent{3}{}\<[3]%
\>[3]{}\Varid{alg}_{\Varid{id}}\;(\Conid{Succ}\;(\anonymous ,\Varid{n})){}\<[24]%
\>[24]{}\mathrel{=}\Conid{In}\;(\Conid{Succ}\;\Varid{n}){}\<[E]%
\ColumnHook
\end{hscode}\resethooks
Better is to use a recursion scheme that captures this common pattern.
\begin{recscheme}\label{scm:para}
Functions given by structured recursion with access
to the original sub-parts of the input are 
called \emph{paramorphisms}, and are described by the
following scheme:
\begin{hscode}\SaveRestoreHook
\column{B}{@{}>{\hspre}l<{\hspost}@{}}%
\column{3}{@{}>{\hspre}l<{\hspost}@{}}%
\column{E}{@{}>{\hspre}l<{\hspost}@{}}%
\>[B]{}\Varid{para}\mathbin{::}\Conid{Functor}\;\Varid{f}\Rightarrow (\Varid{f}\;(\mu\;\Varid{f},\Varid{a})\to \Varid{a})\to \mu\;\Varid{f}\to \Varid{a}{}\<[E]%
\\
\>[B]{}\Varid{para}\;\Varid{alg}\mathrel{=}\Varid{alg}\hsdot{\circ }{.}\Varid{fmap}\;(\Varid{id}\,\triangle\,\Varid{para}\;\Varid{alg})\hsdot{\circ }{.}\Varid{in}^\circ\;\mathbf{where}{}\<[E]%
\\
\>[B]{}\hsindent{3}{}\<[3]%
\>[3]{}(\Varid{f}\,\triangle\,\Varid{g})\;\Varid{x}\mathrel{=}(\Varid{f}\;\Varid{x},\Varid{g}\;\Varid{x}){}\<[E]%
\ColumnHook
\end{hscode}\resethooks
The prefix para- is derived from Greek \textgreek{παρά}, meaning `beside'.
\end{recscheme}

\begin{example}
With \ensuremath{\Varid{para}}, \ensuremath{\Varid{factorial}} is defined neatly:
\begin{hscode}\SaveRestoreHook
\column{B}{@{}>{\hspre}l<{\hspost}@{}}%
\column{3}{@{}>{\hspre}l<{\hspost}@{}}%
\column{23}{@{}>{\hspre}l<{\hspost}@{}}%
\column{E}{@{}>{\hspre}l<{\hspost}@{}}%
\>[B]{}\Varid{factorial''}\mathrel{=}\Varid{para}\;\Varid{alg}\;\mathbf{where}{}\<[E]%
\\
\>[B]{}\hsindent{3}{}\<[3]%
\>[3]{}\Varid{alg}\;\Conid{Zero}{}\<[23]%
\>[23]{}\mathrel{=}\mathrm{1}{}\<[E]%
\\
\>[B]{}\hsindent{3}{}\<[3]%
\>[3]{}\Varid{alg}\;(\Conid{Succ}\;(\Varid{n},\Varid{fn})){}\<[23]%
\>[23]{}\mathrel{=}\Conid{In}\;(\Conid{Succ}\;\Varid{n})\mathbin{*}\Varid{fn}{}\<[E]%
\ColumnHook
\end{hscode}\resethooks
\end{example}

Compared with \ensuremath{\Varid{cata}}, \ensuremath{\Varid{para}} also supplies the original substructures besides their images to the algebra.
However, \ensuremath{\Varid{cata}} and \ensuremath{\Varid{para}} are interdefinable in Haskell.
Every catamorphism is simply a paramorphism that makes no use of the additional information:
\[ \ensuremath{\Varid{cata}\;\Varid{alg}\mathrel{=}\Varid{para}\;(\Varid{alg}\hsdot{\circ }{.}\Varid{fmap}\;\Varid{snd})} \]
Conversely, every paramorphism together with the identity function is a mutumorphism, which in turn is a catamorphism for a pair type \ensuremath{(\Varid{a},\Varid{b})}, or directly:
\[ \ensuremath{\Varid{para}\;\Varid{alg}\mathrel{=}\Varid{snd}\hsdot{\circ }{.}\Varid{cata}\;((\Conid{In}\hsdot{\circ }{.}\Varid{fmap}\;\Varid{fst})\,\triangle\,\Varid{alg})} \]

Sometimes the recursion scheme of paramorphisms is called \emph{primitive recursion}.
However, functions definable with paramorphisms in Haskell are beyond primitive recursive functions in computability theory because of the presence of higher order functions.
Indeed, the canonical example of non-primitive recursive function, the Ackermann function, is definable with \ensuremath{\Varid{cata}} and thus \ensuremath{\Varid{para}}:
\begin{hscode}\SaveRestoreHook
\column{B}{@{}>{\hspre}l<{\hspost}@{}}%
\column{3}{@{}>{\hspre}l<{\hspost}@{}}%
\column{5}{@{}>{\hspre}l<{\hspost}@{}}%
\column{19}{@{}>{\hspre}l<{\hspost}@{}}%
\column{23}{@{}>{\hspre}l<{\hspost}@{}}%
\column{24}{@{}>{\hspre}l<{\hspost}@{}}%
\column{E}{@{}>{\hspre}l<{\hspost}@{}}%
\>[B]{}\Varid{ack}\mathbin{::}\Conid{Nat}\to \Conid{Nat}\to \Conid{Nat}{}\<[E]%
\\
\>[B]{}\Varid{ack}\mathrel{=}\Varid{cata}\;\Varid{alg}\;\mathbf{where}{}\<[E]%
\\
\>[B]{}\hsindent{3}{}\<[3]%
\>[3]{}\Varid{alg}\mathbin{::}\Conid{NatF}\;(\Conid{Nat}\to \Conid{Nat})\to (\Conid{Nat}\to \Conid{Nat}){}\<[E]%
\\
\>[B]{}\hsindent{3}{}\<[3]%
\>[3]{}\Varid{alg}\;\Conid{Zero}{}\<[19]%
\>[19]{}\mathrel{=}\Conid{In}\hsdot{\circ }{.}\Conid{Succ}{}\<[E]%
\\
\>[B]{}\hsindent{3}{}\<[3]%
\>[3]{}\Varid{alg}\;(\Conid{Succ}\;\Varid{a_n}){}\<[19]%
\>[19]{}\mathrel{=}\Varid{cata}\;\Varid{alg'}\;\mathbf{where}{}\<[E]%
\\
\>[3]{}\hsindent{2}{}\<[5]%
\>[5]{}\Varid{alg'}\mathbin{::}\Conid{NatF}\;\Conid{Nat}\to \Conid{Nat}{}\<[E]%
\\
\>[3]{}\hsindent{2}{}\<[5]%
\>[5]{}\Varid{alg'}\;\Conid{Zero}{}\<[23]%
\>[23]{}\mathrel{=}\Varid{a_n}\;(\Conid{In}\;(\Conid{Succ}\;(\Conid{In}\;\Conid{Zero}))){}\<[E]%
\\
\>[3]{}\hsindent{2}{}\<[5]%
\>[5]{}\Varid{alg'}\;(\Conid{Succ}\;\Varid{a_{n+1,m}}){}\<[24]%
\>[24]{}\mathrel{=}\Varid{a_n}\;\Varid{a_{n+1,m}}{}\<[E]%
\ColumnHook
\end{hscode}\resethooks

\subsection{Apomorphisms}
\label{sec:apo}

Paramorphisms can be dualized to corecursion.
The algebra of a paramorphism has type \ensuremath{\Varid{f}\;(\mu\;\Varid{f},\Varid{a})\to \Varid{a}}, in which \ensuremath{\mu\;\Varid{f}} is dual to \ensuremath{\nu\;\Varid{f}}, and the pair type is dual to the \ensuremath{\Conid{Either}} type.
Thus the coalgebra of the dual recursion scheme should have type \ensuremath{\Varid{c}\to \Varid{f}\;(\Conid{Either}\;(\nu\;\Varid{f})\;\Varid{c})}.
\begin{recscheme}\label{scm:apo}
The following recursion scheme gives rise to
\emph{apomorphisms}~\cite{VeU98Fun,UuV99Pri}.
The prefix apo- comes from Greek \textgreek{απο} meaning `apart from'.
\begin{hscode}\SaveRestoreHook
\column{B}{@{}>{\hspre}l<{\hspost}@{}}%
\column{E}{@{}>{\hspre}l<{\hspost}@{}}%
\>[B]{}\Varid{apo}\mathbin{::}\Conid{Functor}\;\Varid{f}\Rightarrow (\Varid{c}\to \Varid{f}\;(\Conid{Either}\;(\nu\;\Varid{f})\;\Varid{c}))\to \Varid{c}\to \nu\;\Varid{f}{}\<[E]%
\\
\>[B]{}\Varid{apo}\;\Varid{coalg}\mathrel{=}\Varid{Out}^\circ\hsdot{\circ }{.}\Varid{fmap}\;(\Varid{either}\;\Varid{id}\;(\Varid{apo}\;\Varid{coalg}))\hsdot{\circ }{.}\Varid{coalg}{}\<[E]%
\ColumnHook
\end{hscode}\resethooks
which is sometimes called \emph{primitive corecursion}.
\end{recscheme}
Similar to anamorphisms, the coalgebra of an apomorphism generates a layer of \ensuremath{\Varid{f}}-structure in each step, but for substructures, it either generates a new seed of type \ensuremath{\Varid{c}} for corecursion as in anamorphisms, or a complete structure of \ensuremath{\nu\;\Varid{f}} and stop the corecursion there.

In the same way that \ensuremath{\Varid{cata}} and \ensuremath{\Varid{para}} are interdefinable, \ensuremath{\Varid{ana}} and \ensuremath{\Varid{apo}} are interdefinable in Haskell too, but \ensuremath{\Varid{apo}} are particularly suitable for corecursive functions in which the future output is fully known at some step.
Consider a function \ensuremath{\Varid{maphd}} from Vene and Uustalu \cite{VeU98Fun} that applies a function \ensuremath{\Varid{f}} to the first element (if there is) of a coinductive list.
\begin{hscode}\SaveRestoreHook
\column{B}{@{}>{\hspre}l<{\hspost}@{}}%
\column{E}{@{}>{\hspre}l<{\hspost}@{}}%
\>[B]{}\Varid{maphd}\mathbin{::}(\Varid{a}\to \Varid{a})\to \nu\;(\Conid{ListF}\;\Varid{a})\to \nu\;(\Conid{ListF}\;\Varid{a}){}\<[E]%
\ColumnHook
\end{hscode}\resethooks
As an anamorphism, it is expressed as
\begin{hscode}\SaveRestoreHook
\column{B}{@{}>{\hspre}l<{\hspost}@{}}%
\column{3}{@{}>{\hspre}l<{\hspost}@{}}%
\column{34}{@{}>{\hspre}l<{\hspost}@{}}%
\column{48}{@{}>{\hspre}l<{\hspost}@{}}%
\column{E}{@{}>{\hspre}l<{\hspost}@{}}%
\>[B]{}\Varid{maphd}\;\Varid{f}\mathrel{=}\Varid{ana}\;\Varid{c}\hsdot{\circ }{.}\Conid{Left}\;\mathbf{where}{}\<[E]%
\\
\>[B]{}\hsindent{3}{}\<[3]%
\>[3]{}\Varid{c}\;(\Conid{Left}\;(\Varid{Out}^\circ\;\Conid{Nil})){}\<[34]%
\>[34]{}\mathrel{=}\Conid{Nil}{}\<[E]%
\\
\>[B]{}\hsindent{3}{}\<[3]%
\>[3]{}\Varid{c}\;(\Conid{Right}\;(\Varid{Out}^\circ\;\Conid{Nil})){}\<[34]%
\>[34]{}\mathrel{=}\Conid{Nil}{}\<[E]%
\\[\blanklineskip]%
\>[B]{}\hsindent{3}{}\<[3]%
\>[3]{}\Varid{c}\;(\Conid{Left}\;(\Varid{Out}^\circ\;(\Conid{Cons}\;\Varid{x}\;\Varid{xs}))){}\<[34]%
\>[34]{}\mathrel{=}\Conid{Cons}\;(\Varid{f}\;\Varid{x})\;{}\<[48]%
\>[48]{}(\Conid{Right}\;\Varid{xs}){}\<[E]%
\\
\>[B]{}\hsindent{3}{}\<[3]%
\>[3]{}\Varid{c}\;(\Conid{Right}\;(\Varid{Out}^\circ\;(\Conid{Cons}\;\Varid{x}\;\Varid{xs}))){}\<[34]%
\>[34]{}\mathrel{=}\Conid{Cons}\;\Varid{x}\;{}\<[48]%
\>[48]{}(\Conid{Right}\;\Varid{xs}){}\<[E]%
\ColumnHook
\end{hscode}\resethooks
in which the seed for generation is of type \ensuremath{\Conid{Either}\;(\nu\;(\Conid{ListF}\;\Varid{a}))\;(\nu\;(\Conid{ListF}\;\Varid{a}))}
to distinguish if the head element has been processed.
This function is more intuitively an apomorphism since the future output is instantly known when the head element gets processed:
\begin{hscode}\SaveRestoreHook
\column{B}{@{}>{\hspre}l<{\hspost}@{}}%
\column{3}{@{}>{\hspre}l<{\hspost}@{}}%
\column{E}{@{}>{\hspre}l<{\hspost}@{}}%
\>[B]{}\Varid{maphd'}\;\Varid{f}\mathrel{=}\Varid{apo}\;\Varid{coalg}\;\mathbf{where}{}\<[E]%
\\
\>[B]{}\hsindent{3}{}\<[3]%
\>[3]{}\Varid{coalg}\;(\Varid{Out}^\circ\;\Conid{Nil})\mathrel{=}\Conid{Nil}{}\<[E]%
\\
\>[B]{}\hsindent{3}{}\<[3]%
\>[3]{}\Varid{coalg}\;(\Varid{Out}^\circ\;(\Conid{Cons}\;\Varid{x}\;\Varid{xs}))\mathrel{=}\Conid{Cons}\;(\Varid{f}\;\Varid{x})\;(\Conid{Left}\;\Varid{xs}){}\<[E]%
\ColumnHook
\end{hscode}\resethooks
Moreover, this definition is more efficient than the previous one because it avoids deconstructing and reconstructing the tail of the input list.

\begin{example}
Another instructive example of apomorphisms is inserting a value into an ordered (coinductive) list:
\begin{hscode}\SaveRestoreHook
\column{B}{@{}>{\hspre}l<{\hspost}@{}}%
\column{3}{@{}>{\hspre}l<{\hspost}@{}}%
\column{5}{@{}>{\hspre}l<{\hspost}@{}}%
\column{18}{@{}>{\hspre}l<{\hspost}@{}}%
\column{E}{@{}>{\hspre}l<{\hspost}@{}}%
\>[B]{}\Varid{insert}\mathbin{::}\Conid{Ord}\;\Varid{a}\Rightarrow \Varid{a}\to \nu\;(\Conid{ListF}\;\Varid{a})\to \nu\;(\Conid{ListF}\;\Varid{a}){}\<[E]%
\\
\>[B]{}\Varid{insert}\;\Varid{y}\mathrel{=}\Varid{apo}\;\Varid{c}\;\mathbf{where}{}\<[E]%
\\
\>[B]{}\hsindent{3}{}\<[3]%
\>[3]{}\Varid{c}\;(\Varid{Out}^\circ\;\Conid{Nil}){}\<[18]%
\>[18]{}\mathrel{=}\Conid{Cons}\;\Varid{y}\;(\Conid{Left}\;(\Varid{Out}^\circ\;\Conid{Nil})){}\<[E]%
\\
\>[B]{}\hsindent{3}{}\<[3]%
\>[3]{}\Varid{c}\;\Varid{xxs}\mathord{@}(\Varid{Out}^\circ\;(\Conid{Cons}\;\Varid{x}\;\Varid{xs})){}\<[E]%
\\
\>[3]{}\hsindent{2}{}\<[5]%
\>[5]{}\mid \Varid{y}\leq \Varid{x}{}\<[18]%
\>[18]{}\mathrel{=}\Conid{Cons}\;\Varid{y}\;(\Conid{Left}\;\Varid{xxs}){}\<[E]%
\\
\>[3]{}\hsindent{2}{}\<[5]%
\>[5]{}\mid \Varid{otherwise}{}\<[18]%
\>[18]{}\mathrel{=}\Conid{Cons}\;\Varid{x}\;(\Conid{Right}\;\Varid{xs}){}\<[E]%
\ColumnHook
\end{hscode}\resethooks
In both cases, an element \ensuremath{\Varid{y}} or \ensuremath{\Varid{x}} is emitted, and \ensuremath{\Conid{Left}\;\Varid{xxs}} makes \ensuremath{\Varid{xxs}} the rest of the output, whereas \ensuremath{\Conid{Right}\;\Varid{xs}} continues the corecursion to insert \ensuremath{\Varid{y}} into \ensuremath{\Varid{xs}}.
\end{example}

\subsection{Zygomorphisms}
When computing a recursive function on a datatype, it is usually the case that some auxiliary information about substructures is needed in addition to the images of substructures under the recursive function being computed.
For instance, when determining if a binary tree is a perfect tree---a tree in which all leaf nodes have the same depth and all interior nodes have two children---by structural recursion, %
besides checking that the left and right subtrees are both perfect, it is also needed to check that they have the same depth:
\begin{hscode}\SaveRestoreHook
\column{B}{@{}>{\hspre}l<{\hspost}@{}}%
\column{28}{@{}>{\hspre}l<{\hspost}@{}}%
\column{E}{@{}>{\hspre}l<{\hspost}@{}}%
\>[B]{}\Varid{perfect}\mathbin{::}\mu\;(\Conid{TreeF}\;\Varid{e}){}\<[28]%
\>[28]{}\to \Conid{Bool}{}\<[E]%
\\
\>[B]{}\Varid{perfect}\;(\Conid{In}\;\Conid{Empty}){}\<[28]%
\>[28]{}\mathrel{=}\Conid{True}{}\<[E]%
\\
\>[B]{}\Varid{perfect}\;(\Conid{In}\;(\Conid{Node}\;\Varid{l}\;\anonymous \;\Varid{r})){}\<[28]%
\>[28]{}\mathrel{=}\Varid{perfect}\;\Varid{l}\mathrel{\wedge}\Varid{perfect}\;\Varid{r}\mathrel{\wedge}(\Varid{depth}\;\Varid{l}\equiv \Varid{depth}\;\Varid{r}){}\<[E]%
\\[\blanklineskip]%
\>[B]{}\Varid{depth}\mathbin{::}\mu\;(\Conid{TreeF}\;\Varid{e}){}\<[28]%
\>[28]{}\to \Conid{Integer}{}\<[E]%
\\
\>[B]{}\Varid{depth}\;(\Conid{In}\;\Conid{Empty}){}\<[28]%
\>[28]{}\mathrel{=}\mathrm{0}{}\<[E]%
\\
\>[B]{}\Varid{depth}\;(\Conid{In}\;(\Conid{Node}\;\Varid{l}\;\anonymous \;\Varid{r})){}\<[28]%
\>[28]{}\mathrel{=}\mathrm{1}\mathbin{+}\Varid{max}\;(\Varid{depth}\;\Varid{l})\;(\Varid{depth}\;\Varid{r}){}\<[E]%
\ColumnHook
\end{hscode}\resethooks
The function \ensuremath{\Varid{perfect}} is not directly a catamorphism because the algebra is not
provided with \ensuremath{\Varid{depth}\;\Varid{l}} and \ensuremath{\Varid{depth}\;\Varid{r}} by the \ensuremath{\Varid{cata}} recursion scheme.
However we can define \ensuremath{\Varid{perfect}} as a paramorphism:
\begin{hscode}\SaveRestoreHook
\column{B}{@{}>{\hspre}l<{\hspost}@{}}%
\column{3}{@{}>{\hspre}l<{\hspost}@{}}%
\column{33}{@{}>{\hspre}l<{\hspost}@{}}%
\column{35}{@{}>{\hspre}l<{\hspost}@{}}%
\column{E}{@{}>{\hspre}l<{\hspost}@{}}%
\>[B]{}\Varid{perfect'}\mathrel{=}\Varid{para}\;\Varid{alg}\;\mathbf{where}{}\<[E]%
\\
\>[B]{}\hsindent{3}{}\<[3]%
\>[3]{}\Varid{alg}\;\Conid{Empty}{}\<[33]%
\>[33]{}\mathrel{=}\Conid{True}{}\<[E]%
\\
\>[B]{}\hsindent{3}{}\<[3]%
\>[3]{}\Varid{alg}\;(\Conid{Node}\;(\Varid{l},\Varid{p_l})\;\anonymous \;(\Varid{r},\Varid{p_r})){}\<[35]%
\>[35]{}\mathrel{=}\Varid{p_l}\mathrel{\wedge}\Varid{p_r}\mathrel{\wedge}(\Varid{depth}\;\Varid{l}\equiv \Varid{depth}\;\Varid{r}){}\<[E]%
\ColumnHook
\end{hscode}\resethooks
But this is inefficient because the depth of a subtree is computed repeatedly at each of its ancestor nodes, despite the fact that \ensuremath{\Varid{depth}} can be computed structurally too.
Thus we need a generalization of paramorphisms in which instead of the original structure being kept and supplied to the algebra, some auxiliary information (that can be computed structurally) is maintained along the recursion and supplied to the algebra, which leads to the following recursion scheme.
\begin{recscheme}\label{scm:zygo}
A structurally recursive function with auxiliary information is called a
\emph{zygomorphism}~\cite{Mal90Alg}:
\begin{hscode}\SaveRestoreHook
\column{B}{@{}>{\hspre}l<{\hspost}@{}}%
\column{20}{@{}>{\hspre}l<{\hspost}@{}}%
\column{E}{@{}>{\hspre}l<{\hspost}@{}}%
\>[B]{}\Varid{zygo}\mathbin{::}\Conid{Functor}\;\Varid{f}{}\<[20]%
\>[20]{}\Rightarrow (\Varid{f}\;(\Varid{a},\Varid{b})\to \Varid{a})\to (\Varid{f}\;\Varid{b}\to \Varid{b})\to \mu\;\Varid{f}\to \Varid{a}{}\<[E]%
\\
\>[B]{}\Varid{zygo}\;\Varid{alg_1}\;\Varid{alg_2}\mathrel{=}\Varid{fst}\;(\Varid{mutu}\;\Varid{alg_1}\;(\Varid{alg_2}\hsdot{\circ }{.}\Varid{fmap}\;\Varid{snd})){}\<[E]%
\ColumnHook
\end{hscode}\resethooks
Here \ensuremath{\Varid{alg_1}} computes the function of interest from the recursive results
together with auxiliary information of type \ensuremath{\Varid{b}}, and \ensuremath{\Varid{alg_2}} maintains the
auxiliary information.
Malcolm \cite{Mal90Alg} called zygomorphisms `yoking together of paramorphisms
and catamorphisms' and prefix `zygo-' is from Greek \textgreek{ζυγόν} meaning
`yoke'.
\end{recscheme}

\begin{example}
As we said, \ensuremath{\Varid{zygo}} is a generalization of paramorphisms: \ensuremath{\Varid{para}\;\Varid{alg}\mathrel{=}\Varid{zygo}\;\Varid{alg}\;\Conid{In}}.
And the above \ensuremath{\Varid{perfect}} is \ensuremath{\Varid{zygo}\;\Varid{p}\;\Varid{d}} where
\begin{hscode}\SaveRestoreHook
\column{B}{@{}>{\hspre}l<{\hspost}@{}}%
\column{23}{@{}>{\hspre}l<{\hspost}@{}}%
\column{36}{@{}>{\hspre}l<{\hspost}@{}}%
\column{E}{@{}>{\hspre}l<{\hspost}@{}}%
\>[B]{}\Varid{p}\mathbin{::}\Conid{TreeF}\;\Varid{e}\;(\Conid{Bool},\Conid{Integer}){}\<[36]%
\>[36]{}\to \Conid{Bool}{}\<[E]%
\\
\>[B]{}\Varid{p}\;\Conid{Empty}{}\<[36]%
\>[36]{}\mathrel{=}\Conid{True}{}\<[E]%
\\
\>[B]{}\Varid{p}\;(\Conid{Node}\;(\Varid{p_l},\Varid{d_l})\;\anonymous \;(\Varid{p_r},\Varid{d_r})){}\<[36]%
\>[36]{}\mathrel{=}\Varid{p_l}\mathrel{\wedge}\Varid{p_r}\mathrel{\wedge}(\Varid{d_l}\equiv \Varid{d_r}){}\<[E]%
\\[\blanklineskip]%
\>[B]{}\Varid{d}\mathbin{::}\Conid{TreeF}\;\Varid{e}\;\Conid{Integer}{}\<[23]%
\>[23]{}\to \Conid{Integer}{}\<[E]%
\\
\>[B]{}\Varid{d}\;\Conid{Empty}{}\<[23]%
\>[23]{}\mathrel{=}\mathrm{0}{}\<[E]%
\\
\>[B]{}\Varid{d}\;(\Conid{Node}\;\Varid{d_l}\;\anonymous \;\Varid{d_r}){}\<[23]%
\>[23]{}\mathrel{=}\mathrm{1}\mathbin{+}(\Varid{max}\;\Varid{d_l}\;\Varid{d_r}){}\<[E]%
\ColumnHook
\end{hscode}\resethooks
\end{example}

Note that although zygomorphisms are special cases of
mutumorphisms, the recursion scheme \ensuremath{\Varid{zygo}} is not a special
case of \ensuremath{\Varid{mutu}}, precisely because of the projection \ensuremath{\Varid{fst}}.
In the unifying framework by means of adjunctions, zygomorphisms arise from an
adjunction between the slice category $C \downarrow b$ and the base category
$C$ \cite{HW16USR}. The same adjunction also leads to the dual of
zygomorphisms---the recursion
scheme in which a seed is unfolded to a value of a recursive datatype that
is defined with some auxiliary datatype.

\section{Course-of-Value (Co)Recursion}
\label{sec:histo}

This section is about the patterns in dynamic programming algorithms, in which
a problem is solved based on solutions to subproblems just as in catamorphisms.
But in dynamic programming algorithms, subproblems are largely shared among
problems, and thus a common implementation technique is to memoize solved
subproblems with a table.
This section shows the recursion scheme for functions that
capture dynamic programming called \emph{histomorphisms}, a generalization called
\emph{dynamorphisms}, the corecursive dual \emph{futumorphisms}, and a further
generalization \emph{chronomorphisms}.
\subsection{Histomorphisms}

A powerful generalization of catamorphisms is to provide the algebra with all
the recursively computed results of direct and indirect substructures rather
than only the \emph{immediate} substructures.
Consider the longest increasing subsequence ({LIS}) problem: given a sequence
of integers, its subsequences are obtained by deleting some (or none) of its
elements and keeping the remaining elements in its original order, and the
problem is to find (the length of) longest subsequences in which the elements
are in increasing order.
For example, the longest increasing subsequences of \ensuremath{[\mskip1.5mu \mathrm{1},\mathrm{6},\mathbin{-}\mathrm{5},\mathrm{4},\mathrm{2},\mathrm{3},\mathrm{9}\mskip1.5mu]}
have length $4$ and one of them is \ensuremath{[\mskip1.5mu \mathrm{1},\mathrm{2},\mathrm{3},\mathrm{9}\mskip1.5mu]}.

A way to find LIS follows the observation that an LIS of \ensuremath{\Varid{x}\mathbin{:}\Varid{xs}} is either an
LIS of \ensuremath{\Varid{xs}}, or a subsequence beginning with the head element \ensuremath{\Varid{x}} and whose
tail is also an LIS (or the whole LIS could be longer).
This idea is implemented by the program below.
\begin{hscode}\SaveRestoreHook
\column{B}{@{}>{\hspre}l<{\hspost}@{}}%
\column{3}{@{}>{\hspre}l<{\hspost}@{}}%
\column{14}{@{}>{\hspre}l<{\hspost}@{}}%
\column{38}{@{}>{\hspre}l<{\hspost}@{}}%
\column{E}{@{}>{\hspre}l<{\hspost}@{}}%
\>[B]{}\Varid{lis}\mathrel{=}\Varid{snd}\hsdot{\circ }{.}\Varid{lis'}{}\<[E]%
\\[\blanklineskip]%
\>[B]{}\Varid{lis'}\mathbin{::}\Conid{Ord}\;\Varid{a}\Rightarrow [\mskip1.5mu \Varid{a}\mskip1.5mu]\to (\Conid{Integer},\Conid{Integer}){}\<[E]%
\\
\>[B]{}\Varid{lis'}\;[\mskip1.5mu \mskip1.5mu]{}\<[14]%
\>[14]{}\mathrel{=}(\mathrm{0},\mathrm{0}){}\<[E]%
\\
\>[B]{}\Varid{lis'}\;(\Varid{x}\mathbin{:}\Varid{xs}){}\<[14]%
\>[14]{}\mathrel{=}(\Varid{a},\Varid{b})\;\mathbf{where}{}\<[E]%
\\
\>[B]{}\hsindent{3}{}\<[3]%
\>[3]{}\Varid{a}\mathrel{=}\mathrm{1}\mathbin{+}\Varid{maximum}\;[\mskip1.5mu \Varid{fst}\;(\Varid{lis'}\;\Varid{sub})\mid {}\<[38]%
\>[38]{}\Varid{sub}\leftarrow \Varid{tails}\;\Varid{xs},\Varid{null}\;\Varid{sub}\mathrel{\vee}\Varid{x}\mathbin{<}\Varid{head}\;\Varid{sub}\mskip1.5mu]{}\<[E]%
\\
\>[B]{}\hsindent{3}{}\<[3]%
\>[3]{}\Varid{b}\mathrel{=}\Varid{max}\;\Varid{a}\;(\Varid{snd}\;(\Varid{lis'}\;\Varid{xs})){}\<[E]%
\ColumnHook
\end{hscode}\resethooks
where the first component of \ensuremath{\Varid{lis'}\;(\Varid{x}\mathbin{:}\Varid{xs})} is the length of the longest increasing subsequence that is restricted to begin with the first element \ensuremath{\Varid{x}}, and the second component is the length of LIS without this restriction and thus \ensuremath{\Varid{lis}\mathrel{=}\Varid{snd}\hsdot{\circ }{.}\Varid{lis'}}.

Unfortunately this implementation is very inefficient because \ensuremath{\Varid{lis'}} is recursively applied to possibly all substructures of the input, leading to exponential running time with respect to the length of the input.
The inefficiency is mainly due to redundant recomputation of \ensuremath{\Varid{lis'}} on substructures:
when computing \ensuremath{\Varid{lis'}\;(\Varid{xs}\plus \Varid{ys})}, for each \ensuremath{\Varid{x}} in \ensuremath{\Varid{xs}}, \ensuremath{\Varid{lis'}\;\Varid{ys}} is recomputed although the results are identical.
A technique to speed up the algorithm is to memoize the results of \ensuremath{\Varid{lis'}} on substructures and skip recomputing the function when identical input is encountered, a technique called \emph{dynamic programming}.

To implement dynamic programming, what we want is a scheme that provides the algebra with a table of the results for all substructures that have been computed.
A table is represented by the \ensuremath{\Conid{Cofree}} comonad
\begin{hscode}\SaveRestoreHook
\column{B}{@{}>{\hspre}l<{\hspost}@{}}%
\column{3}{@{}>{\hspre}l<{\hspost}@{}}%
\column{E}{@{}>{\hspre}l<{\hspost}@{}}%
\>[B]{}\mathbf{data}\;\Conid{Cofree}\;\Varid{f}\;\Varid{a}\;\mathbf{where}{}\<[E]%
\\
\>[B]{}\hsindent{3}{}\<[3]%
\>[3]{}(\triangleleft)\mathbin{::}\Varid{a}\to \Varid{f}\;(\Conid{Cofree}\;\Varid{f}\;\Varid{a})\to \Conid{Cofree}\;\Varid{f}\;\Varid{a}{}\<[E]%
\ColumnHook
\end{hscode}\resethooks
which can be intuitively understood as a (coinductive) tree whose branching structure is determined by functor \ensuremath{\Varid{f}} and all nodes are tagged with a value of type \ensuremath{\Varid{a}}, which can be extracted with
\begin{hscode}\SaveRestoreHook
\column{B}{@{}>{\hspre}l<{\hspost}@{}}%
\column{E}{@{}>{\hspre}l<{\hspost}@{}}%
\>[B]{}\Varid{extract}\mathbin{::}\Conid{Cofree}\;\Varid{f}\;\Varid{a}\to \Varid{a}{}\<[E]%
\\
\>[B]{}\Varid{extract}\;(\Varid{x}\triangleleft\anonymous )\mathrel{=}\Varid{x}{}\<[E]%
\ColumnHook
\end{hscode}\resethooks

\begin{recscheme}\label{scm:histo}
The recursion scheme \emph{histomorphism}~\cite{UuV99Pri} is:
\begin{hscode}\SaveRestoreHook
\column{B}{@{}>{\hspre}l<{\hspost}@{}}%
\column{E}{@{}>{\hspre}l<{\hspost}@{}}%
\>[B]{}\Varid{histo}\mathbin{::}\Conid{Functor}\;\Varid{f}\Rightarrow (\Varid{f}\;(\Conid{Cofree}\;\Varid{f}\;\Varid{a})\to \Varid{a})\to \mu\;\Varid{f}\to \Varid{a}{}\<[E]%
\\
\>[B]{}\Varid{histo}\;\Varid{alg}\mathrel{=}\Varid{extract}\hsdot{\circ }{.}\Varid{cata}\;(\lambda \Varid{x}\to (\Varid{alg}\;\Varid{x})\triangleleft\Varid{x}){}\<[E]%
\ColumnHook
\end{hscode}\resethooks
which is a catamorphism computing a memo-table of type \ensuremath{\Conid{Cofree}\;\Varid{f}\;\Varid{a}} followed by extracting the result for the whole structure.
The name histo- follows that the entire computation history is passed to the algebra.
It is also called \emph{course-of-value} recursion.
\end{recscheme}

\begin{example}
The dynamic programming implementation of \ensuremath{\Varid{lis}} is then:
\begin{hscode}\SaveRestoreHook
\column{B}{@{}>{\hspre}l<{\hspost}@{}}%
\column{3}{@{}>{\hspre}l<{\hspost}@{}}%
\column{15}{@{}>{\hspre}l<{\hspost}@{}}%
\column{21}{@{}>{\hspre}l<{\hspost}@{}}%
\column{E}{@{}>{\hspre}l<{\hspost}@{}}%
\>[B]{}\Varid{lis''}\mathbin{::}\Conid{Ord}\;\Varid{a}\Rightarrow \mu\;(\Conid{ListF}\;\Varid{a})\to \Conid{Integer}{}\<[E]%
\\
\>[B]{}\Varid{lis''}\mathrel{=}\Varid{snd}\hsdot{\circ }{.}\Varid{histo}\;\Varid{alg}{}\<[E]%
\\[\blanklineskip]%
\>[B]{}\Varid{alg}\mathbin{::}\Conid{Ord}\;\Varid{a}{}\<[15]%
\>[15]{}\Rightarrow \Conid{ListF}\;\Varid{a}\;(\Conid{Cofree}\;(\Conid{ListF}\;\Varid{a})\;(\Conid{Integer},\Conid{Integer})){}\<[E]%
\\
\>[15]{}\to (\Conid{Integer},\Conid{Integer}){}\<[E]%
\\
\>[B]{}\Varid{alg}\;\Conid{Nil}{}\<[21]%
\>[21]{}\mathrel{=}(\mathrm{0},\mathrm{0}){}\<[E]%
\\
\>[B]{}\Varid{alg}\;(\Conid{Cons}\;\Varid{x}\;\Varid{table}){}\<[21]%
\>[21]{}\mathrel{=}(\Varid{a},\Varid{b})\;\mathbf{where}{}\<[E]%
\\
\>[B]{}\hsindent{3}{}\<[3]%
\>[3]{}\Varid{a}\mathrel{=}\mathrm{1}\mathbin{+}\Varid{findNext}\;\Varid{x}\;\Varid{table}{}\<[E]%
\\
\>[B]{}\hsindent{3}{}\<[3]%
\>[3]{}\Varid{b}\mathrel{=}\Varid{max}\;\Varid{a}\;(\Varid{snd}\;(\Varid{extract}\;\Varid{table})){}\<[E]%
\ColumnHook
\end{hscode}\resethooks
where \ensuremath{\Varid{findNext}} searches in the rest of the list for the element that is greater than \ensuremath{\Varid{x}} and begins a longest increasing subsequence:
\begin{hscode}\SaveRestoreHook
\column{B}{@{}>{\hspre}l<{\hspost}@{}}%
\column{3}{@{}>{\hspre}l<{\hspost}@{}}%
\column{40}{@{}>{\hspre}l<{\hspost}@{}}%
\column{E}{@{}>{\hspre}l<{\hspost}@{}}%
\>[B]{}\Varid{findNext}\mathbin{::}\Conid{Ord}\;\Varid{a}\Rightarrow \Varid{a}\to \Conid{Cofree}\;(\Conid{ListF}\;\Varid{a})\;(\Conid{Integer},\Conid{Integer})\to \Conid{Integer}{}\<[E]%
\\
\>[B]{}\Varid{findNext}\;\Varid{x}\;((\Varid{a},\anonymous )\triangleleft\Conid{Nil}){}\<[40]%
\>[40]{}\mathrel{=}\Varid{a}{}\<[E]%
\\
\>[B]{}\Varid{findNext}\;\Varid{x}\;((\Varid{a},\anonymous )\triangleleft(\Conid{Cons}\;\Varid{y}\;\Varid{table'})){}\<[40]%
\>[40]{}\mathrel{=}\mathbf{if}\;\Varid{x}\mathbin{<}\Varid{y}\;\mathbf{then}\;\Varid{max}\;\Varid{a}\;\Varid{b}\;\mathbf{else}\;\Varid{b}{}\<[E]%
\\
\>[B]{}\hsindent{3}{}\<[3]%
\>[3]{}\mathbf{where}\;\Varid{b}\mathrel{=}\Varid{findNext}\;\Varid{x}\;\Varid{table'}{}\<[E]%
\ColumnHook
\end{hscode}\resethooks
which improves the time complexity to quadratic time because \ensuremath{\Varid{alg}} runs in linear time for each element and \ensuremath{\Varid{alg}} is computed only once for each element.
\end{example}

In the unifying theory of recursion schemes by adjunctions, histomorphisms arise from the adjunction $\textit{U} \dashv \textit{Cofree}_F$~\cite{HiWu13Histo} where $\textit{Cofree}_F$ sends an object to its cofree coalgebra in the category of $F$-coalgebras, and $\textit{U}$ is the forgetful functor.
As we have seen, cofree coalgebras are used to model the memo-table of computation history in histomorphisms, but an oddity here is that (the carrier of) the cofree coalgebra is a possibly infinite structure, while the computation history is in fact finite because the input is a finite inductive structure.
A remedy for this imprecision is to replace cofree coalgebras with \emph{cofree para-recursive coalgebras} in the construction, and the \ensuremath{\Conid{Cofree}\;\Varid{f}\;\Varid{a}} comonad in \ensuremath{\Varid{histo}} is replaced by its para-recursive counterpart, which is exactly \emph{finite} trees whose branching structure is \ensuremath{\Varid{f}} and nodes are tagged with \ensuremath{\Varid{a}}-values \cite{HWG15Conj}.

\subsection{Dynamorphisms}
Histomorphisms require the input to be an initial algebra, and this is inconvenient in applications whose structure of computation is determined on the fly while computing.
An example is the following program finding the length of \emph{longest common subsequences} (LCS) of two sequences \cite{BHR00LCS}.
\begin{hscode}\SaveRestoreHook
\column{B}{@{}>{\hspre}l<{\hspost}@{}}%
\column{3}{@{}>{\hspre}l<{\hspost}@{}}%
\column{11}{@{}>{\hspre}l<{\hspost}@{}}%
\column{16}{@{}>{\hspre}l<{\hspost}@{}}%
\column{E}{@{}>{\hspre}l<{\hspost}@{}}%
\>[B]{}\Varid{lcs}\mathbin{::}\Conid{Eq}\;\Varid{a}\Rightarrow [\mskip1.5mu \Varid{a}\mskip1.5mu]\to [\mskip1.5mu \Varid{a}\mskip1.5mu]\to \Conid{Integer}{}\<[E]%
\\
\>[B]{}\Varid{lcs}\;[\mskip1.5mu \mskip1.5mu]\;\anonymous {}\<[11]%
\>[11]{}\mathrel{=}\mathrm{0}{}\<[E]%
\\
\>[B]{}\Varid{lcs}\;\anonymous \;[\mskip1.5mu \mskip1.5mu]{}\<[11]%
\>[11]{}\mathrel{=}\mathrm{0}{}\<[E]%
\\
\>[B]{}\Varid{lcs}\;\Varid{xxs}\mathord{@}(\Varid{x}\mathbin{:}\Varid{xs})\;\Varid{yys}\mathord{@}(\Varid{y}\mathbin{:}\Varid{ys}){}\<[E]%
\\
\>[B]{}\hsindent{3}{}\<[3]%
\>[3]{}\mid \Varid{x}\equiv \Varid{y}{}\<[16]%
\>[16]{}\mathrel{=}\Varid{lcs}\;\Varid{xs}\;\Varid{ys}\mathbin{+}\mathrm{1}{}\<[E]%
\\
\>[B]{}\hsindent{3}{}\<[3]%
\>[3]{}\mid \Varid{otherwise}{}\<[16]%
\>[16]{}\mathrel{=}\Varid{max}\;(\Varid{lcs}\;\Varid{xs}\;\Varid{yys})\;(\Varid{lcs}\;\Varid{xxs}\;\Varid{ys}){}\<[E]%
\ColumnHook
\end{hscode}\resethooks
This program runs in exponential time but it is well suited for optimization with dynamic programming because a lot of subproblems are shared across recursion.
However, it is not accommodated by \ensuremath{\Varid{histo}} because the input, a pair of lists, is not an initial algebra.
Therefore it is handy to generalize \ensuremath{\Varid{histo}} by replacing \ensuremath{\Varid{in}^\circ} with a user-supplied recursive coalgebra:

\begin{recscheme}\label{scm:dyna}
A \emph{dynamorphism} (evidently the name is derived from \emph{dyna}mic
programming) introduced by Kabanov and Vene \cite{KV06Rec} is given by:
\begin{hscode}\SaveRestoreHook
\column{B}{@{}>{\hspre}l<{\hspost}@{}}%
\column{20}{@{}>{\hspre}l<{\hspost}@{}}%
\column{E}{@{}>{\hspre}l<{\hspost}@{}}%
\>[B]{}\Varid{dyna}\mathbin{::}\Conid{Functor}\;\Varid{f}{}\<[20]%
\>[20]{}\Rightarrow (\Varid{f}\;(\Conid{Cofree}\;\Varid{f}\;\Varid{a})\to \Varid{a})\to (\Varid{c}\to \Varid{f}\;\Varid{c})\to \Varid{c}\to \Varid{a}{}\<[E]%
\\
\>[B]{}\Varid{dyna}\;\Varid{alg}\;\Varid{coalg}\mathrel{=}\Varid{extract}\hsdot{\circ }{.}\Varid{hylo}\;(\lambda \Varid{x}\to \Varid{alg}\;\Varid{x}\triangleleft\Varid{x})\;\Varid{coalg}{}\<[E]%
\ColumnHook
\end{hscode}\resethooks
in which the recursive coalgebra \ensuremath{\Varid{coalg}} breaks a problem into subproblems, which are recursively solved, and the algebra \ensuremath{\Varid{alg}} solves a problem with solutions to all direct and indirect subproblems.
\end{recscheme}

Because the subproblems of a dynamic programming algorithm together with the dependency relation of subproblems form an acyclic graph, an appealing choice of the functor \ensuremath{\Varid{f}} in \ensuremath{\Varid{dyna}} is \ensuremath{\Conid{ListF}} and the coalgebra \ensuremath{\Varid{c}} generates subproblems in a topological order of the dependency graph of subproblems, so that a subproblem is solved exactly once when it is needed by bigger problems.

\begin{example}
Continuing the example of LCS, the set of subproblems of \ensuremath{\Varid{lcs}\;s_1\;s_2} are all
pairs \ensuremath{(\Varid{x},\Varid{y})} for \ensuremath{\Varid{x}} and \ensuremath{\Varid{y}} being suffixes of \ensuremath{s_1} and \ensuremath{s_2} respectively.
An ordering of subproblems that respects their computing dependency is:
\begin{hscode}\SaveRestoreHook
\column{B}{@{}>{\hspre}l<{\hspost}@{}}%
\column{13}{@{}>{\hspre}l<{\hspost}@{}}%
\column{26}{@{}>{\hspre}l<{\hspost}@{}}%
\column{32}{@{}>{\hspre}l<{\hspost}@{}}%
\column{45}{@{}>{\hspre}l<{\hspost}@{}}%
\column{E}{@{}>{\hspre}l<{\hspost}@{}}%
\>[B]{}\Varid{g}\mathbin{::}([\mskip1.5mu \Varid{a}\mskip1.5mu],[\mskip1.5mu \Varid{a}\mskip1.5mu])\to \Conid{ListF}\;([\mskip1.5mu \Varid{a}\mskip1.5mu],[\mskip1.5mu \Varid{a}\mskip1.5mu])\;([\mskip1.5mu \Varid{a}\mskip1.5mu],[\mskip1.5mu \Varid{a}\mskip1.5mu]){}\<[E]%
\\
\>[B]{}\Varid{g}\;([\mskip1.5mu \mskip1.5mu],[\mskip1.5mu \mskip1.5mu]){}\<[13]%
\>[13]{}\mathrel{=}\Conid{Nil}{}\<[E]%
\\
\>[B]{}\Varid{g}\;(\Varid{x},\Varid{y}){}\<[13]%
\>[13]{}\mathrel{=}\mathbf{if}\;\Varid{null}\;\Varid{y}\;{}\<[26]%
\>[26]{}\mathbf{then}\;{}\<[32]%
\>[32]{}\Conid{Cons}\;(\Varid{x},\Varid{y})\;(\Varid{tail}\;\Varid{x},s_2){}\<[E]%
\\
\>[26]{}\mathbf{else}\;{}\<[32]%
\>[32]{}\Conid{Cons}\;(\Varid{x},\Varid{y})\;{}\<[45]%
\>[45]{}(\Varid{x},\Varid{tail}\;\Varid{y}){}\<[E]%
\ColumnHook
\end{hscode}\resethooks
The algebra \ensuremath{\Varid{a}} solves a problem with solutions to subproblems available:
\begin{hscode}\SaveRestoreHook
\column{B}{@{}>{\hspre}l<{\hspost}@{}}%
\column{3}{@{}>{\hspre}l<{\hspost}@{}}%
\column{13}{@{}>{\hspre}l<{\hspost}@{}}%
\column{24}{@{}>{\hspre}l<{\hspost}@{}}%
\column{31}{@{}>{\hspre}l<{\hspost}@{}}%
\column{E}{@{}>{\hspre}l<{\hspost}@{}}%
\>[B]{}\Varid{a}\mathbin{::}\Conid{ListF}\;([\mskip1.5mu \Varid{a}\mskip1.5mu],[\mskip1.5mu \Varid{a}\mskip1.5mu])\;(\Conid{Cofree}\;(\Conid{ListF}\;([\mskip1.5mu \Varid{a}\mskip1.5mu],[\mskip1.5mu \Varid{a}\mskip1.5mu]))\;\Conid{Integer})\to \Conid{Integer}{}\<[E]%
\\
\>[B]{}\Varid{a}\;\Conid{Nil}\mathrel{=}\mathrm{0}{}\<[E]%
\\
\>[B]{}\Varid{a}\;(\Conid{Cons}\;(\Varid{x},\Varid{y})\;\Varid{table}){}\<[E]%
\\
\>[B]{}\hsindent{3}{}\<[3]%
\>[3]{}\mid \Varid{null}\;\Varid{x}{}\<[13]%
\>[13]{}\mathrel{\vee}\Varid{null}\;\Varid{y}{}\<[24]%
\>[24]{}\mathrel{=}\mathrm{0}{}\<[E]%
\\
\>[B]{}\hsindent{3}{}\<[3]%
\>[3]{}\mid \Varid{head}\;\Varid{x}\equiv \Varid{head}\;\Varid{y}{}\<[24]%
\>[24]{}\mathrel{=}\Varid{index}\;\Varid{table}\;(\Varid{offset}\;\mathrm{1}\;\mathrm{1})\mathbin{+}\mathrm{1}{}\<[E]%
\\
\>[B]{}\hsindent{3}{}\<[3]%
\>[3]{}\mid \Varid{otherwise}{}\<[24]%
\>[24]{}\mathrel{=}\Varid{max}\;{}\<[31]%
\>[31]{}(\Varid{index}\;\Varid{table}\;(\Varid{offset}\;\mathrm{1}\;\mathrm{0}))\;{}\<[E]%
\\
\>[31]{}(\Varid{index}\;\Varid{table}\;(\Varid{offset}\;\mathrm{0}\;\mathrm{1})){}\<[E]%
\ColumnHook
\end{hscode}\resethooks
where \ensuremath{\Varid{index}\;\Varid{t}\;\Varid{n}} extracts the \ensuremath{\Varid{n}}-th entry of the memo-table \ensuremath{\Varid{t}}:
\begin{hscode}\SaveRestoreHook
\column{B}{@{}>{\hspre}l<{\hspost}@{}}%
\column{29}{@{}>{\hspre}l<{\hspost}@{}}%
\column{E}{@{}>{\hspre}l<{\hspost}@{}}%
\>[B]{}\Varid{index}\mathbin{::}\Conid{Cofree}\;(\Conid{ListF}\;\Varid{a})\;\Varid{p}\to \Conid{Integer}\to \Varid{p}{}\<[E]%
\\
\>[B]{}\Varid{index}\;\Varid{t}\;\mathrm{0}{}\<[29]%
\>[29]{}\mathrel{=}\Varid{extract}\;\Varid{t}{}\<[E]%
\\
\>[B]{}\Varid{index}\;(\anonymous \triangleleft(\Conid{Cons}\;\anonymous \;\Varid{t'}))\;\Varid{n}{}\<[29]%
\>[29]{}\mathrel{=}\Varid{index}\;\Varid{t'}\;(\Varid{n}\mathbin{-}\mathrm{1}){}\<[E]%
\ColumnHook
\end{hscode}\resethooks
The tricky part is computing the indices for entries to subproblems in the memo-table.
Because subproblems are enumerated by \ensuremath{\Varid{g}} in the order that reduces the second sequence first, thus the entry for \ensuremath{(\Varid{drop}\;\Varid{n}\;\Varid{x},\Varid{drop}\;\Varid{m}\;\Varid{y})} in the memo-table when computing \ensuremath{(\Varid{x},\Varid{y})} is:
\[ \ensuremath{\Varid{offset}\;\Varid{n}\;\Varid{m}\mathrel{=}\Varid{n}\mathbin{*}(\Varid{length}\;s_2\mathbin{+}\mathrm{1})\mathbin{+}\Varid{m}\mathbin{-}\mathrm{1}} \]
Putting them together, we get the dynamic programming solution to LCS:
\begin{hscode}\SaveRestoreHook
\column{B}{@{}>{\hspre}l<{\hspost}@{}}%
\column{E}{@{}>{\hspre}l<{\hspost}@{}}%
\>[B]{}\Varid{lcs'}\;s_1\;s_2\mathrel{=}\Varid{dyna}\;\Varid{a}\;\Varid{g}\;(s_1,s_2){}\<[E]%
\ColumnHook
\end{hscode}\resethooks
which improves the exponential running time of specification \ensuremath{\Varid{lcs}} to
$\mathcal{O}(|s_1||s_2|^2)$, yet slower than the
$\mathcal{O}(|s_1||s_2|)$ array-based implementation of dynamic programming
because of the cost of indexing the list-structured memo-table.
\end{example}

\subsection{Futumorphisms}
Histomorphisms are generalized catamorphisms that can \emph{inspect the history} of computation.
The dual generalization is anamorphisms that can \emph{control the future}.
As an example, consider the problem of decoding the \emph{run-length encoding} of a sequence:
the input is a list of elements \ensuremath{(\Varid{n},\Varid{x})} of type \ensuremath{(\Conid{Int},\Varid{a})} and $\ensuremath{\Varid{n}} > 0$ for all elements.
The output is a list \ensuremath{[\mskip1.5mu \Varid{a}\mskip1.5mu]} and each \ensuremath{(\Varid{n},\Varid{x})} in the input is interpreted as \ensuremath{\Varid{n}} consecutive copies of \ensuremath{\Varid{x}}.
As an anamorphism, it is expressed as
\begin{hscode}\SaveRestoreHook
\column{B}{@{}>{\hspre}l<{\hspost}@{}}%
\column{3}{@{}>{\hspre}l<{\hspost}@{}}%
\column{5}{@{}>{\hspre}l<{\hspost}@{}}%
\column{18}{@{}>{\hspre}l<{\hspost}@{}}%
\column{E}{@{}>{\hspre}l<{\hspost}@{}}%
\>[B]{}\Varid{rld}\mathbin{::}[\mskip1.5mu (\Conid{Int},\Varid{a})\mskip1.5mu]\to \nu\;(\Conid{ListF}\;\Varid{a}){}\<[E]%
\\
\>[B]{}\Varid{rld}\mathrel{=}\Varid{ana}\;\Varid{c}\;\mathbf{where}{}\<[E]%
\\
\>[B]{}\hsindent{3}{}\<[3]%
\>[3]{}\Varid{c}\;[\mskip1.5mu \mskip1.5mu]\mathrel{=}\Conid{Nil}{}\<[E]%
\\
\>[B]{}\hsindent{3}{}\<[3]%
\>[3]{}\Varid{c}\;((\Varid{n},\Varid{x})\mathbin{:}\Varid{xs}){}\<[E]%
\\
\>[3]{}\hsindent{2}{}\<[5]%
\>[5]{}\mid \Varid{n}\equiv \mathrm{1}{}\<[18]%
\>[18]{}\mathrel{=}\Conid{Cons}\;\Varid{x}\;\Varid{xs}{}\<[E]%
\\
\>[3]{}\hsindent{2}{}\<[5]%
\>[5]{}\mid \Varid{otherwise}{}\<[18]%
\>[18]{}\mathrel{=}\Conid{Cons}\;\Varid{x}\;((\Varid{n}\mathbin{-}\mathrm{1},\Varid{x})\mathbin{:}\Varid{xs}){}\<[E]%
\ColumnHook
\end{hscode}\resethooks
This is slightly awkward because anamorphisms can emit only one layer of the structure in each step, while in this example it is more natural to emit \ensuremath{\Varid{n}} copies of \ensuremath{\Varid{x}} in a batch.
This can be done if the recursion scheme allows the coalgebra to generate more than one layer in a single step---in a sense controlling the future of the computation.

Multiple layers of a structure given by a functor \ensuremath{\Varid{f}} are represented by the \ensuremath{\Conid{Free}} monad:
\begin{hscode}\SaveRestoreHook
\column{B}{@{}>{\hspre}l<{\hspost}@{}}%
\column{E}{@{}>{\hspre}l<{\hspost}@{}}%
\>[B]{}\mathbf{data}\;\Conid{Free}\;\Varid{f}\;\Varid{a}\mathrel{=}\Conid{Ret}\;\Varid{a}\mid \Conid{Op}\;(\Varid{f}\;(\Conid{Free}\;\Varid{f}\;\Varid{a})){}\<[E]%
\ColumnHook
\end{hscode}\resethooks
which is the type of (inductive) trees whose branching is determined by \ensuremath{\Varid{f}} and leaf nodes are \ensuremath{\Varid{a}}-values. 
Free algebras subsume initial algebras as $\ensuremath{\Conid{Free}\;\Varid{f}\;\Conid{Void}} \cong \ensuremath{\mu\;\Varid{f}}$ where \ensuremath{\Conid{Void}} is the bottom type, and \ensuremath{\Varid{cata}} for \ensuremath{\mu\;\Varid{f}} is replaced by
\begin{hscode}\SaveRestoreHook
\column{B}{@{}>{\hspre}l<{\hspost}@{}}%
\column{21}{@{}>{\hspre}l<{\hspost}@{}}%
\column{E}{@{}>{\hspre}l<{\hspost}@{}}%
\>[B]{}\Varid{eval}\mathbin{::}\Conid{Functor}\;\Varid{f}\Rightarrow (\Varid{f}\;\Varid{b}\to \Varid{b})\to (\Varid{a}\to \Varid{b})\to \Conid{Free}\;\Varid{f}\;\Varid{a}\to \Varid{b}{}\<[E]%
\\
\>[B]{}\Varid{eval}\;\Varid{alg}\;\Varid{g}\;(\Conid{Ret}\;\Varid{a}){}\<[21]%
\>[21]{}\mathrel{=}\Varid{g}\;\Varid{a}{}\<[E]%
\\
\>[B]{}\Varid{eval}\;\Varid{alg}\;\Varid{g}\;(\Conid{Op}\;\Varid{k}){}\<[21]%
\>[21]{}\mathrel{=}\Varid{alg}\;(\Varid{fmap}\;(\Varid{eval}\;\Varid{alg}\;\Varid{g})\;\Varid{k}){}\<[E]%
\ColumnHook
\end{hscode}\resethooks

\begin{recscheme}\label{scm:futu}
With these constructions, the recursion scheme for \emph{futumorphisms}~\cite{UuV99Pri} is defined by:
\begin{hscode}\SaveRestoreHook
\column{B}{@{}>{\hspre}l<{\hspost}@{}}%
\column{3}{@{}>{\hspre}l<{\hspost}@{}}%
\column{19}{@{}>{\hspre}l<{\hspost}@{}}%
\column{E}{@{}>{\hspre}l<{\hspost}@{}}%
\>[B]{}\Varid{futu}\mathbin{::}\Conid{Functor}\;\Varid{f}\Rightarrow (\Varid{c}\to \Varid{f}\;(\Conid{Free}\;\Varid{f}\;\Varid{c}))\to \Varid{c}\to \nu\;\Varid{f}{}\<[E]%
\\
\>[B]{}\Varid{futu}\;\Varid{coalg}\mathrel{=}\Varid{ana}\;\Varid{coalg'}\hsdot{\circ }{.}\Conid{Ret}\;\mathbf{where}{}\<[E]%
\\
\>[B]{}\hsindent{3}{}\<[3]%
\>[3]{}\Varid{coalg'}\;(\Conid{Ret}\;\Varid{a}){}\<[19]%
\>[19]{}\mathrel{=}\Varid{coalg}\;\Varid{a}{}\<[E]%
\\
\>[B]{}\hsindent{3}{}\<[3]%
\>[3]{}\Varid{coalg'}\;(\Conid{Op}\;\Varid{k}){}\<[19]%
\>[19]{}\mathrel{=}\Varid{k}{}\<[E]%
\ColumnHook
\end{hscode}\resethooks
\end{recscheme}

\begin{example}
We can redefine \ensuremath{\Varid{rld}} as a futumorphism:
\begin{hscode}\SaveRestoreHook
\column{B}{@{}>{\hspre}l<{\hspost}@{}}%
\column{3}{@{}>{\hspre}l<{\hspost}@{}}%
\column{10}{@{}>{\hspre}l<{\hspost}@{}}%
\column{19}{@{}>{\hspre}l<{\hspost}@{}}%
\column{46}{@{}>{\hspre}l<{\hspost}@{}}%
\column{E}{@{}>{\hspre}l<{\hspost}@{}}%
\>[B]{}\Varid{rld'}\mathbin{::}[\mskip1.5mu (\Conid{Int},\Varid{a})\mskip1.5mu]\to \nu\;(\Conid{ListF}\;\Varid{a}){}\<[E]%
\\
\>[B]{}\Varid{rld'}\mathrel{=}\Varid{futu}\;\Varid{dec}{}\<[E]%
\\[\blanklineskip]%
\>[B]{}\Varid{dec}\;[\mskip1.5mu \mskip1.5mu]{}\<[19]%
\>[19]{}\mathrel{=}\Conid{Nil}{}\<[E]%
\\
\>[B]{}\Varid{dec}\;((\Varid{n},\Varid{c})\mathbin{:}\Varid{xs}){}\<[19]%
\>[19]{}\mathrel{=}\mathbf{let}\;(\Conid{Op}\;\Varid{g})\mathrel{=}\Varid{rep}\;\Varid{n}\;\mathbf{in}\;\Varid{g}\;{}\<[46]%
\>[46]{}\mathbf{where}{}\<[E]%
\\
\>[B]{}\hsindent{3}{}\<[3]%
\>[3]{}\Varid{rep}\;\mathrm{0}{}\<[10]%
\>[10]{}\mathrel{=}\Conid{Ret}\;\Varid{xs}{}\<[E]%
\\
\>[B]{}\hsindent{3}{}\<[3]%
\>[3]{}\Varid{rep}\;\Varid{m}{}\<[10]%
\>[10]{}\mathrel{=}\Conid{Op}\;(\Conid{Cons}\;\Varid{c}\;(\Varid{rep}\;(\Varid{m}\mathbin{-}\mathrm{1}))){}\<[E]%
\ColumnHook
\end{hscode}\resethooks
Note that \ensuremath{\Varid{dec}} assumes \ensuremath{\Varid{n}\mathbin{>}\mathrm{0}} because \ensuremath{\Varid{futu}} demands that the coalgebra generate at least one layer of \ensuremath{\Varid{f}}-structure.
\end{example}

Theoretically, futumorphisms are adjoint unfolds from the adjunction $\textit{Free}_F \dashv \textit{U}$ where $\textit{Free}_F$ maps object $a$ to the free algebra generated by $a$ in the category of $F$-algebras.
In the same way that dynamorphisms generalize histomorphisms, futumorphisms can be generalized by replacing \ensuremath{(\nu\;\Conid{F},\Varid{Out}^\circ)} with a user-supplied corecursive $F$-algebra.

A broader generalization is to combine futumorphisms and histomorphisms in a
similar way to hylomorphisms combining anamorphisms and catamorphisms:
\begin{hscode}\SaveRestoreHook
\column{B}{@{}>{\hspre}l<{\hspost}@{}}%
\column{3}{@{}>{\hspre}l<{\hspost}@{}}%
\column{9}{@{}>{\hspre}l<{\hspost}@{}}%
\column{19}{@{}>{\hspre}l<{\hspost}@{}}%
\column{23}{@{}>{\hspre}l<{\hspost}@{}}%
\column{E}{@{}>{\hspre}l<{\hspost}@{}}%
\>[B]{}\Varid{chrono}{}\<[9]%
\>[9]{}\mathbin{::}\Conid{Functor}\;\Varid{f}{}\<[23]%
\>[23]{}\Rightarrow (\Varid{f}\;(\Conid{Cofree}\;\Varid{f}\;\Varid{b})\to \Varid{b}){}\<[E]%
\\
\>[23]{}\to (\Varid{a}\to \Varid{f}\;(\Conid{Free}\;\Varid{f}\;\Varid{a})){}\<[E]%
\\
\>[23]{}\to \Varid{a}\to \Varid{b}{}\<[E]%
\\
\>[B]{}\Varid{chrono}\;\Varid{alg}\;\Varid{coalg}\mathrel{=}\Varid{extract}\hsdot{\circ }{.}\Varid{hylo}\;\Varid{alg'}\;\Varid{coalg'}\hsdot{\circ }{.}\Conid{Ret}\;\mathbf{where}{}\<[E]%
\\
\>[B]{}\hsindent{3}{}\<[3]%
\>[3]{}\Varid{alg'}\;\Varid{x}\mathrel{=}\Varid{alg}\;\Varid{x}\triangleleft\Varid{x}{}\<[E]%
\\
\>[B]{}\hsindent{3}{}\<[3]%
\>[3]{}\Varid{coalg'}\;(\Conid{Ret}\;\Varid{a}){}\<[19]%
\>[19]{}\mathrel{=}\Varid{coalg}\;\Varid{a}{}\<[E]%
\\
\>[B]{}\hsindent{3}{}\<[3]%
\>[3]{}\Varid{coalg'}\;(\Conid{Op}\;\Varid{k}){}\<[19]%
\>[19]{}\mathrel{=}\Varid{k}{}\<[E]%
\ColumnHook
\end{hscode}\resethooks
These were dubbed \emph{chronomorphisms} by Kmett \cite{Kmt08Tim} (prefix chrono-
from Greek \textgreek{χρόνος} meaning `time'), because they subsume both
histo- and futumorphisms.

\section{Monadic Structural Recursion}\label{sec:monadic}
Up to now we have been working in the world of pure functions. 
It is certainly possible to extend the recursion schemes to the non-pure world where computational effects are modelled with monads.

\subsection{Monadic Catamorphism}
Let us start with a straightforward example of printing a tree with the \ensuremath{\Conid{IO}} monad:
\begin{hscode}\SaveRestoreHook
\column{B}{@{}>{\hspre}l<{\hspost}@{}}%
\column{30}{@{}>{\hspre}l<{\hspost}@{}}%
\column{E}{@{}>{\hspre}l<{\hspost}@{}}%
\>[B]{}\Varid{printTree}\mathbin{::}\Conid{Show}\;\Varid{a}\Rightarrow \mu\;(\Conid{TreeF}\;\Varid{a})\to \Conid{IO}\;(){}\<[E]%
\\
\>[B]{}\Varid{printTree}\;(\Conid{In}\;\Conid{Empty}){}\<[30]%
\>[30]{}\mathrel{=}\Varid{return}\;(){}\<[E]%
\\
\>[B]{}\Varid{printTree}\;(\Conid{In}\;(\Conid{Node}\;\Varid{l}\;\Varid{a}\;\Varid{r})){}\<[30]%
\>[30]{}\mathrel{=}\mathbf{do}\;\Varid{printTree}\;\Varid{l};\Varid{printTree}\;\Varid{r};\Varid{print}\;\Varid{a}{}\<[E]%
\ColumnHook
\end{hscode}\resethooks
The reader may have recognized that it is already a catamorphism:
\begin{hscode}\SaveRestoreHook
\column{B}{@{}>{\hspre}l<{\hspost}@{}}%
\column{3}{@{}>{\hspre}l<{\hspost}@{}}%
\column{28}{@{}>{\hspre}l<{\hspost}@{}}%
\column{E}{@{}>{\hspre}l<{\hspost}@{}}%
\>[B]{}\Varid{printTree'}\mathbin{::}\Conid{Show}\;\Varid{a}\Rightarrow \mu\;(\Conid{TreeF}\;\Varid{a})\to \Conid{IO}\;(){}\<[E]%
\\
\>[B]{}\Varid{printTree'}\mathrel{=}\Varid{cata}\;\Varid{printAlg}\;\mathbf{where}{}\<[E]%
\\
\>[B]{}\hsindent{3}{}\<[3]%
\>[3]{}\Varid{printAlg}\mathbin{::}\Conid{Show}\;\Varid{a}\Rightarrow \Conid{TreeF}\;\Varid{a}\;(\Conid{IO}\;())\to \Conid{IO}\;(){}\<[E]%
\\
\>[B]{}\hsindent{3}{}\<[3]%
\>[3]{}\Varid{printAlg}\;\Conid{Empty}{}\<[28]%
\>[28]{}\mathrel{=}\Varid{return}\;(){}\<[E]%
\\
\>[B]{}\hsindent{3}{}\<[3]%
\>[3]{}\Varid{printAlg}\;(\Conid{Node}\;\Varid{ml}\;\Varid{a}\;\Varid{mr}){}\<[28]%
\>[28]{}\mathrel{=}\mathbf{do}\;\Varid{ml};\Varid{mr};\Varid{print}\;\Varid{a}{}\<[E]%
\ColumnHook
\end{hscode}\resethooks
Thus a straightforward way of abstracting `monadic catamorphisms' is to restrict \ensuremath{\Varid{cata}} to monadic values.
\begin{recscheme}[\ensuremath{\Varid{cataM}}] \label{scm:catam}
We call the morphisms given by the following recursion scheme \emph{catamorphisms on monadic values}:
\begin{hscode}\SaveRestoreHook
\column{B}{@{}>{\hspre}l<{\hspost}@{}}%
\column{E}{@{}>{\hspre}l<{\hspost}@{}}%
\>[B]{}\Varid{cataM}\mathbin{::}(\Conid{Functor}\;\Varid{f},\Conid{Monad}\;\Varid{m})\Rightarrow (\Varid{f}\;(\Varid{m}\;\Varid{a})\to \Varid{m}\;\Varid{a})\to \mu\;\Varid{f}\to \Varid{m}\;\Varid{a}{}\<[E]%
\\
\>[B]{}\Varid{cataM}\;\Varid{algM}\mathrel{=}\Varid{cata}\;\Varid{algM}{}\<[E]%
\ColumnHook
\end{hscode}\resethooks
which is the second approach to monadic catamorphisms in~\cite{Par05Com}.
\end{recscheme}

However, \ensuremath{\Varid{cataM}} does not fully capture our intuition for `monadic catamorphism' because the algebra \ensuremath{\Varid{algM}\mathbin{::}\Varid{f}\;(\Varid{m}\;\Varid{a})\to \Varid{m}\;\Varid{a}} is allowed to combine computations from subparts arbitrarily.
For a more precise characterization, we decompose \ensuremath{\Varid{algM}\mathbin{::}\Varid{f}\;(\Varid{m}\;\Varid{a})\to \Varid{m}\;\Varid{a}} in \ensuremath{\Varid{cataM}} into two parts:
a function \ensuremath{\Varid{alg}\mathbin{::}\Varid{f}\;\Varid{a}\to \Varid{m}\;\Varid{a}} which (monadically) computes the result for the whole structure given the results of substructures, and a polymorphic function
\[\ensuremath{\Varid{seq}\mathbin{::}\forall \Varid{x}\hsforall \hsdot{\circ }{.}\Varid{f}\;(\Varid{m}\;\Varid{x})\to \Varid{m}\;(\Varid{f}\;\Varid{x})}\]
called a \emph{sequencing} of \ensuremath{\Varid{f}} over \ensuremath{\Varid{m}}, which combines computations for substructures into one monadic computation.
The decomposition reflects the intuition that a monadic catamorphism processes substructures (in the order determined by \ensuremath{\Varid{seq}}) and combines their results (by \ensuremath{\Varid{alg}}) to process the root structure: 
\[\ensuremath{\Varid{algM}\;\Varid{r}\mathrel{=}\Varid{seq}\;\Varid{r}\bind \Varid{alg}}.\]

\begin{example}\label{eg:dist}
Binary trees \ensuremath{\Conid{TreeF}} can be sequenced from left to right:
\begin{hscode}\SaveRestoreHook
\column{B}{@{}>{\hspre}l<{\hspost}@{}}%
\column{22}{@{}>{\hspre}l<{\hspost}@{}}%
\column{E}{@{}>{\hspre}l<{\hspost}@{}}%
\>[B]{}\Varid{lToR}\mathbin{::}\Conid{Monad}\;\Varid{m}\Rightarrow \Conid{TreeF}\;\Varid{a}\;(\Varid{m}\;\Varid{x})\to \Varid{m}\;(\Conid{TreeF}\;\Varid{a}\;\Varid{x}){}\<[E]%
\\
\>[B]{}\Varid{lToR}\;\Conid{Empty}{}\<[22]%
\>[22]{}\mathrel{=}\Varid{return}\;\Conid{Empty}{}\<[E]%
\\
\>[B]{}\Varid{lToR}\;(\Conid{Node}\;\Varid{ml}\;\Varid{a}\;\Varid{mr}){}\<[22]%
\>[22]{}\mathrel{=}\mathbf{do}\;\Varid{l}\leftarrow \Varid{ml};\Varid{r}\leftarrow \Varid{mr};\Varid{return}\;(\Conid{Node}\;\Varid{l}\;\Varid{a}\;\Varid{r}){}\<[E]%
\ColumnHook
\end{hscode}\resethooks
and also from right to left:
\begin{hscode}\SaveRestoreHook
\column{B}{@{}>{\hspre}l<{\hspost}@{}}%
\column{22}{@{}>{\hspre}l<{\hspost}@{}}%
\column{E}{@{}>{\hspre}l<{\hspost}@{}}%
\>[B]{}\Varid{rToL}\mathbin{::}\Conid{Monad}\;\Varid{m}\Rightarrow \Conid{TreeF}\;\Varid{a}\;(\Varid{m}\;\Varid{x})\to \Varid{m}\;(\Conid{TreeF}\;\Varid{a}\;\Varid{x}){}\<[E]%
\\
\>[B]{}\Varid{rToL}\;\Conid{Empty}{}\<[22]%
\>[22]{}\mathrel{=}\Varid{return}\;\Conid{Empty}{}\<[E]%
\\
\>[B]{}\Varid{rToL}\;(\Conid{Node}\;\Varid{ml}\;\Varid{a}\;\Varid{mr}){}\<[22]%
\>[22]{}\mathrel{=}\mathbf{do}\;\Varid{r}\leftarrow \Varid{mr};\Varid{l}\leftarrow \Varid{ml};\Varid{return}\;(\Conid{Node}\;\Varid{l}\;\Varid{a}\;\Varid{r}){}\<[E]%
\ColumnHook
\end{hscode}\resethooks
\end{example}

\begin{recscheme}[\ensuremath{\Varid{mcata}}] \label{scm:mcata}
A \emph{monadic catamorphism}~\cite{Fok94Monadic,Par05Com} is given by the following recursion scheme:
\begin{hscode}\SaveRestoreHook
\column{B}{@{}>{\hspre}l<{\hspost}@{}}%
\column{8}{@{}>{\hspre}l<{\hspost}@{}}%
\column{E}{@{}>{\hspre}l<{\hspost}@{}}%
\>[B]{}\Varid{mcata}{}\<[8]%
\>[8]{}\mathbin{::}(\Conid{Monad}\;\Varid{m},\Conid{Functor}\;\Varid{f})\Rightarrow (\forall \Varid{x}\hsforall \hsdot{\circ }{.}\Varid{f}\;(\Varid{m}\;\Varid{x})\to \Varid{m}\;(\Varid{f}\;\Varid{x})){}\<[E]%
\\
\>[8]{}\to (\Varid{f}\;\Varid{a}\to \Varid{m}\;\Varid{a})\to \mu\;\Varid{f}\to \Varid{m}\;\Varid{a}{}\<[E]%
\\
\>[B]{}\Varid{mcata}\;\Varid{seq}\;\Varid{alg}\mathrel{=}\Varid{cata}\;((\bind \Varid{alg})\hsdot{\circ }{.}\Varid{seq}){}\<[E]%
\ColumnHook
\end{hscode}\resethooks
\end{recscheme}

\begin{example}
The program \ensuremath{\Varid{printTree}} above is a monadic catamorphism:
\begin{hscode}\SaveRestoreHook
\column{B}{@{}>{\hspre}l<{\hspost}@{}}%
\column{3}{@{}>{\hspre}l<{\hspost}@{}}%
\column{27}{@{}>{\hspre}l<{\hspost}@{}}%
\column{E}{@{}>{\hspre}l<{\hspost}@{}}%
\>[B]{}\Varid{printTree''}\mathbin{::}\Conid{Show}\;\Varid{a}\Rightarrow \mu\;(\Conid{TreeF}\;\Varid{a})\to \Conid{IO}\;(){}\<[E]%
\\
\>[B]{}\Varid{printTree''}\mathrel{=}\Varid{mcata}\;\Varid{lToR}\;\Varid{printElem}\;\mathbf{where}{}\<[E]%
\\
\>[B]{}\hsindent{3}{}\<[3]%
\>[3]{}\Varid{printElem}\;\Conid{Empty}{}\<[27]%
\>[27]{}\mathrel{=}\Varid{return}\;(){}\<[E]%
\\
\>[B]{}\hsindent{3}{}\<[3]%
\>[3]{}\Varid{printElem}\;(\Conid{Node}\;\anonymous \;\Varid{a}\;\anonymous ){}\<[27]%
\>[27]{}\mathrel{=}\Varid{print}\;\Varid{a}{}\<[E]%
\ColumnHook
\end{hscode}\resethooks
Note that \ensuremath{\Varid{mcata}} is strictly less expressive than \ensuremath{\Varid{cataM}} because \ensuremath{\Varid{mcata}}
requires all subtrees processed before the root.
\end{example}

\subsubsection{Distributive Conditions}
In the literature~\cite{Fok94Monadic,Par05Com,HW16USR}, the sequencing of a
monadic catamorphism is required to satisfy a \emph{distributive law} of functor \ensuremath{\Varid{f}}
over monad \ensuremath{\Varid{m}}, which means that \ensuremath{\Varid{seq}\mathbin{::}\forall \Varid{x}\hsforall \hsdot{\circ }{.}\Varid{f}\;(\Varid{m}\;\Varid{x})\to \Varid{m}\;(\Varid{f}\;\Varid{x})}
satisfies two conditions:
\begin{align}
\ensuremath{\Varid{seq}\hsdot{\circ }{.}\Varid{fmap}\;\Varid{return}} &= \ensuremath{\Varid{return}} \label{eq:dist1}\\
\ensuremath{\Varid{seq}\hsdot{\circ }{.}\Varid{fmap}\;\Varid{join}}   &= \ensuremath{\Varid{join}\hsdot{\circ }{.}\Varid{fmap}\;\Varid{seq}\hsdot{\circ }{.}\Varid{seq}}\label{eq:dist2}
\end{align}
Intuitively, condition (\ref{eq:dist1}) prohibits \ensuremath{\Varid{seq}} from inserting additional computational effects when combining computations for substructures, which is a reasonable requirement.
Condition (\ref{eq:dist2}) requires \ensuremath{\Varid{seq}} to be commutative with monadic sequencing.
These requirements are theoretically elegant, because they allow functor \ensuremath{\Varid{f}} to be lifted to the Kleisli category of \ensuremath{\Varid{m}} and consequently \ensuremath{\Varid{mcata}\;\Varid{seq}\;\Varid{alg}} is also a catamorphism in the Kleisli category (\ensuremath{\Varid{mcata}} by definition is a catamorphism in the base category)---giving us nicer calculational properties. 

Unfortunately, condition (\ref{eq:dist2}) is usually too strong in practice.
For example, neither \ensuremath{\Varid{lToR}} nor \ensuremath{\Varid{rToL}} in \autoref{eg:dist} satisfies condition (\ref{eq:dist2}) when \ensuremath{\Varid{m}} is the \ensuremath{\Conid{IO}} monad.
To see this, let 
\begin{hscode}\SaveRestoreHook
\column{B}{@{}>{\hspre}l<{\hspost}@{}}%
\column{11}{@{}>{\hspre}l<{\hspost}@{}}%
\column{E}{@{}>{\hspre}l<{\hspost}@{}}%
\>[B]{}\Varid{c}\mathrel{=}\Conid{Node}\;{}\<[11]%
\>[11]{}(\Varid{putStr}\;\text{\ttfamily \char34 A\char34}\sequ \Varid{return}\;(\Varid{putStr}\;\text{\ttfamily \char34 C\char34}))\;()\;{}\<[E]%
\\
\>[11]{}(\Varid{putStr}\;\text{\ttfamily \char34 B\char34}\sequ \Varid{return}\;(\Varid{putStr}\;\text{\ttfamily \char34 D\char34})){}\<[E]%
\ColumnHook
\end{hscode}\resethooks
Then \ensuremath{(\Varid{lToR}\hsdot{\circ }{.}\Varid{fmap}\;\Varid{join})\;\Varid{c}} prints \ensuremath{\text{\ttfamily \char34 ACBD\char34}} but \ensuremath{(\Varid{join}\hsdot{\circ }{.}\Varid{fmap}\;\Varid{lToR}\hsdot{\circ }{.}\Varid{lToR})\;\Varid{c}} prints \ensuremath{\text{\ttfamily \char34 ABCD\char34}}.
In fact, there is no distributive law of \ensuremath{\Conid{TreeF}\;\Varid{a}} over a monad unless it is commutative, excluding the \ensuremath{\Conid{IO}} monad and \ensuremath{\Conid{State}} monad.
Thus we drop the requirement for \ensuremath{\Varid{seq}} being a distributive law in our definition of monadic catamorphism.

\subsection{More Monadic Recursion Schemes}
As we mentioned above, \ensuremath{\Varid{mcata}} is the catamorphism in the Kleisli category provided \ensuremath{\Varid{seq}} is a distributive law.
No doubt, we can replay our development of recursion schemes in the Kleisli category to get the monadic version of more recursion schemes.
For example, we have \emph{monadic hylomorphisms}~\cite{Par98Mon,Par05Com}:
\begin{hscode}\SaveRestoreHook
\column{B}{@{}>{\hspre}l<{\hspost}@{}}%
\column{8}{@{}>{\hspre}l<{\hspost}@{}}%
\column{29}{@{}>{\hspre}l<{\hspost}@{}}%
\column{E}{@{}>{\hspre}l<{\hspost}@{}}%
\>[B]{}\Varid{mhylo}{}\<[8]%
\>[8]{}\mathbin{::}(\Conid{Monad}\;\Varid{m},\Conid{Functor}\;\Varid{f})\Rightarrow (\forall \Varid{x}\hsforall \hsdot{\circ }{.}\Varid{f}\;(\Varid{m}\;\Varid{x})\to \Varid{m}\;(\Varid{f}\;\Varid{x})){}\<[E]%
\\
\>[8]{}\to (\Varid{f}\;\Varid{a}\to \Varid{m}\;\Varid{a})\to (\Varid{c}\to \Varid{m}\;(\Varid{f}\;\Varid{c}))\to \Varid{c}\to \Varid{m}\;\Varid{a}{}\<[E]%
\\
\>[B]{}\Varid{mhylo}\;\Varid{seq}\;\Varid{alg}\;\Varid{coalg}\;\Varid{c}\mathrel{=}\mathbf{do}\;{}\<[29]%
\>[29]{}\Varid{x}\leftarrow \Varid{coalg}\;\Varid{c}{}\<[E]%
\\
\>[29]{}\Varid{y}\leftarrow \Varid{seq}\;(\Varid{fmap}\;(\Varid{mhylo}\;\Varid{seq}\;\Varid{alg}\;\Varid{coalg})\;\Varid{x}){}\<[E]%
\\
\>[29]{}\Varid{alg}\;\Varid{y}{}\<[E]%
\ColumnHook
\end{hscode}\resethooks
which specializes to \ensuremath{\Varid{mcata}} by \ensuremath{\Varid{mhylo}\;\Varid{seq}\;\Varid{alg}\;(\Varid{return}\hsdot{\circ }{.}\Varid{in}^\circ)} and \emph{monadic anamorphisms} by
\begin{hscode}\SaveRestoreHook
\column{B}{@{}>{\hspre}l<{\hspost}@{}}%
\column{7}{@{}>{\hspre}l<{\hspost}@{}}%
\column{32}{@{}>{\hspre}l<{\hspost}@{}}%
\column{E}{@{}>{\hspre}l<{\hspost}@{}}%
\>[B]{}\Varid{mana}{}\<[7]%
\>[7]{}\mathbin{::}(\Conid{Monad}\;\Varid{m},\Conid{Functor}\;\Varid{f}){}\<[32]%
\>[32]{}\Rightarrow (\forall \Varid{x}\hsforall \hsdot{\circ }{.}\Varid{f}\;(\Varid{m}\;\Varid{x})\to \Varid{m}\;(\Varid{f}\;\Varid{x})){}\<[E]%
\\
\>[32]{}\to (\Varid{c}\to \Varid{m}\;(\Varid{f}\;\Varid{c}))\to \Varid{c}\to \Varid{m}\;(\nu\;\Varid{f}){}\<[E]%
\\
\>[B]{}\Varid{mana}\;\Varid{seq}\;\Varid{coalg}\mathrel{=}\Varid{mhylo}\;\Varid{seq}\;(\Varid{return}\hsdot{\circ }{.}\Varid{Out}^\circ)\;\Varid{coalg}{}\<[E]%
\ColumnHook
\end{hscode}\resethooks
Other recursion schemes discussed in this paper can be devised in the same way.

\begin{example}
Generating a random binary tree of some depth with \ensuremath{\Varid{randomIO}\mathbin{::}\Conid{IO}\;\Conid{Int}} is a monadic anamorphism:
\begin{hscode}\SaveRestoreHook
\column{B}{@{}>{\hspre}l<{\hspost}@{}}%
\column{3}{@{}>{\hspre}l<{\hspost}@{}}%
\column{15}{@{}>{\hspre}l<{\hspost}@{}}%
\column{E}{@{}>{\hspre}l<{\hspost}@{}}%
\>[B]{}\Varid{ranTree}\mathbin{::}\Conid{Integer}\to \Conid{IO}\;(\nu\;(\Conid{TreeF}\;\Conid{Int})){}\<[E]%
\\
\>[B]{}\Varid{ranTree}\mathrel{=}\Varid{mana}\;\Varid{lToR}\;\Varid{gen}\;\mathbf{where}{}\<[E]%
\\
\>[B]{}\hsindent{3}{}\<[3]%
\>[3]{}\Varid{gen}\mathbin{::}\Conid{Integer}\to \Conid{IO}\;(\Conid{TreeF}\;\Conid{Int}\;\Conid{Integer}){}\<[E]%
\\
\>[B]{}\hsindent{3}{}\<[3]%
\>[3]{}\Varid{gen}\;\mathrm{0}\mathrel{=}\Varid{return}\;\Conid{Empty}{}\<[E]%
\\
\>[B]{}\hsindent{3}{}\<[3]%
\>[3]{}\Varid{gen}\;\Varid{n}\mathrel{=}\mathbf{do}\;{}\<[15]%
\>[15]{}\Varid{a}\leftarrow \Varid{randomIO}\mathbin{::}\Conid{IO}\;\Conid{Int}{}\<[E]%
\\
\>[15]{}\Varid{return}\;(\Conid{Node}\;(\Varid{n}\mathbin{-}\mathrm{1})\;\Varid{a}\;(\Varid{n}\mathbin{-}\mathrm{1})){}\<[E]%
\ColumnHook
\end{hscode}\resethooks
\end{example}

\section{Structural Recursion on \textsc{gadt}s}\label{sec:GADTs}

So far we have worked exclusively with (co)inductive datatypes, but they do not cover all algebraic datatypes and \emph{generalized algebraic datatypes} (\textsc{gadt}s).
An example of algebraic datatypes that is not (co)inductive is the datatype for purely functional \emph{random-access lists}~\cite{Oka95Pur}:
\begin{hscode}\SaveRestoreHook
\column{B}{@{}>{\hspre}l<{\hspost}@{}}%
\column{E}{@{}>{\hspre}l<{\hspost}@{}}%
\>[B]{}\mathbf{data}\;\Conid{RList}\;\Varid{a}\mathrel{=}\Conid{Null}\mid \Conid{Zero}\;(\Conid{RList}\;(\Varid{a},\Varid{a}))\mid \Conid{One}\;\Varid{a}\;(\Conid{RList}\;(\Varid{a},\Varid{a})){}\<[E]%
\ColumnHook
\end{hscode}\resethooks
The recursive occurrences of \ensuremath{\Conid{RList}} in constructor \ensuremath{\Conid{Zero}} and \ensuremath{\Conid{One}} are \ensuremath{\Conid{RList}\;(\Varid{a},\Varid{a})} rather than \ensuremath{\Conid{RList}\;\Varid{a}}, and consequently we cannot model \ensuremath{\Conid{RList}\;\Varid{a}} as \ensuremath{\mu\;\Varid{f}} for some functor \ensuremath{\Varid{f}} as we did for lists.
Algebraic datatypes such as \ensuremath{\Conid{RList}} whose defining equation has on the right-hand side any occurrence of the declared type applied to parameters different from those on the left-hand side are called \emph{non-regular} datatypes or \emph{nested} datatypes~\cite{BM98Nes,PN07Ini,PN2009}.

Nested datatypes are covered by a broader range of datatypes called \emph{generalized algebraic datatypes} (\textsc{gadt}s) \cite{PN08Fou,hinze2003fun}.
In terms of the \ensuremath{\mathbf{data}} syntax in Haskell, the generalization of \textsc{gadt}s is to allow the parameters \ensuremath{\Conid{P}} supplied to the declared type \ensuremath{\Conid{D}} on the left-hand side of an defining equation
\ensuremath{\mathbf{data}\;\Conid{D}\;\Conid{P}\mathrel{=}\mathbin{...}}
to be more complex than type variables.
\textsc{gadt}s have a different syntax from that of \textsc{adt}s in Haskell\footnote{Support of \textsc{gadt}s is turned on by the extension \texttt{\textsc{gadt}s} in \textsc{ghc}.}.
For example, as a \textsc{gadt}, \ensuremath{\Conid{RList}} is 
\begin{hscode}\SaveRestoreHook
\column{B}{@{}>{\hspre}l<{\hspost}@{}}%
\column{3}{@{}>{\hspre}l<{\hspost}@{}}%
\column{9}{@{}>{\hspre}l<{\hspost}@{}}%
\column{E}{@{}>{\hspre}l<{\hspost}@{}}%
\>[B]{}\mathbf{data}\;\Conid{RList}\mathbin{::}\mathbin{*}\to \mathbin{*}\;\mathbf{where}{}\<[E]%
\\
\>[B]{}\hsindent{3}{}\<[3]%
\>[3]{}\Conid{Null}{}\<[9]%
\>[9]{}\mathbin{::}\Conid{RList}\;\Varid{a}{}\<[E]%
\\
\>[B]{}\hsindent{3}{}\<[3]%
\>[3]{}\Conid{Zero}{}\<[9]%
\>[9]{}\mathbin{::}\Conid{RList}\;(\Varid{a},\Varid{a})\to \Conid{RList}\;\Varid{a}{}\<[E]%
\\
\>[B]{}\hsindent{3}{}\<[3]%
\>[3]{}\Conid{One}{}\<[9]%
\>[9]{}\mathbin{::}\Varid{a}\to \Conid{RList}\;(\Varid{a},\Varid{a})\to \Conid{RList}\;\Varid{a}{}\<[E]%
\ColumnHook
\end{hscode}\resethooks
in which each constructor is directly declared with a type signature.
With this syntax, allowing parameters on the left-hand side of an ADT equation to be not just variables means that the finally returned type of constructors of a \textsc{gadt} \ensuremath{\Conid{G}} can be more complex than \ensuremath{\Conid{G}\;\Varid{a}} where \ensuremath{\Varid{a}} is a type variable.
A classic example is fixed length vectors of \ensuremath{\Varid{a}}-values:
first we define two datatypes \ensuremath{\mathbf{data}\;\Conid{Z'}} and \ensuremath{\mathbf{data}\;\Conid{S'}\;\Varid{n}} with no constructors, then the 
\textsc{gadt} for vectors is
\begin{hscode}\SaveRestoreHook
\column{B}{@{}>{\hspre}l<{\hspost}@{}}%
\column{3}{@{}>{\hspre}l<{\hspost}@{}}%
\column{9}{@{}>{\hspre}l<{\hspost}@{}}%
\column{E}{@{}>{\hspre}l<{\hspost}@{}}%
\>[B]{}\mathbf{data}\;\Conid{Vec}\;(\Varid{a}\mathbin{::}\mathbin{*})\mathbin{::}\mathbin{*}\to \mathbin{*}\;\mathbf{where}{}\<[E]%
\\
\>[B]{}\hsindent{3}{}\<[3]%
\>[3]{}\Conid{Nil}{}\<[9]%
\>[9]{}\mathbin{::}\Conid{Vec}\;\Varid{a}\;\Conid{Z'}{}\<[E]%
\\
\>[B]{}\hsindent{3}{}\<[3]%
\>[3]{}\Conid{Cons}{}\<[9]%
\>[9]{}\mathbin{::}\Varid{a}\to \Conid{Vec}\;\Varid{a}\;\Varid{n}\to \Conid{Vec}\;\Varid{a}\;(\Conid{S'}\;\Varid{n}){}\<[E]%
\ColumnHook
\end{hscode}\resethooks
in which types \ensuremath{\Conid{Z'}} and \ensuremath{\Conid{S'}\;\Varid{n}} encode natural numbers at the type level, thus 
it does not matter what their term constructors are.

\hyphenation{data-types}

\textsc{gadt}s are a powerful tool to ensure program correctness by indexing datatypes with sophisticated properties of data, such as the size or shape of data, and then the type checker can check these properties statically.
For example, the following program extracting the first element of a vector is always safe because the type of its argument guarantees it is non-empty.
\begin{hscode}\SaveRestoreHook
\column{B}{@{}>{\hspre}l<{\hspost}@{}}%
\column{27}{@{}>{\hspre}l<{\hspost}@{}}%
\column{E}{@{}>{\hspre}l<{\hspost}@{}}%
\>[B]{}\Varid{safeHead}\mathbin{::}\Conid{Vec}\;\Varid{a}\;(\Conid{S'}\;\Varid{n}){}\<[27]%
\>[27]{}\to \Varid{a}{}\<[E]%
\\
\>[B]{}\Varid{safeHead}\;(\Conid{Cons}\;\Varid{a}\;\anonymous ){}\<[27]%
\>[27]{}\mathrel{=}\Varid{a}{}\<[E]%
\ColumnHook
\end{hscode}\resethooks

\subsubsection{\textsc{gadt}s as Fixed Points}
As we mentioned earlier, nested datatypes and \textsc{gadt}s cannot be modelled as fixed
points of Haskell functors in general, making them out of the reach of
the recursion schemes that we have seen so far.
However, there are other ways to view them as fixed points.
Let us look at the \ensuremath{\Conid{RList}} datatype again,
\[\ensuremath{\mathbf{data}\;\Conid{RList}\;\Varid{a}\mathrel{=}\Conid{Null}\mid \Conid{Zero}\;(\Conid{RList}\;(\Varid{a},\Varid{a}))\mid \Conid{One}\;\Varid{a}\;(\Conid{RList}\;(\Varid{a},\Varid{a}))}\]
instead of viewing it as defining a type \ensuremath{\Conid{RList}\;\Varid{a}\mathbin{::}\mathbin{*}}, we can alternatively
understand it as defining a functor \ensuremath{\Conid{RList}\mathbin{::}\mathbin{*}\to \mathbin{*}}, where \ensuremath{\mathbin{*}} is the
category of Haskell types, such that \ensuremath{\Conid{RList}} satisfies the fixed point equation
$\ensuremath{\Conid{RList}} \cong \ensuremath{\Conid{RListF}\;\Conid{RList}}$ for a \emph{higher-order} functor \ensuremath{\Conid{RListF}\mathbin{::}(\mathbin{*}\to \mathbin{*})\to (\mathbin{*}\to \mathbin{*})} defined as
\begin{hscode}\SaveRestoreHook
\column{B}{@{}>{\hspre}l<{\hspost}@{}}%
\column{E}{@{}>{\hspre}l<{\hspost}@{}}%
\>[B]{}\mathbf{data}\;\Conid{RListF}\;\Varid{f}\;\Varid{a}\mathrel{=}\Conid{NullF}\mid \Conid{ZeroF}\;(\Varid{f}\;(\Varid{a},\Varid{a}))\mid \Conid{OneF}\;\Varid{a}\;(\Varid{f}\;(\Varid{a},\Varid{a})){}\<[E]%
\ColumnHook
\end{hscode}\resethooks
In this way, nested datatypes are still fixed points, but of higher-order
functors, rather than usual Haskell functors \cite{BM98Nes,PN07Ini}.

This idea applies to \textsc{gadt}s as well, but with a caveat:
consider the \textsc{gadt} \ensuremath{\Conid{G}} defined as follows:
\begin{hscode}\SaveRestoreHook
\column{B}{@{}>{\hspre}l<{\hspost}@{}}%
\column{3}{@{}>{\hspre}l<{\hspost}@{}}%
\column{9}{@{}>{\hspre}l<{\hspost}@{}}%
\column{E}{@{}>{\hspre}l<{\hspost}@{}}%
\>[B]{}\mathbf{data}\;\Conid{G}\;\Varid{a}\;\mathbf{where}{}\<[E]%
\\
\>[B]{}\hsindent{3}{}\<[3]%
\>[3]{}\Conid{Leaf}{}\<[9]%
\>[9]{}\mathbin{::}\Varid{a}\to \Conid{G}\;\Varid{a}{}\<[E]%
\\
\>[B]{}\hsindent{3}{}\<[3]%
\>[3]{}\Conid{Prod}{}\<[9]%
\>[9]{}\mathbin{::}\Conid{G}\;\Varid{a}\to \Conid{G}\;\Varid{b}\to \Conid{G}\;(\Varid{a},\Varid{b}){}\<[E]%
\ColumnHook
\end{hscode}\resethooks
then \ensuremath{\Conid{G}} \emph{cannot} be a functor at all, let alone a fixed point of
some higher-order functor.
The problem is defining \ensuremath{\Varid{fmap}} for the \ensuremath{\Conid{Prod}} constructor:
\[\ensuremath{\Varid{fmap}\;\Varid{f}\;(\Conid{Prod}\;\Varid{ga}\;\Varid{gb})\mathrel{=}\anonymous \mathbin{::}\Conid{G}\;\Varid{c}}\]
However, we have no way to construct a \ensuremath{\Conid{G}\;\Varid{c}} given \ensuremath{\Varid{f}\mathbin{::}(\Varid{a},\Varid{b})\to \Varid{c}}, \ensuremath{\Varid{ga}\mathbin{::}\Conid{G}\;\Varid{a}}
and \ensuremath{\Varid{gb}\mathbin{::}\Conid{G}\;\Varid{b}}.
Luckily, Johann and Ghani \cite{PN08Fou} shows how to fix this problem. 
In fact, all we need to do is to give up the expectation that a \textsc{gadt} \ensuremath{\Conid{G}\mathbin{::}\mathbin{*}\to \mathbin{*}}
is functorial in its domain.
In categorical terminology, we view \textsc{gadt}s as functors from the \emph{discrete category}
$\disccat{\ensuremath{\mathbin{*}}}$ of Haskell types to the category \ensuremath{\mathbin{*}} of Haskell types, rather than
functors from \ensuremath{\mathbin{*}} to \ensuremath{\mathbin{*}}.
In other words, a \textsc{gadt} \ensuremath{\Conid{G}\mathbin{::}\mathbin{*}\to \mathbin{*}} is then merely a type constructor in Haskell,
without necessarily a \ensuremath{\Conid{Functor}} instance.
A \emph{natural transformation} between two functors \ensuremath{\Varid{a}} and \ensuremath{\Varid{b}} from
$\disccat{\ensuremath{\mathbin{*}}}$ to \ensuremath{\mathbin{*}} is a polymorphic function \ensuremath{\forall \Varid{i}\hsforall \hsdot{\circ }{.}\Varid{a}\;\Varid{i}\to \Varid{b}\;\Varid{i}}, which
we give a type synonym \ensuremath{\Varid{a}\;\dot{\rightarrow}\;\Varid{b}}\footnote{It requires the \texttt{RankNTypes}
extension of \textsc{ghc}.}:
\begin{hscode}\SaveRestoreHook
\column{B}{@{}>{\hspre}l<{\hspost}@{}}%
\column{E}{@{}>{\hspre}l<{\hspost}@{}}%
\>[B]{}\mathbf{type}\;(\;\dot{\rightarrow}\;)\;\Varid{a}\;\Varid{b}\mathrel{=}\forall \Varid{i}\hsforall \hsdot{\circ }{.}\Varid{a}\;\Varid{i}\to \Varid{b}\;\Varid{i}{}\<[E]%
\ColumnHook
\end{hscode}\resethooks
And a higher-order endofunctor (on the functor category
$\ensuremath{\mathbin{*}}^{\disccat{\ensuremath{\mathbin{*}}}}$) is \ensuremath{\Varid{f}} instantiating the following type class, which is analogous
to the \ensuremath{\Conid{Functor}} type class of Haskell:
\begin{hscode}\SaveRestoreHook
\column{B}{@{}>{\hspre}l<{\hspost}@{}}%
\column{3}{@{}>{\hspre}l<{\hspost}@{}}%
\column{E}{@{}>{\hspre}l<{\hspost}@{}}%
\>[B]{}\mathbf{class}\;\Conid{HFunctor}\;(\Varid{f}\mathbin{::}(\mathbin{*}\to \mathbin{*})\to (\mathbin{*}\to \mathbin{*}))\;\mathbf{where}{}\<[E]%
\\
\>[B]{}\hsindent{3}{}\<[3]%
\>[3]{}\Varid{hfmap}\mathbin{::}(\Varid{a}\;\dot{\rightarrow}\;\Varid{b})\to (\Varid{f}\;\Varid{a}\;\dot{\rightarrow}\;\Varid{f}\;\Varid{b}){}\<[E]%
\ColumnHook
\end{hscode}\resethooks
in which \ensuremath{\Varid{fmap}}'s counterpart \ensuremath{\Varid{hfmap}} maps a natural transformation \ensuremath{\Varid{a}\;\dot{\rightarrow}\;\Varid{b}}
to another natural transformation \ensuremath{\Varid{f}\;\Varid{a}\;\dot{\rightarrow}\;\Varid{f}\;\Varid{b}}.
On top of these, the least-fixed-point operator for an \ensuremath{\Conid{HFunctor}} is
\begin{hscode}\SaveRestoreHook
\column{B}{@{}>{\hspre}l<{\hspost}@{}}%
\column{3}{@{}>{\hspre}l<{\hspost}@{}}%
\column{E}{@{}>{\hspre}l<{\hspost}@{}}%
\>[B]{}\mathbf{data}\;\dot{\mu}\mathbin{::}((\mathbin{*}\to \mathbin{*})\to (\mathbin{*}\to \mathbin{*}))\to (\mathbin{*}\to \mathbin{*})\;\mathbf{where}{}\<[E]%
\\
\>[B]{}\hsindent{3}{}\<[3]%
\>[3]{}\dot{\Varid{In}}\mathbin{::}\Varid{f}\;(\dot{\mu}\;\Varid{f})\;\Varid{i}\to \dot{\mu}\;\Varid{f}\;\Varid{i}{}\<[E]%
\ColumnHook
\end{hscode}\resethooks

\begin{example} Fixed-length vectors \ensuremath{\Conid{Vec}\;\Varid{e}} are isomorphic to \ensuremath{\dot{\mu}\;(\Conid{VecF}\;\Varid{e})} where
\begin{hscode}\SaveRestoreHook
\column{B}{@{}>{\hspre}l<{\hspost}@{}}%
\column{3}{@{}>{\hspre}l<{\hspost}@{}}%
\column{10}{@{}>{\hspre}l<{\hspost}@{}}%
\column{E}{@{}>{\hspre}l<{\hspost}@{}}%
\>[B]{}\mathbf{data}\;\Conid{VecF}\mathbin{::}\mathbin{*}\to (\mathbin{*}\to \mathbin{*})\to (\mathbin{*}\to \mathbin{*})\;\mathbf{where}{}\<[E]%
\\
\>[B]{}\hsindent{3}{}\<[3]%
\>[3]{}\Conid{NilF}{}\<[10]%
\>[10]{}\mathbin{::}\Conid{VecF}\;\Varid{e}\;\Varid{f}\;\Conid{Z'}{}\<[E]%
\\
\>[B]{}\hsindent{3}{}\<[3]%
\>[3]{}\Conid{ConsF}{}\<[10]%
\>[10]{}\mathbin{::}\Varid{e}\to \Varid{f}\;\Varid{n}\to \Conid{VecF}\;\Varid{e}\;\Varid{f}\;(\Conid{S'}\;\Varid{n}){}\<[E]%
\ColumnHook
\end{hscode}\resethooks
which has \ensuremath{\Conid{HFunctor}} instance
\begin{hscode}\SaveRestoreHook
\column{B}{@{}>{\hspre}l<{\hspost}@{}}%
\column{3}{@{}>{\hspre}l<{\hspost}@{}}%
\column{27}{@{}>{\hspre}l<{\hspost}@{}}%
\column{E}{@{}>{\hspre}l<{\hspost}@{}}%
\>[B]{}\mathbf{instance}\;\Conid{HFunctor}\;(\Conid{VecF}\;\Varid{e})\;\mathbf{where}{}\<[E]%
\\
\>[B]{}\hsindent{3}{}\<[3]%
\>[3]{}\Varid{hfmap}\;\Varid{phi}\;\Conid{NilF}{}\<[27]%
\>[27]{}\mathrel{=}\Conid{NilF}{}\<[E]%
\\
\>[B]{}\hsindent{3}{}\<[3]%
\>[3]{}\Varid{hfmap}\;\Varid{phi}\;(\Conid{ConsF}\;\Varid{e}\;\Varid{es}){}\<[27]%
\>[27]{}\mathrel{=}\Conid{ConsF}\;\Varid{e}\;(\Varid{phi}\;\Varid{es}){}\<[E]%
\ColumnHook
\end{hscode}\resethooks
\end{example}

\begin{recscheme}[\ensuremath{\Varid{icata}}]\label{scm:icata}
With the machinery above, we can devise the structural recursion scheme for \ensuremath{\dot{\mu}}, 
which we call \emph{indexed catamorphisms}:
\begin{hscode}\SaveRestoreHook
\column{B}{@{}>{\hspre}l<{\hspost}@{}}%
\column{20}{@{}>{\hspre}l<{\hspost}@{}}%
\column{E}{@{}>{\hspre}l<{\hspost}@{}}%
\>[B]{}\Varid{icata}\mathbin{::}\Conid{HFunctor}\;\Varid{f}\Rightarrow (\Varid{f}\;\Varid{a}\;\dot{\rightarrow}\;\Varid{a})\to \dot{\mu}\;\Varid{f}\;\dot{\rightarrow}\;\Varid{a}{}\<[E]%
\\
\>[B]{}\Varid{icata}\;\Varid{alg}\;(\dot{\Varid{In}}\;\Varid{x}){}\<[20]%
\>[20]{}\mathrel{=}\Varid{alg}\;(\Varid{hfmap}\;(\Varid{icata}\;\Varid{alg})\;\Varid{x}){}\<[E]%
\ColumnHook
\end{hscode}\resethooks
\end{recscheme}

\begin{example}
Just like list processing functions such as \ensuremath{\Varid{map}} are catamorphisms,
their counterparts for vectors can also be written as indexed catamorphisms:
\begin{hscode}\SaveRestoreHook
\column{B}{@{}>{\hspre}l<{\hspost}@{}}%
\column{3}{@{}>{\hspre}l<{\hspost}@{}}%
\column{21}{@{}>{\hspre}l<{\hspost}@{}}%
\column{E}{@{}>{\hspre}l<{\hspost}@{}}%
\>[B]{}\Varid{vmap}\mathbin{::}\forall \Varid{a}\hsforall \;\Varid{b}\hsdot{\circ }{.}(\Varid{a}\to \Varid{b})\to \dot{\mu}\;(\Conid{VecF}\;\Varid{a})\;\dot{\rightarrow}\;\dot{\mu}\;(\Conid{VecF}\;\Varid{b}){}\<[E]%
\\
\>[B]{}\Varid{vmap}\;\Varid{f}\mathrel{=}\Varid{icata}\;\Varid{alg}\;\mathbf{where}{}\<[E]%
\\
\>[B]{}\hsindent{3}{}\<[3]%
\>[3]{}\Varid{alg}\mathbin{::}\Conid{VecF}\;\Varid{a}\;(\dot{\mu}\;(\Conid{VecF}\;\Varid{b}))\;\dot{\rightarrow}\;\dot{\mu}\;(\Conid{VecF}\;\Varid{b}){}\<[E]%
\\
\>[B]{}\hsindent{3}{}\<[3]%
\>[3]{}\Varid{alg}\;\Conid{NilF}{}\<[21]%
\>[21]{}\mathrel{=}\dot{\Varid{In}}\;\Conid{NilF}{}\<[E]%
\\
\>[B]{}\hsindent{3}{}\<[3]%
\>[3]{}\Varid{alg}\;(\Conid{ConsF}\;\Varid{a}\;\Varid{bs}){}\<[21]%
\>[21]{}\mathrel{=}\dot{\Varid{In}}\;(\Conid{ConsF}\;(\Varid{f}\;\Varid{a})\;\Varid{bs}){}\<[E]%
\ColumnHook
\end{hscode}\resethooks
\end{example}

\begin{example}
Terms of untyped lambda calculus with de Bruijn indices can be modelled as the fixed point of the following higher-order
functor \cite{Bird_paterson_1999}:
\begin{hscode}\SaveRestoreHook
\column{B}{@{}>{\hspre}l<{\hspost}@{}}%
\column{3}{@{}>{\hspre}l<{\hspost}@{}}%
\column{E}{@{}>{\hspre}l<{\hspost}@{}}%
\>[B]{}\mathbf{data}\;\Conid{LambdaF}\mathbin{::}(\mathbin{*}\to \mathbin{*})\to (\mathbin{*}\to \mathbin{*})\;\mathbf{where}{}\<[E]%
\\
\>[B]{}\hsindent{3}{}\<[3]%
\>[3]{}\Conid{Var}\mathbin{::}\Varid{a}\to \Conid{LambdaF}\;\Varid{f}\;\Varid{a}{}\<[E]%
\\
\>[B]{}\hsindent{3}{}\<[3]%
\>[3]{}\Conid{App}\mathbin{::}\Varid{f}\;\Varid{a}\to \Varid{f}\;\Varid{a}\to \Conid{LambdaF}\;\Varid{f}\;\Varid{a}{}\<[E]%
\\
\>[B]{}\hsindent{3}{}\<[3]%
\>[3]{}\Conid{Abs}\mathbin{::}\Varid{f}\;(\Conid{Maybe}\;\Varid{a})\to \Conid{LambdaF}\;\Varid{f}\;\Varid{a}{}\<[E]%
\ColumnHook
\end{hscode}\resethooks
Letting \ensuremath{\Varid{a}} be some type, inhabitants of \ensuremath{\dot{\mu}\;\Conid{LambdaF}\;\Varid{a}} are precisely the lambda terms in which free variables range over \ensuremath{\Varid{a}}.
Thus \ensuremath{\dot{\mu}\;\Conid{LambdaF}\;\Conid{Void}} is the type of closed lambda terms where \ensuremath{\Conid{Void}} is the type has no inhabitants.
Note that the constructor \ensuremath{\Conid{Abs}} applies the recursive placeholder \ensuremath{\Varid{f}} to \ensuremath{\Conid{Maybe}\;\Varid{a}}, providing the inner term with exactly one more fresh variable \ensuremath{\Conid{Nothing}}.

The free variables of a lambda term can be extracted into a list
quite easily, by converting from \ensuremath{\dot{\mu}\;\Conid{LambdaF}\;\Varid{a}} to \ensuremath{[\mskip1.5mu \Varid{a}\mskip1.5mu]} for any
type \ensuremath{\Varid{a}}:
\begin{hscode}\SaveRestoreHook
\column{B}{@{}>{\hspre}l<{\hspost}@{}}%
\column{3}{@{}>{\hspre}l<{\hspost}@{}}%
\column{22}{@{}>{\hspre}l<{\hspost}@{}}%
\column{E}{@{}>{\hspre}l<{\hspost}@{}}%
\>[B]{}\Varid{vars}\mathbin{::}\dot{\mu}\;\Conid{LambdaF}\;\dot{\rightarrow}\;[\mskip1.5mu \mskip1.5mu]{}\<[E]%
\\
\>[B]{}\Varid{vars}\mathrel{=}\Varid{icata}\;\Varid{alg}\;\mathbf{where}{}\<[E]%
\\
\>[B]{}\hsindent{3}{}\<[3]%
\>[3]{}\Varid{alg}\mathbin{::}\Conid{LambdaF}\;[\mskip1.5mu \mskip1.5mu]\;\dot{\rightarrow}\;[\mskip1.5mu \mskip1.5mu]{}\<[E]%
\\
\>[B]{}\hsindent{3}{}\<[3]%
\>[3]{}\Varid{alg}\;(\Conid{Var}\;\Varid{v}){}\<[22]%
\>[22]{}\mathrel{=}[\mskip1.5mu \Varid{v}\mskip1.5mu]{}\<[E]%
\\
\>[B]{}\hsindent{3}{}\<[3]%
\>[3]{}\Varid{alg}\;(\Conid{App}\;\Varid{fvs}\;\Varid{xvs}){}\<[22]%
\>[22]{}\mathrel{=}\Varid{fvs}\plus \Varid{xvs}{}\<[E]%
\\
\>[B]{}\hsindent{3}{}\<[3]%
\>[3]{}\Varid{alg}\;(\Conid{Abs}\;\Varid{vs}){}\<[22]%
\>[22]{}\mathrel{=}[\mskip1.5mu \Varid{v}\mid \Conid{Just}\;\Varid{v}\leftarrow \Varid{vs}\mskip1.5mu]{}\<[E]%
\ColumnHook
\end{hscode}\resethooks
This obtains all the free variables without attempting to remove
duplicates.

The size of a lambda term can also be computed structurally.
However, what we get from \ensuremath{\Varid{icata}} is always an arrow \ensuremath{\dot{\mu}\;\Conid{LambdaF}\;\dot{\rightarrow}\;\Varid{a}} for some \ensuremath{\Varid{a}\mathbin{::}\mathbin{*}\to \mathbin{*}}.
If we want to compute just an integer, we need to wrap it in a constant functor:
\begin{hscode}\SaveRestoreHook
\column{B}{@{}>{\hspre}l<{\hspost}@{}}%
\column{E}{@{}>{\hspre}l<{\hspost}@{}}%
\>[B]{}\mathbf{newtype}\;\Conid{K}\;\Varid{a}\;\Varid{x}\mathrel{=}\Conid{K}\;\{\mskip1.5mu \Varid{unwrap}\mathbin{::}\Varid{a}\mskip1.5mu\}{}\<[E]%
\ColumnHook
\end{hscode}\resethooks
Computing the size of a term is done by
\begin{hscode}\SaveRestoreHook
\column{B}{@{}>{\hspre}l<{\hspost}@{}}%
\column{3}{@{}>{\hspre}l<{\hspost}@{}}%
\column{26}{@{}>{\hspre}l<{\hspost}@{}}%
\column{E}{@{}>{\hspre}l<{\hspost}@{}}%
\>[B]{}\Varid{size}\mathbin{::}\dot{\mu}\;\Conid{LambdaF}\;\dot{\rightarrow}\;\Conid{K}\;\Conid{Integer}{}\<[E]%
\\
\>[B]{}\Varid{size}\mathrel{=}\Varid{icata}\;\Varid{alg}\;\mathbf{where}{}\<[E]%
\\
\>[B]{}\hsindent{3}{}\<[3]%
\>[3]{}\Varid{alg}\mathbin{::}\Conid{LambdaF}\;(\Conid{K}\;\Conid{Integer})\;\dot{\rightarrow}\;\Conid{K}\;\Conid{Integer}{}\<[E]%
\\
\>[B]{}\hsindent{3}{}\<[3]%
\>[3]{}\Varid{alg}\;(\Conid{Var}\;\anonymous ){}\<[26]%
\>[26]{}\mathrel{=}\Conid{K}\;\mathrm{1}{}\<[E]%
\\
\>[B]{}\hsindent{3}{}\<[3]%
\>[3]{}\Varid{alg}\;(\Conid{App}\;(\Conid{K}\;\Varid{n})\;(\Conid{K}\;\Varid{m})){}\<[26]%
\>[26]{}\mathrel{=}\Conid{K}\;(\Varid{n}\mathbin{+}\Varid{m}\mathbin{+}\mathrm{1}){}\<[E]%
\\
\>[B]{}\hsindent{3}{}\<[3]%
\>[3]{}\Varid{alg}\;(\Conid{Abs}\;(\Conid{K}\;\Varid{n})){}\<[26]%
\>[26]{}\mathrel{=}\Conid{K}\;(\Varid{n}\mathbin{+}\mathrm{1}){}\<[E]%
\ColumnHook
\end{hscode}\resethooks

\end{example}

\begin{example}
An indexed catamorphism \ensuremath{\Varid{icata}\;\Varid{alg}} is a function \ensuremath{\forall \Varid{i}\hsforall \hsdot{\circ }{.}\dot{\mu}\;\Varid{f}\;\Varid{i}\to \Varid{a}\;\Varid{i}}
polymorphic in index \ensuremath{\Varid{i}}.
However, we might be interested in \textsc{gadt}s and nested datatypes applied to some monomorphic index.
Consider the following program summing up a random-access list of integers.
\begin{hscode}\SaveRestoreHook
\column{B}{@{}>{\hspre}l<{\hspost}@{}}%
\column{22}{@{}>{\hspre}l<{\hspost}@{}}%
\column{E}{@{}>{\hspre}l<{\hspost}@{}}%
\>[B]{}\Varid{sumRList}\mathbin{::}\Conid{RList}\;\Conid{Integer}\to \Conid{Integer}{}\<[E]%
\\
\>[B]{}\Varid{sumRList}\;\Conid{Null}{}\<[22]%
\>[22]{}\mathrel{=}\mathrm{0}{}\<[E]%
\\
\>[B]{}\Varid{sumRList}\;(\Conid{Zero}\;\Varid{xs}){}\<[22]%
\>[22]{}\mathrel{=}\Varid{sumRList}\;(\Varid{fmap}\;(\Varid{uncurry}\;(\mathbin{+}))\;\Varid{xs}){}\<[E]%
\\
\>[B]{}\Varid{sumRList}\;(\Conid{One}\;\Varid{x}\;\Varid{xs}){}\<[22]%
\>[22]{}\mathrel{=}\Varid{x}\mathbin{+}\Varid{sumRList}\;(\Varid{fmap}\;(\Varid{uncurry}\;(\mathbin{+}))\;\Varid{xs}){}\<[E]%
\ColumnHook
\end{hscode}\resethooks
Does it fit into an indexed catamorphism from \ensuremath{\dot{\mu}\;\Conid{RListF}}?
The answer is yes, with the clever choice of the continuation monad \ensuremath{\Conid{Cont}\;\Conid{Integer}\;\Varid{a}} as the result type of \ensuremath{\Varid{icata}}.
\begin{hscode}\SaveRestoreHook
\column{B}{@{}>{\hspre}l<{\hspost}@{}}%
\column{3}{@{}>{\hspre}l<{\hspost}@{}}%
\column{5}{@{}>{\hspre}l<{\hspost}@{}}%
\column{21}{@{}>{\hspre}l<{\hspost}@{}}%
\column{E}{@{}>{\hspre}l<{\hspost}@{}}%
\>[B]{}\mathbf{newtype}\;\Conid{Cont}\;\Varid{r}\;\Varid{a}\mathrel{=}\Conid{Cont}\;\{\mskip1.5mu \Varid{runCont}\mathbin{::}(\Varid{a}\to \Varid{r})\to \Varid{r}\mskip1.5mu\}{}\<[E]%
\\[\blanklineskip]%
\>[B]{}\Varid{sumRList'}\mathbin{::}\dot{\mu}\;\Conid{RListF}\;\Conid{Integer}\to \Conid{Integer}{}\<[E]%
\\
\>[B]{}\Varid{sumRList'}\;\Varid{x}\mathrel{=}\Varid{runCont}\;(\Varid{h}\;\Varid{x})\;\Varid{id}\;\mathbf{where}{}\<[E]%
\\
\>[B]{}\hsindent{3}{}\<[3]%
\>[3]{}\Varid{h}\mathbin{::}\dot{\mu}\;\Conid{RListF}\;\dot{\rightarrow}\;\Conid{Cont}\;\Conid{Integer}{}\<[E]%
\\
\>[B]{}\hsindent{3}{}\<[3]%
\>[3]{}\Varid{h}\mathrel{=}\Varid{icata}\;\Varid{sum}\;\mathbf{where}{}\<[E]%
\\
\>[3]{}\hsindent{2}{}\<[5]%
\>[5]{}\Varid{sum}\mathbin{::}\Conid{RListF}\;(\Conid{Cont}\;\Conid{Integer})\;\dot{\rightarrow}\;\Conid{Cont}\;\Conid{Integer}{}\<[E]%
\\
\>[3]{}\hsindent{2}{}\<[5]%
\>[5]{}\Varid{sum}\;\Conid{NullF}{}\<[21]%
\>[21]{}\mathrel{=}\Conid{Cont}\;(\lambda \Varid{k}\to \mathrm{0}){}\<[E]%
\\
\>[3]{}\hsindent{2}{}\<[5]%
\>[5]{}\Varid{sum}\;(\Conid{ZeroF}\;\Varid{s}){}\<[21]%
\>[21]{}\mathrel{=}\Conid{Cont}\;(\lambda \Varid{k}\to \Varid{runCont}\;\Varid{s}\;(\Varid{fork}\;\Varid{k})){}\<[E]%
\\
\>[3]{}\hsindent{2}{}\<[5]%
\>[5]{}\Varid{sum}\;(\Conid{OneF}\;\Varid{a}\;\Varid{s}){}\<[21]%
\>[21]{}\mathrel{=}\Conid{Cont}\;(\lambda \Varid{k}\to \Varid{k}\;\Varid{a}\mathbin{+}\Varid{runCont}\;\Varid{s}\;(\Varid{fork}\;\Varid{k})){}\<[E]%
\\[\blanklineskip]%
\>[3]{}\hsindent{2}{}\<[5]%
\>[5]{}\Varid{fork}\mathbin{::}(\Varid{y}\to \Conid{Integer})\to (\Varid{y},\Varid{y})\to \Conid{Integer}{}\<[E]%
\\
\>[3]{}\hsindent{2}{}\<[5]%
\>[5]{}\Varid{fork}\;\Varid{k}\;(\Varid{a},\Varid{b})\mathrel{=}\Varid{k}\;\Varid{a}\mathbin{+}\Varid{k}\;\Varid{b}{}\<[E]%
\ColumnHook
\end{hscode}\resethooks
Historically, structural recursion on nested datatypes applied to a monomorphic type was thought as falling out of \ensuremath{\Varid{icata}} and led to the development of \emph{generalized folds}~\cite{BiP98Gen,AMU05Ite}.
Later, Johann and Ghani \cite{PN07Ini} showed \ensuremath{\Varid{icata}} is in fact expressive enough by using right Kan extensions as the result type of \ensuremath{\Varid{icata}}, of which \ensuremath{\Conid{Cont}} used in this example is a special case.
\end{example}

\section{Equational Reasoning with Recursion Schemes}\label{sec:calc}

We have talked about a handful of recursion schemes, which are recognized common patterns in recursive functions.
Recognizing common patterns help programmers understand a new problem and communicate their solutions with others.
Better still, recursion schemes offer rigorous and formal \emph{calculational properties} with which the programmer can manipulate programs in a way similar to manipulate standard mathematical objects such as numbers and polynomials.
In this section, we briefly show some of the properties and an example of reasoning about programs using them.
We refer to Bird and de Moor \cite{BdM97Alg} for a comprehensive introduction to this subject and Bird \cite{Bir10Pea} for more examples of reasoning about and optimizing algorithms in this approach.

We focus on \emph{hylomorphisms}, as almost all recursion schemes are a hylomorphism in a certain category.
The fundamental property is the unique existence of the solution to a hylo equation given a recursive coalgebra \ensuremath{\Varid{c}} (or dually, a corecursive algebra \ensuremath{\Varid{a}}):
for any \ensuremath{\Varid{x}},
\begin{equation}\label{law:uniq}
\ensuremath{\Varid{x}\mathrel{=}\Varid{a}\hsdot{\circ }{.}\Varid{fmap}\;\Varid{x}\hsdot{\circ }{.}\Varid{c}} \iff \ensuremath{\Varid{x}\mathrel{=}\Varid{hylo}\;\Varid{a}\;\Varid{c}}  \tag{\textsc{HyloUniq}}
\end{equation}
which directly follows the definition of a recursive coalgebra.
Instantiating \ensuremath{\Varid{x}} to \ensuremath{\Varid{hylo}\;\Varid{a}\;\Varid{c}}, we get the defining equation of \ensuremath{\Varid{hylo}}
\begin{equation}
\ensuremath{\Varid{hylo}\;\Varid{a}\;\Varid{c}\mathrel{=}\Varid{a}\hsdot{\circ }{.}\Varid{fmap}\;(\Varid{hylo}\;\Varid{a}\;\Varid{c})\hsdot{\circ }{.}\Varid{c}}       \tag{\textsc{HyloComp}}
\end{equation}
which is sometimes called the \emph{computation law}, because it tells how to compute \ensuremath{\Varid{hylo}\;\Varid{a}\;\Varid{c}} recursively.
Instantiating \ensuremath{\Varid{x}} to \ensuremath{\Varid{id}}, we get
\begin{equation}\label{law:refl}
\ensuremath{\Varid{id}\mathrel{=}\Varid{a}\hsdot{\circ }{.}\Varid{c}} \iff \ensuremath{\Varid{id}\mathrel{=}\Varid{hylo}\;\Varid{a}\;\Varid{c}}         \tag{\textsc{HyloRefl}}
\end{equation}
called the \emph{reflection law}, which gives a necessary and sufficient
condition for \ensuremath{\Varid{hylo}\;\Varid{a}\;\Varid{c}} being the identity function.
Note that in this law, \ensuremath{\Varid{c}\mathbin{::}\Varid{r}\to \Varid{f}\;\Varid{r}} and \ensuremath{\Varid{a}\mathbin{::}\Varid{f}\;\Varid{r}\to \Varid{r}} share the same
carrier type \ensuremath{\Varid{r}}. Furthermore this law entails that the algebra \ensuremath{\Varid{a}} is surjective,
and the coalgebra \ensuremath{\Varid{c}} is injective.
A direct consequence of \ref{law:refl} is \ensuremath{\Varid{cata}\;\Conid{In}\mathrel{=}\Varid{id}} because \ensuremath{\Varid{cata}\;\Varid{a}\mathrel{=}\Varid{hylo}\;\Varid{a}\;\Varid{in}^\circ} and \ensuremath{\Varid{id}\mathrel{=}\Conid{In}\hsdot{\circ }{.}\Varid{in}^\circ}. Dually, we also have \ensuremath{\Varid{ana}\;\Conid{Out}\mathrel{=}\Varid{id}}.

An important consequence of \ref{law:uniq} is the following \emph{fusion law}.
It is easier to describe diagrammatically:
The \ref{law:uniq} law states that there is exactly one \ensuremath{\Varid{x}}, i.e.\ \ensuremath{\Varid{hylo}\;\Varid{a}\;\Varid{c}}, such that the following diagram commutes (i.e.\ all paths with the same start and end points give the same result when their edges are composed together):
\begin{equation*}
\begin{tikzcd}
\ensuremath{\Varid{ta}}                  & \ensuremath{\Varid{tc}} \arrow[l, "\ensuremath{\Varid{x}}"'] \arrow[d, "\ensuremath{\Varid{c}}"] \\
\ensuremath{\Varid{f}\;\Varid{ta}} \arrow[u, "\ensuremath{\Varid{a}}"] & \ensuremath{\Varid{f}\;\Varid{tc}} \arrow[l, "\ensuremath{\Varid{fmap}\;\Varid{x}}"]      
\end{tikzcd}
\end{equation*}
If we put another \emph{commuting} square beside it,
\begin{equation}
\begin{tikzcd}
\ensuremath{\Varid{tb}}                  & \ensuremath{\Varid{ta}} \arrow[l, "\ensuremath{\Varid{h}}"']                      & \ensuremath{\Varid{tc}} \arrow[l, "x"'] \arrow[d, "c"] \\
\ensuremath{\Varid{f}\;\Varid{tb}} \arrow[u, "\ensuremath{\Varid{b}}"] & \ensuremath{\Varid{f}\;\Varid{ta}} \arrow[u, "a"] \arrow[l, "\ensuremath{\Varid{fmap}\;\Varid{h}}"] & \ensuremath{\Varid{f}\;\Varid{tc}} \arrow[l, "\ensuremath{\Varid{fmap}\;\Varid{x}}"]      
\end{tikzcd}
\end{equation}
the outer rectangle (with top edge \ensuremath{\Varid{h}\hsdot{\circ }{.}\Varid{x}}) also commutes, and it is also an instance of \ref{law:uniq} with coalgebra \ensuremath{\Varid{c}} and algebra \ensuremath{\Varid{b}}.
Because \ref{law:uniq} states \ensuremath{\Varid{hylo}\;\Varid{c}\;\Varid{b}} is the only arrow making the outer rectangle commute, thus \ensuremath{\Varid{hylo}\;\Varid{c}\;\Varid{b}\mathrel{=}\Varid{h}\hsdot{\circ }{.}\Varid{x}\mathrel{=}\Varid{h}\hsdot{\circ }{.}\Varid{hylo}\;\Varid{a}\;\Varid{c}}.
In summary, the fusion law is:
\begin{equation}\label{law:fusion}
\ensuremath{\Varid{h}\hsdot{\circ }{.}\Varid{hylo}\;\Varid{a}\;\Varid{c}\mathrel{=}\Varid{hylo}\;\Varid{b}\;\Varid{c}} \ \Longleftarrow\  \ensuremath{\Varid{h}\hsdot{\circ }{.}\Varid{a}\mathrel{=}\Varid{b}\hsdot{\circ }{.}\Varid{fmap}\;\Varid{h}}  \tag{\textsc{HyloFusion}}
\end{equation}
and its dual version for corecursive algebra \ensuremath{\Varid{a}} is
\begin{equation*}
\ensuremath{\Varid{hylo}\;\Varid{a}\;\Varid{c}\hsdot{\circ }{.}\Varid{h}\mathrel{=}\Varid{hylo}\;\Varid{a}\;\Varid{d}} \ \Longleftarrow\  \ensuremath{\Varid{c}\hsdot{\circ }{.}\Varid{h}\mathrel{=}\Varid{fmap}\;\Varid{h}\hsdot{\circ }{.}\Varid{d}}   \tag{\textsc{HyloFusionCo}}
\end{equation*}
where \ensuremath{\Varid{d}\mathbin{::}\Varid{td}\to \Varid{f}\;\Varid{td}}.
Fusion laws combine a function after or before a hylomorphism into one
hylomorphism, and thus they are is widely used for
optimization~\cite{CLS07Str}.

We demonstrate how these calculational properties can be used to reason about programs with an example.
\begin{example}
Suppose some \ensuremath{\Varid{f}\mathbin{::}\Conid{Integer}\to \Conid{Integer}} such that for all \ensuremath{\Varid{a},\Varid{b}\mathbin{::}\Conid{Integer}},
\begin{equation}\label{eq:f}
\ensuremath{\Varid{f}\;(\Varid{a}\mathbin{+}\Varid{b})\mathrel{=}\Varid{f}\;\Varid{a}\mathbin{+}\Varid{f}\;\Varid{b}} \quad \wedge \quad \ensuremath{\Varid{f}\;\mathrm{0}\mathrel{=}\mathrm{0}} 
\end{equation}
and \ensuremath{\Varid{sum}} and \ensuremath{\Varid{map}} are the familiar Haskell functions defined with \ensuremath{\Varid{hylo}}:\\
\begin{minipage}[t]{.5\linewidth} 
\vspace{-\abovedisplayskip}
\begin{hscode}\SaveRestoreHook
\column{B}{@{}>{\hspre}l<{\hspost}@{}}%
\column{3}{@{}>{\hspre}l<{\hspost}@{}}%
\column{20}{@{}>{\hspre}l<{\hspost}@{}}%
\column{E}{@{}>{\hspre}l<{\hspost}@{}}%
\>[B]{}\mathbf{type}\;\Conid{List}\;\Varid{a}\mathrel{=}\mu\;(\Conid{ListF}\;\Varid{a}){}\<[E]%
\\[\blanklineskip]%
\>[B]{}\Varid{sum}\mathbin{::}\Conid{List}\;\Conid{Integer}\to \Conid{Integer}{}\<[E]%
\\
\>[B]{}\Varid{sum}\mathrel{=}\Varid{hylo}\;\Varid{plus}\;\Varid{in}^\circ\;\mathbf{where}{}\<[E]%
\\
\>[B]{}\hsindent{3}{}\<[3]%
\>[3]{}\Varid{plus}\;\Conid{Nil}{}\<[20]%
\>[20]{}\mathrel{=}\mathrm{0}{}\<[E]%
\\
\>[B]{}\hsindent{3}{}\<[3]%
\>[3]{}\Varid{plus}\;(\Conid{Cons}\;\Varid{a}\;\Varid{b}){}\<[20]%
\>[20]{}\mathrel{=}\Varid{a}\mathbin{+}\Varid{b}{}\<[E]%
\ColumnHook
\end{hscode}\resethooks
\end{minipage}%
\begin{minipage}[t]{.5\linewidth} 
\begin{hscode}\SaveRestoreHook
\column{B}{@{}>{\hspre}l<{\hspost}@{}}%
\column{3}{@{}>{\hspre}l<{\hspost}@{}}%
\column{20}{@{}>{\hspre}l<{\hspost}@{}}%
\column{E}{@{}>{\hspre}l<{\hspost}@{}}%
\>[B]{}\Varid{map}\mathbin{::}(\Varid{a}\to \Varid{b})\to \Conid{List}\;\Varid{a}\to \Conid{List}\;\Varid{b}{}\<[E]%
\\
\>[B]{}\Varid{map}\;\Varid{f}\mathrel{=}\Varid{hylo}\;\Varid{app}\;\Varid{in}^\circ\;\mathbf{where}{}\<[E]%
\\
\>[B]{}\hsindent{3}{}\<[3]%
\>[3]{}\Varid{app}\;\Conid{Nil}{}\<[20]%
\>[20]{}\mathrel{=}\Conid{In}\;\Conid{Nil}{}\<[E]%
\\
\>[B]{}\hsindent{3}{}\<[3]%
\>[3]{}\Varid{app}\;(\Conid{Cons}\;\Varid{a}\;\Varid{bs}){}\<[20]%
\>[20]{}\mathrel{=}\Conid{In}\;(\Conid{Cons}\;(\Varid{f}\;\Varid{a})\;\Varid{bs}){}\<[E]%
\ColumnHook
\end{hscode}\resethooks
\end{minipage}
Let us prove \ensuremath{\Varid{sum}\hsdot{\circ }{.}\Varid{map}\;\Varid{f}\mathrel{=}\Varid{f}\hsdot{\circ }{.}\Varid{sum}} with the properties of \ensuremath{\Varid{hylo}}.

\begin{proof}
Both \ensuremath{\Varid{sum}\hsdot{\circ }{.}\Varid{map}\;\Varid{f}} and \ensuremath{\Varid{f}\hsdot{\circ }{.}\Varid{sum}} are in the form of a function after a hylomorphism, and thus we can try to use the fusion law to establish
\[\ensuremath{\Varid{sum}\hsdot{\circ }{.}\Varid{map}\;\Varid{f}\mathrel{=}\Varid{hylo}\;\Varid{g}\;\Varid{in}^\circ\mathrel{=}\Varid{f}\hsdot{\circ }{.}\Varid{sum}}\]
for some \ensuremath{\Varid{g}}.
The correct choice of \ensuremath{\Varid{g}} is
\begin{hscode}\SaveRestoreHook
\column{B}{@{}>{\hspre}l<{\hspost}@{}}%
\column{15}{@{}>{\hspre}l<{\hspost}@{}}%
\column{E}{@{}>{\hspre}l<{\hspost}@{}}%
\>[B]{}\Varid{g}\mathbin{::}\Conid{ListF}\;\Conid{Integer}\to \Conid{Integer}{}\<[E]%
\\
\>[B]{}\Varid{g}\;\Conid{Nil}{}\<[15]%
\>[15]{}\mathrel{=}\Varid{f}\;\mathrm{0}{}\<[E]%
\\
\>[B]{}\Varid{g}\;(\Conid{Cons}\;\Varid{x}\;\Varid{y}){}\<[15]%
\>[15]{}\mathrel{=}\Varid{f}\;\Varid{x}\mathbin{+}\Varid{y}{}\<[E]%
\ColumnHook
\end{hscode}\resethooks
First, \ensuremath{\Varid{sum}\hsdot{\circ }{.}\Varid{map}\;\Varid{f}\mathrel{=}\Varid{sum}\hsdot{\circ }{.}\Varid{hylo}\;\Varid{app}\;\Varid{in}^\circ}, and by \ref{law:fusion}, 
\[\ensuremath{\Varid{sum}\hsdot{\circ }{.}\Varid{hylo}\;\Varid{app}\;\Varid{in}^\circ\mathrel{=}\Varid{hylo}\;\Varid{g}\;\Varid{in}^\circ}\]
is implied by
\begin{equation} \label{eq:fuseCond1} 
\ensuremath{\Varid{sum}\hsdot{\circ }{.}\Varid{app}\mathrel{=}\Varid{g}\hsdot{\circ }{.}\Varid{fmap}\;\Varid{sum}}
\end{equation}
Expanding \ensuremath{\Varid{sum}} on the left-hand side, it is equivalent to
\begin{equation} \label{eq:fuseCond2}
\ensuremath{(\Varid{plus}\hsdot{\circ }{.}\Varid{fmap}\;\Varid{sum}\hsdot{\circ }{.}\Varid{in}^\circ)\hsdot{\circ }{.}\Varid{app}\mathrel{=}\Varid{g}\hsdot{\circ }{.}\Varid{fmap}\;\Varid{sum}}
\end{equation}
which is an equation of functions
\[\ensuremath{\Conid{ListF}\;\Conid{Integer}\;(\mu\;(\Conid{ListF}\;\Conid{Integer}))\to \Conid{Integer}}\]
and it can be shown by a case analysis on the input.
For \ensuremath{\Conid{Nil}}, the left-hand side of (\ref{eq:fuseCond2}) equals to 
\begin{align*}
    & \ensuremath{\Varid{plus}\;(\Varid{fmap}\;\Varid{sum}\;(\Varid{in}^\circ\;(\Varid{app}\;\Conid{Nil})))}  \\
 =\  & \ensuremath{\Varid{plus}\;(\Varid{fmap}\;\Varid{sum}\;(\Varid{in}^\circ\;(\Conid{In}\;\Conid{Nil})))}  \\
 =\  & \ensuremath{\Varid{plus}\;(\Varid{fmap}\;\Varid{sum}\;\Conid{Nil})}              \\
 =\  & \ensuremath{\Varid{plus}\;\Conid{Nil}}                            && \text{(by definition of \ensuremath{\Varid{fmap}} for \ensuremath{\Conid{ListF}})} \\
 =\  & \ensuremath{\mathrm{0}}
\end{align*}
and the right-hand side of (\ref{eq:fuseCond2}) equals to
\begin{align*}
    \ensuremath{\Varid{g}\;(\Varid{fmap}\;\Varid{sum}\;\Conid{Nil})} = \ensuremath{\Varid{g}\;\Conid{Nil}} = \ensuremath{\Varid{g}\;\mathrm{0}} = \ensuremath{\Varid{f}\;\mathrm{0}}
\end{align*}
and by assumption (\ref{eq:f}) about \ensuremath{\Varid{f}}, \ensuremath{\Varid{f}\;\mathrm{0}\mathrel{=}\mathrm{0}}.
Similarly when the input is \ensuremath{\Conid{Cons}\;\Varid{a}\;\Varid{b}}, we can calculate that both sides equal to \ensuremath{\Varid{f}\;\Varid{a}\mathbin{+}\Varid{sum}\;\Varid{b}}.
Thus we have shown (\ref{eq:fuseCond1}), and therefore \ensuremath{\Varid{sum}\hsdot{\circ }{.}\Varid{hylo}\;\Varid{app}\;\Varid{in}^\circ\mathrel{=}\Varid{hylo}\;\Varid{g}\;\Varid{in}^\circ}.

Similarly, by \ref{law:fusion}, \ensuremath{\Varid{f}\hsdot{\circ }{.}\Varid{sum}\mathrel{=}\Varid{hylo}\;\Varid{g}\;\Varid{in}^\circ} is implied by
\[\ensuremath{\Varid{f}\hsdot{\circ }{.}\Varid{plus}\mathrel{=}\Varid{g}\hsdot{\circ }{.}\Varid{fmap}\;\Varid{f}}\]
which can be verified by case analysis on the input:
When the input is \ensuremath{\Conid{Nil}}, both sides equal to \ensuremath{\Varid{f}\;\mathrm{0}}.
When the input is \ensuremath{\Conid{Cons}\;\Varid{a}\;\Varid{b}}, the left-hand side equals to \ensuremath{\Varid{f}\;(\Varid{a}\mathbin{+}\Varid{b})} and the right-hand side is \ensuremath{\Varid{f}\;\Varid{a}\mathbin{+}\Varid{f}\;\Varid{b}}.
By assumption (\ref{eq:f}) on \ensuremath{\Varid{f}}, \ensuremath{\Varid{f}\;(\Varid{a}\mathbin{+}\Varid{b})\mathrel{=}\Varid{f}\;\Varid{a}\mathbin{+}\Varid{f}\;\Varid{b}}.
\end{proof}
\end{example}

\section{Closing Remarks and Further Reading}
\label{sec:more}

We have shown a handful of structural recursion schemes and their applications
by examples.
We hope that this paper can be an accessible introduction to this subject and a
quick reference when functional programmers hear about some \emph{morphism}
with an obscure Greek prefix.
We end this paper with some remarks on general approaches to find more
\emph{fantastic morphisms} and some pointers to further reading about the theory
and applications of recursion schemes.

\paragraph{From Categories and Adjunctions}
As we have seen, recursion schemes live with categories and adjunctions, so
whenever we see a new category, it is a good idea to think about catamorphisms
and anamorphisms in this category, as we did for the Kleisli category, where we
obtained \ensuremath{\Varid{mcata}}, and the functor category $\ensuremath{\mathbin{*}}^{\disccat{\ensuremath{\mathbin{*}}}}$, where we
obtained \ensuremath{\Varid{icata}}, etc.
Also, whenever we encounter an adjunction $L \dashv R$, we can think about if
functions of type \ensuremath{\Conid{L}\;\Varid{c}\to \Varid{a}}, especially \ensuremath{\Conid{L}\;(\mu\;\Varid{f})\to \Varid{a}}, are anything
interesting.
If they are, there might be interesting conjugate hylomorphisms from this adjunction.

\paragraph{Composing Recursion Schemes}
Up to now we have considered recursion schemes in isolation, each of which
provides an extra functionality compared with \ensuremath{\Varid{cata}} or \ensuremath{\Varid{ana}}, such as mutual
recursion, accessing the original structure, accessing the computation history.
However, when writing larger programs in practice, we probably want to combine
the functionalities of recursion schemes.
For example, if we want to define two mutually recursive functions with
historical information, we need a recursion scheme of type
\begin{hscode}\SaveRestoreHook
\column{B}{@{}>{\hspre}l<{\hspost}@{}}%
\column{24}{@{}>{\hspre}l<{\hspost}@{}}%
\column{E}{@{}>{\hspre}l<{\hspost}@{}}%
\>[B]{}\Varid{mutuHist}\mathbin{::}\Conid{Functor}\;\Varid{f}{}\<[24]%
\>[24]{}\Rightarrow (\Varid{f}\;(\Conid{Cofree}\;\Varid{f}\;(\Varid{a},\Varid{b}))\to \Varid{a}){}\<[E]%
\\
\>[24]{}\to (\Varid{f}\;(\Conid{Cofree}\;\Varid{f}\;(\Varid{a},\Varid{b}))\to \Varid{b})\to (\mu\;\Varid{f}\to \Varid{a},\mu\;\Varid{f}\to \Varid{b}){}\<[E]%
\ColumnHook
\end{hscode}\resethooks
Theoretically, \ensuremath{\Varid{mutuHist}} is the composite of \ensuremath{\Varid{mutu}} and \ensuremath{\Varid{accu}} in the sense
that the adjunction $\textit{U} \dashv \textit{Cofree}_F$ underlying \ensuremath{\Varid{hist}} and
the adjunction $\Delta \dashv \times$ underlying \ensuremath{\Varid{mutu}} can be composed to an
adjunction inducing \ensuremath{\Varid{mutuHist}}~\cite{Hin12Adj}.
Unfortunately, our Haskell implementations of \ensuremath{\Varid{mutu}} and \ensuremath{\Varid{hist}} are not
composable.
A composable library of recursion schemes in Haskell would require considerable
machinery for doing category theory in Haskell, and how to do it with good
usability is a question worth exploring.

\paragraph{Further Reading}
The examples in this paper are fairly small ones, but recursion schemes are surely
useful in real-world programs and algorithms. 
For the reader who wants to see recursion schemes in real-world algorithms, we
recommend books by Bird \cite{Bir10Pea} and Bird and Gibbons \cite{adwh}.
Their books provide a great deal of examples of proving correctness of
algorithms using properties of recursion schemes, which we only briefly
showcased in \autoref{sec:calc}.

We have only glossed over the category theory of the unifying theories of
recursion schemes.
For the reader interested in them, a good place to start is
Hinze's lecture notes \cite{Hinze2012LN} on adjoint folds and unfolds, and then
Uustalu et al.'s paper~\cite{UVP01Rec} on recursion schemes from comonads, which are less
general than adjoint folds, but they have generic implementations in Haskell
\cite{recursionSchemes}. 
Finally, Hinze et al.'s conjugate hylomorphisms \cite{HWG15Conj} are the most general
framework of recursion schemes so far.

\subsubsection*{Acknowledgements}
Particular thanks are due to Jeremy Gibbons for his numerous suggestions and
comments. We would also like to thank the reviewers for the efforts in helping
us to improve this paper.

\medskip

We dedicate this paper to the memory of Richard Bird.

\bibliography{references}

\label{lastpage01}
\end{document}